\documentclass[mnsc]{informs4}
\pdfoutput=1
\RequirePackage{tgtermes}
\RequirePackage{newtxtext}
\RequirePackage{newtxmath}
\RequirePackage{bm}
\RequirePackage{endnotes}

\OneAndAHalfSpacedXII 

\usepackage{algorithm}
\usepackage{algpseudocode}
\usepackage{tikz}
\usepackage{natbib}
 \bibpunct[, ]{(}{)}{,}{a}{}{,}%
 \def\bibfont{\small}%

\EquationsNumberedThrough    

\TheoremsNumberedThrough     
\ECRepeatTheorems  %

\MANUSCRIPTNO{MNSC-0001-2024.00}

\usepackage{amssymb}
\usepackage{bm,bbm}
\usepackage{dsfont}
\usepackage{mathtools}
\usepackage{enumerate}
\usepackage{amsmath}
\usepackage{comment}
\usepackage{caption}[size=normalsize]
\usepackage{subcaption}
\usepackage{multirow}
\usepackage{hyperref}
\usepackage{booktabs}

\newcommand{\lrp}[1]{\left(#1\right)}

\newcommand{\lrc}[1]{\left[#1\right]}
\newcommand{\lrl}[1]{\left\{#1\right\}}
\newcommand{\lra}[1]{\left|#1\right|}

\makeatletter
\def\BState{\State\hskip-\ALG@thistlm}
\makeatother

\def\ZZ{{\mathbb Z}}

\def\C{{\mathcal C}}

\newcommand{\cP}{\mathcal{P}}
\newcommand{\cQ}{\mathcal{Q}}
\newcommand{\cR}{\mathcal{R}}
\newcommand{\cF}{\mathcal{F}}

\newcommand{\cV}{\mathcal{V}}

\def\ZZ{{\mathbb Z}}

\definecolor{mygreen}{RGB}{0,128,0}

\definecolor{mysilver}{RGB}{220,220,220}

\newcommand{\indicator}[1]{1_{\lrl{#1}}}

\newcommand{\students}{\mathcal{S}}

\newcommand{\schools}{\mathcal{C}}
\newcommand{\levels}{\mathcal{L}}
\newcommand{\enrollees}{\bar{\mathcal{S}}}
\newcommand{\pairs}{\mathcal{V}}
\newcommand{\groups}{\mathcal{G}}
\newcommand{\lotteries}{\bm{p}}
\newcommand{\families}{\mathcal{F}}
\renewcommand{\theARTICLETOP}{}

\begin{document}

\RUNAUTHOR{Rios et al.}
\RUNTITLE{Stable Matching with Contingent Priorities}
\TITLE{Stable Matching with Contingent Priorities}

\ARTICLEAUTHORS{
\AUTHOR{Ignacio Rios}
\AFF{School of Management,
The University of Texas at Dallas, 
\EMAIL{ignacio.riosuribe@utdallas.edu}}


\AUTHOR{Federico Bobbio, Margarida Carvalho}
\AFF{CERC, CIRRELT and DIRO, Université de Montréal,
\EMAIL{federico.bobbio@umontreal.ca}, \EMAIL{carvalho@iro.umontreal.ca}}

\AUTHOR{Alfredo Torrico}
\AFF{CDSES, Cornell University,
\EMAIL{alfredo.torrico@cornell.edu}}
}

\ABSTRACT{%
Using school choice as a motivating example, we introduce a stylized model of a many-to-one matching market where the clearinghouse aims to implement contingent priorities, i.e., priorities that depend on the current assignment, to prioritize students with siblings and match them together. We provide a series of guidelines and introduce two natural approaches to implement them: (i) absolute, whereby a prioritized student can displace any student without siblings assigned to the school, and (ii) partial, whereby prioritized students can only displace students that have a less favorable lottery than their priority provider. We study several properties of the corresponding mechanisms, including the existence of a stable assignment under \emph{contingent} priorities, the complexity of finding one if it exists, and its incentive properties. Furthermore, we introduce a soft version of these priorities to guarantee existence, and we provide mathematical programming formulations to find such stable matching or certify that one does not exist. Finally, using data from the Chilean school choice system, we show that our framework can significantly increase the number of students assigned to their top preference and the number of siblings assigned together relative to current practice.  
}%




\maketitle

\section{Introduction}\label{sec: introduction}

The theory of two-sided many-to-one matching markets, introduced by~\citet{gale1962college}, provides a framework for solving many large-scale real-life matching problems. Examples include entry-level labor markets for doctors and teachers, education markets ranging from daycare to college admissions, and other applications such as refugee resettlement. 

In many of these markets, the clearinghouse may be interested in finding a stable allocation to guarantee that no coalition of agents has incentives to circumvent the match, while individual agents may care about their matching and that of other agents. For instance, in the hospital-resident problem, couples jointly participate and must coordinate to find two positions that complement each other. In refugee resettlement, agencies may prioritize allocating families with similar backgrounds (e.g., from the same region or speaking the same language) to the same cities. In our primary motivating example, school choice, students may prefer to be assigned with their siblings.

A common approach to accommodate these joint preferences is to provide priorities, such as sibling priorities in school choice, increasing the chances of specific agents being matched together.\footnote{In refugee resettlement, families may get higher priority in localities where they have relatives based on \emph{family reunification}. This type of priority does not exist in the residency matching problems, as couples must participate together to be considered as such, and candidates do not receive priority (at least explicitly) if their partner already works at a given hospital.} 
However, most clearinghouses assume that priorities are fixed and known before the matching process and thus cannot accommodate settings in which priorities depend on the current matching.
For instance, Boston Public Schools (BPS) prioritizes students who have a sibling currently enrolled for the next academic year (most clearinghouses know this by the time they perform the allocation) but provides no special treatment for families involving multiple applicants (e.g., twins, triplets, or siblings applying to different levels) participating in the system. As a result, many families end up being separated, resulting in higher transportation costs, emotional distress, and logistical challenges for parents, among other undesirable outcomes. To tackle this issue, some school districts have introduced special rules for multiples, whereby they try to accommodate them in the same school provided some requirements (e.g., both siblings must submit the same preference list, they must apply to the same level/program, among others).\footnote{New York City (NYC) considers a special treatment of multiples in 6th grade (entry level of middle school) starting from the 2022-2023 school year, and it also considers multiples for 3-K and Pre-K (see 
\href{https://www.chalkbeat.org/newyork/2022/1/12/22880099/nyc-middle-school-admissions-twins-sibling-priority}{link} 
for more details). 
New Orleans (NOLA) uses a unique placement process for multiples, i.e., it is not part of their matching mechanism, and they solve it ``manually''. Wake County Public Schools (WCPS) goes one step further and only deems feasible those matchings where multiples are assigned to the same school.}
Other clearinghouses, such as the Chilean school choice system, provide sibling priority to applicants if they have a sibling (i) enrolled in the school for the next year or (ii) concurrently participating in the admissions process and assigned to a higher level;  nevertheless, they do not consider special treatment of multiples nor flexibility in the direction of priorities.
Hence, none of the practical approaches mentioned above entirely solves the problem. Moreover, from a theoretical standpoint, most definitions of stability and justified-envy assume that priorities are fixed and known, and there are no guidelines for how to account for priorities that depend on the matching or their potential consequences. Thus, the theory of stable matching also fails to capture and provide solutions to these settings.

In this paper, our primary goals are (i) to provide a conceptual framework to incorporate \emph{contingent} priorities, i.e., priorities that depend on the current matching, and (ii) to design methodologies to find rank-optimal allocations (i.e., that optimize the rankings of the students assigned) while accommodating these priorities. 
To accomplish this, we introduce a stylized model of a many-to-one matching market where students belong to (potentially different) levels and may have siblings concurrently applying to the system (potentially in different levels). 
Each student has preferences over schools, while schools prioritize students by combining group memberships and random tie-breakers. Importantly, we assume that schools seek to prioritize students with siblings assigned to the school, aiming to increase the number of siblings attending the same institution. 
The ultimate goal of the clearinghouse is to achieve a rank-optimal stable matching incorporating these contingent priorities.

\subsection{Contributions}
Our work makes several contributions that we now describe in detail.

\paragraph{Framework.} Our primary contribution is the introduction and formalization of contingent priorities. To accomplish that, we provide guidelines that delimit the implementation of contingent priorities to prevent undesirable outcomes. Namely, we assume that the clearinghouse breaks ties within each group using random tie-breakers, that providers of priority must be able to rightfully claim a seat without being prioritized, and that contingent priorities take one of two forms: (i) \emph{absolute}, whereby a prioritized applicant can displace any other student with no siblings assigned to the school; and (ii) \emph{partial}, whereby a prioritized applicant can only displace another with no siblings if the tie-breaker of the sibling providing them with the priority is better than that of the displaced student. Finally, we define the notions of contingent justified-envy, contingent stability, and we discuss how these priorities can be incorporated as \emph{hard} and \emph{soft} requirements.

\paragraph{Properties.} We analyze several properties of the mechanism to find a 
stable matching for each variant of contingent priorities. First, we show that a stable matching with absolute priorities may not exist and that {the decision problem of whether there exists such a matching is NP-complete.}  We find a similar result for the partial case. However, when combined with lotteries at the family level,\footnote{Under family lotteries, each student has two random tie-breakers for each school: one shared with their siblings and one individual. The order of students within the same group at each school is determined lexicographically using the family tie-breaker first, followed by the individual tie-breaker. This tie-breaking rule is equivalent to using a single tie-breaker per student per school obtained from adding sufficiently small perturbations to the family tie-breakers (to preserve the order of families). Conversely, under individual lotteries, each student only has their individual tie-breaker, which is independent of that of their siblings.} the problem is equivalent to having no contingent priorities so we can always find a stable matching with partial priorities in polynomial time.
{Then, we study each mechanism's incentive properties. For absolute, we show that the mechanism to find a rank-optimal matching is not strategy-proof for families under any tie-breaking rule. For partial, we show that this mechanism is not strategy-proof for families under individual lotteries but is strategy-proof under family lotteries. Lastly, we show that the mechanism to find a contingent stable matching, under either type of priority, is strategy-proof in the large.}

\paragraph{Formulations.} We provide mathematical programming formulations that enable us to either find a stable matching for each type of contingent priority (i.e., absolute/partial and hard/soft) or show infeasibility. Moreover, our formulations are flexible enough to accommodate several practical concerns, including hard vs. soft priorities, static priorities and secured enrollment for students currently enrolled but looking to transfer to another school, among others. Finally, we introduce a novel mathematical programming formulation modeling the problem of finding a stable matching that maximizes the number of siblings assigned together under the standard notion of stability (i.e., without contingent priorities).

\paragraph{Impact.} To illustrate the benefits of our framework, we apply it to the Chilean school choice system and compare its results against sensitive benchmarks, including the student-optimal stable matching under the standard concept of stability and the mechanism currently used in Chile to perform the allocation. 
First, we empirically show that enforcing absolute priorities as a hard requirement increases the number of students assigned to their top choice (particularly among students with siblings) and the number of siblings assigned together compared to all benchmarks. However, this approach faces the potential drawback of the non-existence of a stable matching. In contrast, its soft version ensures the existence of a stable matching but offers limited benefits compared to current practice. Notably, a hybrid approach that combines soft priorities with minimum requirements on the number of effective providers effectively balances existence and enhancing sibling assignments.
Second, we demonstrate that the standard notion of stability is inadequate for increasing the number of siblings placed together, as the differences between the stable matching that maximizes sibling placement and the student-optimal matching are negligible. 
Finally, we analyze the sensitivity of the results to various tie-breaking rules and find that different combinations of contingent priorities and tie-breaking rules lead to significantly different outcomes. Thus, policymakers should carefully choose the right combination depending on their goals.

\subsection{Organization}
The remainder of this paper is organized as follows. In Section~\ref{sec: literature}, we discuss the relevant literature. In Section~\ref{sec: model}, we introduce our model. In Section~\ref{sec: properties}, we discuss several properties of the mechanisms to find stable matchings under contingent priorities. In Section~\ref{sec: formulations}, we provide mathematical programming formulations to identify them. In Section~\ref{sec: application}, we illustrate the potential benefits of our framework using data from the Chilean school choice system. Finally, in Section~\ref{sec: conclusions}, we conclude.

\section{Literature}\label{sec: literature}

Our paper is related to several strands of the literature.

\paragraph{Matching with families.} A recent strand of the literature has extended the classic school choice model~\citep{Abdulkadiroglu2003} to incorporate families. \citet{Dur_2022} consider a setting where siblings report the same preferences, and assignments are feasible if and only if all family members are assigned to the same school (or all of them are unassigned). The authors argue that justified envy is not an adequate criterion for the problem. Thus, they propose a new solution concept (suitability), show that a suitable matching always exists, and introduce a new class of strategy-proof mechanisms that finds a suitable matching. \citet{Correa_2022} also consider a model with siblings applying to potentially different levels, but assume that each sibling submits their own (potentially different) preference list. In addition, the authors assume that the clearinghouse aims to prioritize the joint assignment of siblings, but they model it as a soft requirement, i.e., an assignment may be feasible even if siblings are not assigned to the same school. To prioritize the joint assignment of siblings, \citet{Correa_2022} introduce (i) the use of lotteries at the family level; (ii) a heuristic that processes levels sequentially in decreasing order, updating priorities in each step to capture siblings' priorities that result from the assignment of higher levels; and (iii) the option for families to report that they prefer their siblings to be assigned to the same school rather than following their individual reported preferences. This last feature, called \emph{family application}, prioritizes the joint assignment of siblings by updating the preferences of younger siblings by adding the school of assignment of their older siblings. The authors show that all these features significantly increase the probability that families get assigned together. {\citet{sun2023daycare} define rationality and stability in a daycare matching model with siblings. In their model, already matched siblings may seek transfers to more preferred daycares, and families can express joint preference lists. Although stable matchings are not guaranteed to exist, their experiments on real-world instances using integer programming demonstrate that a stable matching  always exists. Importantly, each level at each daycare is treated as an independent daycare that only admits children of that level.}

\paragraph{Matching with couples.} Our paper is also related to the matching with couples literature, which is commonly motivated by labor markets such as the matching for medical residents. In this setting, couples wish to be matched in the same hospital and, thus, they report a joint preference list of pairs of hospitals. For an extension of the stability concept with couples, \citet{roth1984evolution} shows that a stable matching may not exist if couples participate. Furthermore,~\cite{ronn1990} shows that determining the existence of a stable matching in this context is NP-complete, and~\citet{mcdermid2010keeping} show that the problem remains NP-complete even if the members of each couple report their preferences individually in a consistent way. To overcome the existence limitation, \citet{klaus2005stable} introduce the property of weak responsive preferences and show that this guarantees the existence of a stable assignment.~\citet{kojima13} provide conditions under which a stable matching exists with high probability in large markets, and introduce an algorithm that finds a stable matching with high probability which is approximately strategy-proof.~\citet{Ashlagi_2014_couples} find a similar result, as they show that a stable matching exists with high probability if the number of couples grows slower than the size of the market. However, the authors also show that a stable matching may not exist if the number of couples grows linearly. Finally,~\citet{nguyen2018near} show that the existence of a stable matching is guaranteed if the capacity of the market is expanded by at most a fixed number of spots. {For more details on the literature of matching with couples, we refer to~\citep{biro2013matching}.}

\paragraph{Matching with complementarities.} Beyond families and couples, the matching literature has studied other settings with complementarities. For instance, \citet{Ashlagi_2014} show that correlating lotteries can increase community cohesion by increasing the probability of neighbors going to the same schools. \citet{Dur_2019} also study the matching problem with neighbors and show that a stable matching may not exist if students have preferences over joint assignments with their neighbors. Moreover, the authors show that the student-proposing deferred acceptance algorithm is not strategy-proof and propose a new algorithm to address these issues. \citet{kamada15} study matching markets where the clearinghouse cares about the composition of the match and, thus, imposes distributional constraints. The authors show that existing mechanisms suffer from inefficiency and instability and propose a mechanism that addresses these issues while respecting the distributional constraints. \citet{nguyen19} also study the problem with distributional concerns but consider these constraints as soft bounds and provide ex-post guarantees on how close the constraints are satisfied while preserving stability. \citet{nguyen21} introduce a new model of many-to-one matching where agents with multi-unit demand maximize a cardinal linear objective subject to multidimensional knapsack constraints, capturing settings such as refugee resettlement, day-care matching, and school choice/college admissions with diversity concerns. The authors show that a pairwise stable matching may not exist and provide a new algorithm that finds a group-stable matching that approximately satisfies all the multidimensional knapsack constraints. Finally, motivated by labor markets, \citet{dooley2020affiliate} and \citet{knittel2022dichotomous} study the ``affiliate matching problem'', in which firms (universities) have preferences over the applicants for their positions but also over the placement of their own workers (job-market candidates).

\section{Model}\label{sec: model}
\allowdisplaybreaks

We now introduce a two-sided matching market model that includes a priority system. To facilitate the exposition, we use school choice with sibling priorities as a concrete application of the model.

\paragraph{Students.} Let \(\students\) be a finite set of students and
$\cF\subseteq 2^{\students}$ be a partition of $\students$ where $f\in\cF$ is called a \emph{family} and its size is denoted as $|f|$. Given two distinct students, $s$ and $s'$, we say that they are \emph{siblings} if there is $f \in \cF$ such that $s,s' \in f$. If $f\in\cF$ is such that $f=\{s\}$, then we say that $s$ has no siblings.
With a slight abuse of notation, we define the function \(f:\students\to \cF\) that maps a student into their specific family, i.e., each student \(s\in\students\) belongs to family \(f(s)\in\cF\). Note that students \(s\) and \(s'\) are siblings if \(f(s) = f(s')\).

\paragraph{Levels.} Let \(\levels\) be a finite set of levels (e.g., from Pre-K to 12th grade). With a slight abuse of notation, we define the function \(\ell:\students\to\levels\) that maps each student \(s\in\students\) into the level \(\ell(s)\) they belong. Moreover, we denote by \(\students^{\ell} \subseteq \students\) the set of students that belong to level \(\ell\in \levels\); thus, the collection of sets $\lrl{\students^\ell}_{\ell\in \levels}$ also defines a partition over $\students$. 

\paragraph{Schools.} Let \(\schools\) be a finite set of schools.
We assume that each school \(c\in\schools\) offers \(q_{c}^\ell\in\ZZ_+\) seats on level \(\ell\in \levels\), where \(q_{c}^\ell = 0\) means that school \(c\) does not offer level \(\ell\). When clear from the context, we extend the set of schools to include the outside option of being unassigned, which we denote by $\emptyset$ and whose capacity in each level we assume to be sufficiently large (e.g., $q^\ell_\emptyset = \lra{\students^\ell}$) so every student can potentially be unassigned.

\paragraph{Preferences.} Let $\succ_s$ be the strict preference order of student $s$ over schools in $\schools\cup \lrl{\emptyset}$.
Then, we use (i) \(c \succ_s c'\) to represent that student $s$ \emph{strictly prefers} school $c$ over school $c'$, (ii) \(c \succeq_s c'\) to capture that $s$ \emph{weakly prefers} $c$ over $c'$, and (iii) \(\emptyset \succ_s c\) to denote that $s$ prefers to be unassigned over attending school $c$. With a slight abuse of notation, we use $\lra{\succ_s}$ to represent the number of schools that student $s$ strictly prefers over being unassigned, i.e., $\lra{\succ_s} = \lra{\lrl{c\in \schools \;:\; c \succ_c \emptyset}}$.

\paragraph{Priorities.} 
Let $\groups$ be a finite set of groups that schools use to prioritize students,\footnote{For instance, students who live in the schools' walk-zone, who have a sibling currently enrolled, who are disadvantaged, etc.} and let $g:\students \times \schools \rightarrow \groups$ be the function that maps each pair $(s, c) \in \students \times \schools$ to a group $g(s, c) \in \groups$. For simplicity, we assume $\groups = \lrl{1, \ldots, |\groups|}$, where students with lower values of $g(\cdot, c)$ are strictly preferred over those with higher values for all $c\in \schools$ and, thus, $g(s,c) = |\groups|$ corresponds to students with no priority.\footnote{The group mapping is an input of the problem and serves as a labeling system set in advance by the clearinghouse based on various student attributes. For example, low-income students or those living closer to the school may be assigned a lower group number, indicating higher priority. The simplest scenario occurs when all students belong to the same group, such as the no-priority group.}
Additionally, let $\lotteries \in \mathbb{R}_+^{\students \times \schools}$ be the vector of random tie-breakers (lotteries) that the clearinghouse uses to resolve ties among students within the same group, i.e., $p_{s, c} \in \mathbb{R}_+$ represents the random tie-breaker value for student $s$ at school $c$. As we later discuss,  we consider two different tie-breaking rules: (i) at the \emph{family level}, where all family members are initially assigned the same tie-breaker, with slight perturbations introduced afterward to break ties within the family while preserving the overall family order, and (ii) at the \emph{individual level}, where each student is independently assigned a different random tie-breaker. 
Finally, the combination of groups and random tie-breakers defines a unique \emph{priority order} $\succ_c$ for each school $c \in \schools$, which we refer to as \emph{initial}, where $s \succ_c s'$ if (i) $g(s, c) < g(s', c)$, or (ii) $g(s, c) = g(s', c)$ and $p_{s, c} < p_{s', c}$ for any $s,s'\in \students$. In the remainder of the paper, we say that a student $s$ has higher priority than $s'$ at school $c$ if $s\succ_c s'$ and, similarly, we say that groups with lower values in $\groups$ have higher priority.

\paragraph{Matching.} Let \(\pairs \subseteq \students \times (\schools\cup \lrl{\emptyset})\) be the set of feasible pairs, i.e., \((s,c) \in \pairs\) implies that $c \succeq_s \emptyset$ and \(q_c^{\ell(s)} > 0\).
A matching is an assignment \(\mu\subseteq \pairs\) such that (i) each student is assigned to at most one school in \(\schools\), and (ii) each school is assigned at most its capacity in each level. Formally, for \(\mu\subseteq \pairs\), let \(\mu(s) \in \schools\cup \lrl{\emptyset}\) be the school that student \(s\) was assigned to, \(\mu(c)\subseteq\students\) be the set of students assigned to school \(c\), and $\mu^\ell(c)$ be the set of students assigned to school $c$ in level $\ell\in\levels$. Then, a matching satisfies that (i) \(\mu(s) \in \schools\cup \lrl{\emptyset}\) for all students \(s\in \students\) and (ii) \(\lra{\mu^\ell(c)} \leq q_{c}^{\ell}\) for all schools \(c\in \schools\) and levels \(\ell\in \levels\).\footnote{Notice that the model captures other single-level applications such as refugee resettlement, college admissions and the hospital-resident problem.}

\paragraph{Stability.}
As \citet{roth02} discusses, a desirable property of any matching is stability, i.e., that there is no group of agents that prefer to circumvent their current match and be matched to each other. To distinguish it from our notion of contingent stability that we define later, we call it \emph{initial} stability. Formally, given a matching \(\mu\subseteq\pairs\), we say that student \(s\) has \emph{initial justified envy} towards another student \(s'\) assigned to school \(c\) if (i) \(\ell(s)=\ell(s')\), (ii) \(c \succ_s \mu(s)\),  and (iii) \(s \succ_c s'\). In words, the first condition states that both students belong to the same level; the second condition implies that student $s$ prefers school \(c\) rather than $\mu(s)$; and the third condition states that student $s$ has higher priority than $s'$ in school \(c\). In addition, we say that a matching \(\mu\) is \emph{non-wasteful} if there is no student \(s\in \students\) and school \(c \in \schools\) such that \(c \succ_s \mu(s)\) and \(\lra{ \mu^\ell(c)} < q_c^\ell\). Finally, we say that a matching is \emph{initially stable} if it is \emph{non-wasteful} and no student has \emph{initial justified envy}.

\subsection{Contingent Priorities}\label{sec: contingentprio}
{Many school districts strive to prioritize students based on the  matching of their siblings who are concurrently seeking admission to a new school. To accomplish this, most school districts either (i) process siblings applying to the same level manually (outside the primary matching mechanism), as in NYC and NOLA, or (ii) define an arbitrary order among levels and process them sequentially using the deferred acceptance algorithm, updating school priorities (to account for the sibling priorities arising from the levels processed thus far as an additional group) before moving to the next level, as it is the case in Chile~\citep{Correa_2022}.
These solutions face several limitations, including limited scalability, reach, and arbitrariness in prioritization (as discussed in Appendix~\ref{sec:extra_discussion}). Additionally, these solutions often have a limited impact, failing to effectively address the broader needs of the student population.

This section aims to provide a framework to incorporate \emph{contingent} priorities in a general and scalable way. To this end, we start formalizing the concept of providing \emph{contingent priority}.}

\begin{definition}[Contingent Siblings Priority]\label{def: contingent priority}
{Given a matching $\mu$, a student $s$ provides contingent priority in school $c$ if $\mu(s) = c$ and there exists a student $s'\in \students \setminus \lrl{s}$ such that (i) $f(s) = f(s')$ and (ii) $\mu(s') \preceq_{s'} c$.\footnote{Note that condition (ii) implies that $(s',c)\in\pairs$.}}
\end{definition}




{Based on this definition, a student $s$ provides contingent priority to their siblings in school $c$ if they are assigned to it in the current matching $\mu$ {and one of their siblings may benefit from it}. Note, however, that Definition~\ref{def: contingent priority} does not specify how \emph{receiving} priority affects schools' (initial) priority orders. For example, clearinghouses could implement {contingent} priorities by either (i) modifying the set of initial groups by creating an additional group that corresponds to students with contingent priority, in a similar fashion to the siblings priority implemented in Chile; or (ii) updating the random tie-breakers of the students with siblings assigned to the school, in the same spirit of admissions systems that process multiples as a batch---with all their members having the same lottery and priority---demanding more than one seat. To formalize these two approaches, we assume that the action of \emph{receiving contingent priority} in a given matching $\mu$ is captured by the \emph{contingent priority order} $\succ^\mu$, which is uniquely defined by (i) a function $g^\mu: \students \times \schools \rightarrow  \groups^\mu$ that maps each pair $(s,c) \in \students \times \schools$ to their \emph{contingent group} in $\groups^\mu $ , where $\mathbb{Z}_+\supseteq\groups^\mu \supseteq \groups$, and (ii) a vector $\lotteries^\mu \in \mathbb{R}_+^{\students \times \schools}$ of \emph{contingent tie-breakers}, both of which depend on their {initial} counterparts $g$ and $\lotteries$. Then, we use $s\succ_c^\mu s'$ to represent that $s$ has higher contingent priority than $s'$ in school $c$, which holds if either (i) $g^\mu(s,c) < g^\mu(s',c)$ or (ii) $g^\mu(s,c) = g^\mu(s',c)$ and $p_{s,c}^\mu < p_{s',c}^\mu$.}

{The inclusion of {contingent} priorities imposes several challenges. First, as shown in Example~\ref{ex: no unique priority order} in Appendix~\ref{app: additional examples}, there is no longer a unique ordering of students in schools since they may change from one priority group to the other depending on their siblings' (current) matching. Consequently, we cannot apply the notion of initial stability (as defined earlier) and, thus, use the deferred acceptance algorithm since this algorithm requires priority orders to be fixed and known. Similarly, we cannot use algorithms that permanently reject students in each iteration (as in the mechanism used by the NRMP to solve the matching with couples problem or the Immediate Acceptance algorithm) because contingent priorities may change as the matching process progresses and, thus, initially rejected students may later get prioritized. 
Second, it is technically challenging to effectively define the concept of {providing} contingent priority and, thus, operationalize the action of {receiving} it through {contingent} groups and lotteries. This will ultimately determine how to obtain a matching that successfully incorporates {contingent} priorities while at the same time leading to desirable outcomes, such as existence, fairness, and others.
Finally, multiple matchings may account for contingent priorities, so there may not be a unique one that is Pareto-dominant for students. Consequently, finding the best stable matching under contingent priorities may require defining an appropriate goal and exploring several tools to identify the best. In the coming sections, we aim to tackle each of these challenges.}


\subsection{Refining Contingent Priorities}\label{sec: refining contingent priorities}

Although finding a stable matching with the highest number of siblings assigned to the same school may be a desirable outcome, {contingent} priorities as defined in the previous section may lead to ``undesirable'' matchings, as the following example illustrates.

\begin{example}\label{ex: unlawful siblings priority}
    Consider an instance with a set of students \(\students = \lrl{s_1, s_2, s_3, f_1,f_2, f'_1, f'_2}\) where \(f=\lrl{f_1, f_2}\) and \(f'=\lrl{f'_1, f'_2}\) are siblings, and a single school $c$ with capacity 4 and a single level. Moreover, suppose that every student initially belongs to the same group, i.e., $g(s,c) = g $ for all $(s,c)\in\pairs$ and that the initial random-tie breakers in school $c$ are such that
    \(p_{s_1,c} < p_{s_2,c} < p_{s_3,c} < p_{f_1,c} < p_{f'_1,c} < p_{f_2,c} < p_{f'_2,c}.\) 
    Then, one possible matching is 
    \[\mu = \lrl{
            (s_1, c), (s_2, c), (s_3, c), 
            (f_1, c), (f_2, \emptyset), (f_1', \emptyset), (f_2', \emptyset)
        }.
    \] 
    However, the alternative matchings 
    \[
    \begin{aligned}
        \mu' &= \lrl{
            (s_1, \emptyset), (s_2, \emptyset), (s_3, \emptyset), 
            (f_1, c), (f_2,c), (f_1', c), (f_2',c)
        }\\
    \mu'' &= \lrl{
            (s_1, c), (s_2, c), (s_3, \emptyset), 
            (f_1, c), (f_2,c), (f_1', \emptyset), (f_2',\emptyset)
        }
    \end{aligned}
    \]         
    are also feasible in terms of capacity and differ in how students with siblings are prioritized. 
    
    \noindent The matching \(\mu'\) may not be desirable since neither \(f_1'\) nor \(f_2'\) would be matched to school $c$ without contingent priority, so it may be ``unfair'' to allow them to skip everyone else in the initial ordering. 
    This differs from $\mu''$ since $f_1$ would have also been matched to $c$ in the absence of contingent priority and, consequently, could potentially provide contingent priority to $f_2$. \hfill \(\square\)
\end{example}

To rule out the issue shown in Example~\ref{ex: unlawful siblings priority}, we refine who can {provide} contingent siblings priority.

\begin{definition}[Contingent Sibling Priority, Refined]\label{def: provider of contingent priority}
{Given a matching $\mu$, a student $s$ provides contingent priority in school $c$ if $\mu(s) = c$ and there exists a student $s'\in \students \setminus \lrl{s}$ such that (i) $f(s) = f(s')$, (ii) $\mu(s') \preceq_{s'} c$, and (iii) $\lra{\lrl{s'' \in \students^{\ell(s)}\;:\; s'' \succ_c s, \; c \succeq_{s''} \mu(s'') }} \leq q_c^{\ell(s)}-1.$}    
\end{definition}
\vspace{0.5em}
{The first two conditions coincide with Definition~\ref{def: contingent priority}, requiring that the students involved are siblings and that the potential receiver may benefit from the contingent priority. 
}
{The last condition states that student $s$ can rightfully claim a spot in school $c$ based on the initial priority order, as the number of students who may have initial justified-envy (i.e., who belong to the same level, 
have higher initial priority than $s$, and weakly prefer $c$ over their assignment in $\mu$) is less than the school's capacity.}
As we illustrate in Example~\ref{ex: provider of priority} in Appendix~\ref{app: additional examples}, this condition is not equivalent to requiring that $(s,c)$ belongs to some {initially stable} matching (i.e., under $\succ_{c'}$ for all $c' \in \schools$), as we allow students with siblings to potentially benefit from changes in the priority orders of other schools resulting from the contingent priorities.

{In the remainder of this paper, we use $z^\mu(s,c) = 1$ to capture that student $s$ has the most favorable tie-breaker among the members of their family who provide contingent priority in school $c$ as in Definition~\ref{def: provider of contingent priority}---in which case we say that $s$ is the \emph{effective priority provider} for their family in school $c$---, and $z^\mu(s,c) = 0$ otherwise.
As a result, note that (i) $\sum_{s\in f} z^\mu(s,c) \leq 1$ for all $f\in \cF$ and $c\in \schools$, i.e., each family has at most one {effective provider} of contingent priority in each school, and (ii) $z^\mu(s,c) = 0$ for all $s$ such that $f(s) = \lrl{s}$ {or \((s',c) \notin \pairs\) for all $s'\in f(s)\setminus \lrl{s}$}, i.e., {students with no siblings applying to school $c$ cannot provide contingent priority in that school}.} 

\noindent {\sffamily\mdseries\scshape Example~\ref{ex: unlawful siblings priority} (revisited).}
Note that \(\mu''\) and \(\mu\) satisfy Definition~\ref{def: provider of contingent priority} but differ in how a prioritized student compares to another without priority and how they implement {contingent} priorities. 
On the one hand, $\mu''$ is the resulting matching if {contingent} priorities allow prioritized students to displace any student without priority, which can be implemented by considering students with {contingent} priority as an additional group with higher priority. For instance, the matching $\mu''$ can be obtained considering $g^{\mu''}$ such that $g^{\mu''}(f_1,c) = g^{\mu''}(f_2,c) < g^{\mu''}(s,c), \; \forall s\in \students \setminus f$, enabling student $f_2$ to displace $s_3$ in school $c$.
On the other hand, $\mu$ is the resulting matching if {contingent} priorities allow prioritized students to displace others with worse random tie-breakers than their priority provider, which could be captured by the following $\lotteries^\mu$: $p^\mu_{s_1,c} < p^\mu_{s_2,c} < p^\mu_{s_3,c} < p^\mu_{f_1,c} = p^\mu_{f_2,c} < p^\mu_{f'_1,c} < p^\mu_{f'_2,c}$. With these lotteries, we observe that $f_1$'s tie-breaker does not help $f_2$ to displace $s_1$, $s_2$ and $s_3$.  \hfill\Halmos

Both approaches discussed above differ in how much they prioritize the joint assignment of siblings relative to sticking to the {initial} priority order $\succ_c$ given by the initial groups and random tie-breakers. 
The first approach, which we refer to as \emph{absolute} contingent priority, provides full priority to students with siblings, as they can displace any student without priority. 
The second approach, which we refer to as \emph{partial} contingent priority, provides moderate priority to students with siblings, as they can only displace students with worse random tie-breakers than their priority provider. 
To account for these two cases and provide a flexible framework, we introduce two types of \emph{contingent} priority in Definitions~\ref{def: absolute contingent priority} and~\ref{def: partial contingent priority}.

\begin{definition}[Absolute Contingent Priority]\label{def: absolute contingent priority}
    Contingent priorities are \emph{absolute} if, given a matching $\mu$, contingent lotteries and groups are such that, for any $(s,c) \in \pairs$, \(p^\mu_{s,c} = p_{s,c}\) and 
{    \begin{equation}\label{eq: groups under absolute priorities}
        g^\mu(s,c) = 
        \begin{cases}
            g(s,c) & { \text{if }} g(s,c) < \lra{\groups} \\
            g(s,c) & { \text{if }} \lrp{z^\mu(s,c) = 1 \text{ and } \lra{ f(s)\cap \mu(c)} \geq 2} \\
            & \quad { \text{ or }} \lrp{\exists s' \in f(s)\setminus \lrl{s} \text{ with } z^\mu(s',c) = 1} \\
            \lra{\groups} + 1 & \text{otherwise.}
        \end{cases}       
    \end{equation}  }  
\end{definition}


\begin{definition}[Partial Contingent Priority]\label{def: partial contingent priority}
    Contingent priorities are \emph{partial} if, given a matching $\mu$, contingent lotteries and groups are such that, for each $(s,c) \in \pairs$, $g^\mu(s,c) = g(s,c)$ and 
    \begin{equation}\label{eq: lotteries under partial priorities}
         p^\mu_{s,c} = \begin{cases}
             p_{s,c} & \text{if } g(s,c) < \lra{\groups} \\
             p_{s,c} - \epsilon & \text{if } g(s,c) = \lra{\groups} \text{ and } z^\mu(s,c) = 1 \\
             \min\lrl{p_{s',c} \,:\, {s'\in f(s)\text{ with } z^\mu(s',c) = 1} \;\text{or}\; s'=s} & \text{if } g(s,c) = \lra{\groups} \text{ and } z^\mu(s,c) = 0,\\
         \end{cases}
    \end{equation}
    where $\epsilon$ is sufficiently small so the contingent tie-breaker of an effective priority provider does not break the initial order. Moreover, if two or more siblings receive the same contingent tie-breaker at school $c$ due to an effective priority provider $s$, we resolve ties by assigning different contingent tie-breakers within the interval $(p_{s,c}-\epsilon, p_{s,c}]$, ensuring that their initial order is preserved.\footnote{For instance, $0 < \epsilon < \min\lrl{\lra{p_{s,c}-p_{a,c'}}: (s,c), (a,c')\in \pairs}$. Then, if $s',s''\in f(s)\setminus\{s\}$ and $p_{s',c}<p_{s'',c}$, we set  $p^\mu_{s',c}=p_{s,c}-\epsilon'$ and $p^\mu_{s'',c}=p_{s,c}-\epsilon''$ such that $\epsilon'>\epsilon''$ and  
    $\epsilon',\epsilon''\in[0,\epsilon)$. Same logic applies for three or more siblings.}
\end{definition}

{In both definitions, students who initially belong to a higher priority group (i.e., $g(s,c) < \lra{\groups}$) retain their group and random tie-breaker. Consequently, students from these higher priority groups do not gain additional benefits from contingent priorities, as is common in practice.\footnote{In most school choice systems, students who belong to multiple groups (e.g., walk-zone and siblings priority) are not additionally prioritized and are treated as belonging to the higher priority group only.}

The two definitions differ in their treatment of priority recipients and students without priority, capturing two ``extremes''.
Under {absolute} priorities, groups are updated while maintaining the initial lotteries so that students who receive contingent priority (and who have initially no priority, i.e., $g(s,c) = \lra{\groups}$) stay in their initial group with their initial lottery, while students without siblings or initial priority (i.e., $g(s,c)=|\groups|$) move to a new group with the lowest priority (i.e., $g^\mu(s,c)=|\groups|+1$). {Note that effective providers of priority only get prioritized if they have a sibling assigned to the school; otherwise, they are considered as a single student, and thus, they have no priority.}
Conversely, under {partial} priorities, groups remain unchanged while lotteries are updated so that priority recipients inherit the initial lottery of their effective priority provider (if beneficial), and students without priority retain their initial lotteries.
As a result, a {contingently} prioritized student can displace any other student without priority under {absolute} priorities (as in the Chilean school choice system). In contrast, under partial priorities, they can only displace students in their group with a worse random tie-breaker than theirs or their effective provider (if any). 
}

We emphasize that Definitions~\ref{def: absolute contingent priority} and~\ref{def: partial contingent priority} imply that students who receive {contingent} priority cannot displace students from other groups (e.g., students with \emph{secured enrollment} or \emph{static} siblings priorities, as we discuss in Appendix~\ref{app: additional groups}) which is determined by the initial group mapping $g$. However, this can be easily adjusted if the clearinghouse intends to allow students with {contingent} priority to displace those from other groups.  Finally, observe that the updated groups and lotteries resulting from a matching $\mu$ under both types of contingent priority induce a unique strict order $\succ_c^\mu$ for each school $c\in \schools$. As a result, we can use the {contingent} priority orders to define the concepts of \emph{contingent justified-envy} and \emph{contingent stability}.

\begin{definition}[Contingent Justified-Envy and Stability]\label{def: contingent justified envy and stability}
    Given a matching $\mu$, a student $s$ has \emph{contingent justified-envy} towards another student $s'$ assigned to school $c$ if (i) $\ell(s) = \ell(s')$, (ii) $c \succ_s \mu(s)$, and (iii) $s \succ_c^\mu s'$. A matching is \emph{contingent stable} if it is non-wasteful and if no student has \emph{contingent justified-envy}. 
\end{definition}



\begin{remark}
Our definition of contingent stability (in particular, the absolute case) significantly broadens the set of stable matchings to capture more complex scenarios involving multiple levels or sibling applicants. In Appendix~\ref{appendix:stability_comparisons}, we provide a series of instances illustrating the differences between contingent stability and other notions of stability in the literature of matching with couples.
\end{remark}

\section{Properties}\label{sec: properties}

    In this section, we discuss several properties of the proposed mechanism, including (i) the (un)existence of contingent stable matchings and the computational complexity of determining whether one exists, (ii) the non-existence of a student-optimal contingent stable matching, and (iii) the incentive properties of the mechanism. {Furthermore, we discuss how these properties change when considering a ``softer'' notion of contingent priorities, whereby the clearinghouse can decide whether or not it incorporates the contingent priorities given by some providers. } 
    We defer all the proofs to the Appendix. 
    
    \subsection{Existence and Complexity}\label{sec:existence}
    
        As discussed in~\cite{roth02}, stability is a desirable property since it correlates with the long-term success of the matching process. Unfortunately, as we show in Proposition~\ref{thm: inexistence absolute envy} and Theorem~\ref{theorem:absolute_complexity}, a contingent stable matching under absolute priorities may not exist and the decision problem of certifying the (un)existence is NP-complete.
        
        \begin{proposition}\label{thm: inexistence absolute envy}
            A contingent stable matching under absolute priorities may not exist {regardless of the tie-breaking rule}, even if families are of size at most two and the initial tie-breakers are at the family level.
        \end{proposition}

        \begin{theorem}\label{theorem:absolute_complexity}
            The problem of determining whether a contingent stable matching with {absolute} priorities exists is NP-complete, even if each family is of size at most 2, there are at most 2 levels and initial tie-breakers are at the family level. 
        \end{theorem}
        
        The intuition behind this result is that a ``cycle'' may arise when a student receives contingent priority and gets matched to some school, triggering a chain of displacements that enables their priority provider to move to a more preferred school, which in turn causes the initially prioritized student to lose their priority. However, as detailed later in Section~\ref{subsec: hard vs soft}, such cycles can be mitigated by allowing the clearinghouse to use a relaxed version of contingent priorities.


        Under partial priorities, existence and complexity heavily depend on the tie-breaking rule. Specifically, as we show in Proposition~\ref{thm: existence partial} and in Theorem~\ref{theorem:partial_complexity}, a contingent stable matching may not exist, and the decision problem of certifying the (un)existence is NP-complete under tie-breakers at the individual level. 
        In contrast, if tie-breakers are at the family level, we show that partial contingent stability is equivalent to initial stability and, consequently, a stable matching always exists and can be found in polynomial time using the Deferred Acceptance algorithm.
        
        \begin{proposition}\label{thm: existence partial}
             {A contingent stable matching under partial priorities may not exist when the initial tie-breakers are at the individual level}, even if families are of size at most two and there at most two levels. In contrast, the existence is guaranteed when tie-breakers are at the family level. 
        \end{proposition}    

        \begin{theorem}\label{theorem:partial_complexity}
            The problem of determining whether a contingent stable matching with {partial} priorities under individual tie-breakers exists is NP-complete, even if the size of each family is at most 2 and there are at most 2 levels. 
        \end{theorem}   
                

    
    \subsection{Student-Optimality}\label{sec: student optimality}
        Most school districts use some variant of the student-proposing Deferred Acceptance algorithm, which is known to return the (unique) student-optimal stable assignment under the notion of (initial) stability~\citep{gale1962college}. Moreover, the Rural Hospital Theorem~\citep{roth1986allocation} implies that the set of students matched is the same at every stable matching. As Example~\ref{ex: non-uniqueness} in Appendix~\ref{app: additional examples}  illustrates, these properties do not hold under contingent priorities.

    The varying cardinality of the set of matched students requires a more precise basis for comparing different contingent stable matchings. For example, school districts may be legally obligated to guarantee each applicant a seat, leading them to prefer matchings with maximum cardinality. In other cases, such as in Chile, the clearinghouse may prioritize maximizing the number of siblings placed together. Consequently, throughout this paper, we will assume that the clearinghouse seeks to find a stable matching under contingent priorities that optimizes students' preferences (rank-optimal), with the understanding that being unassigned is preferable to being assigned to a school not listed in their preferences.\footnote{Our framework is flexible enough to accommodate other objective functions, such as maximizing the number of students matched with their siblings, minimizing the number of unassigned students, among others.}  {Note that a rank-optimal matching under the standard notion of stability coincides with the student-optimal stable matching (see e.g.~\cite{bobbio22}).} 

    \subsection{Incentives}\label{sec:incentives}
        A desired property of any mechanism is strategy-proofness, i.e., that students have no incentive to misreport their preferences to improve their allocation. \citet{roth1982economics} and~\citet{dubins1981machiavelli} show that, under the standard concept of stability, the student-proposing version of DA is strategy-proof for students.
        Unfortunately, the mechanisms to find rank-optimal contingent stable matchings are not strategy-proof unless we consider the partial case under family lotteries, as we show in Propositions~\ref{prop: incentives absolute} and~\ref{prop: incentives partial}.
                
        \begin{proposition}\label{prop: incentives absolute}
            The mechanism to find a rank-optimal contingent stable matching with absolute priorities is not strategy-proof for the families, regardless of the tie-breaking rule. 
        \end{proposition}

    
        \begin{proposition}\label{prop: incentives partial}
            The mechanism to find a contingent {rank-optimal} stable matching with partial priorities is not strategy-proof for the families under individual initial tie-breakers, but {it} is strategy-proof under family tie-breaking rule.
        \end{proposition}

        Although strategy-proofness is desirable, the required knowledge about others' preferences and priorities to make a profitable deviation makes these unlikely to happen in practice. Moreover, as we show in Proposition~\ref{prop: SP-L}, the mechanisms to find contingent stable matching are strategy-proof in the large (see~\citet{azevedo18}), i.e., it is approximately optimal for students to report their true preferences for any i.i.d. distribution of students' reports. Hence, in large markets such as the ones motivating this work, the lack of strategy-proofness is not a major concern. 

        \begin{proposition}\label{prop: SP-L}
            The mechanism to find a contingent stable matching (under either {absolute} or partial priorities) is strategy-proof in the large.
        \end{proposition}

    \subsection{Hard vs. Soft Priorities}\label{subsec: hard vs soft}
        Some of the ``negative'' results discussed in this section, such as hardness and lack of existence, arise because the clearinghouse is compelled to grant contingent priorities to every student with an effective priority provider. However, the clearinghouse could have greater flexibility in deciding which priority providers to consider and which to disregard.  By selectively applying contingent priorities, the clearinghouse could alleviate these issues and ensure the existence of a stable matching. 
        
        We say that contingent priorities are \emph{hard} if every student with an effective priority provider must be prioritized according to Definitions~\ref{def: absolute contingent priority} and~\ref{def: partial contingent priority} for absolute and partial, respectively. 
        Conversely, contingent priorities are \emph{soft} if the clearinghouse can ignore some effective priority providers and treat their siblings as students without priority. Formally, whenever a pair $(s,c)\in \pairs$ satisfies the conditions in Definition~\ref{def: provider of contingent priority} for a given matching $\mu$, hard priorities imply that $z^\mu(s,c)=1$, while soft priorities imply that $z^\mu(s,c)\in \lrl{0,1}$, with the clearinghouse being able to choose either depending on its goals. For example, if the clearinghouse sets $z^\mu(s,c)=0$ for every effective priority provider, then the clearinghouse can entirely ignore contingent priorities and target an initial stable matching.       
        
        In line with previous work comparing hard and soft quotas or diversity constraints~\citep{kojima12,Ehlers14}, the enhanced flexibility that results from soft contingent priorities allows the clearinghouse access to a broader space of feasible stable matchings. At one extreme, this includes the set of initial stable matchings where contingent priorities are ignored; at the other, it encompasses the set of contingent stable matchings where these priorities are fully enforced. 
        This expanded space of stable matchings is always non-empty, as the clearinghouse can choose to disregard all providers of contingent priority and use the student-proposing Deferred Acceptance algorithm to find the initially stable student-optimal matching in polynomial time. Furthermore, any initially stable matching belongs to the set of stable matchings under soft contingent priorities, whereas the opposite is not true, as illustrated by the examples in the previous section.

        Ultimately, the choice between hard and soft contingent priorities depends on the policymakers' goals and their capacity to discriminate among students. Soft priorities may be a desirable solution for clearinghouses aiming to prioritize the joint assignment of siblings while ensuring the existence of a stable allocation. However, this approach may not be feasible if the clearinghouse must treat all students equally for fairness reasons.
        
\section{Formulations}\label{sec: formulations}
    The results in Section~\ref{sec:existence} imply that there is no hope of designing a polynomial-time approach to finding a stable matching with contingent priorities, unless NP = P. This motivates our use of integer linear programming to obtain the rank-optimal stable matchings for each type of {contingent} priority (if it exists), which is the focus of this section.   
    
    The formulations we present in Sections~\ref{subsec: formulation absolute} and~\ref{subsec: formulation partial} (for {absolute} and {partial}, respectively) extend that in~\citet{baiou2000stable} to find a rank-optimal stable matching that accounts for contingent priorities. Specifically, let $r_{s,c}$ be the position of school $c\in \schools$ in student $s$'s preference list, and let $r_{s, \emptyset}$ be a parameter that captures the cost of having student $s$ unassigned. Then, a rank-optimal (initial) stable matching\footnote{Under standard stability, it is also the student-optimal stable matching.} corresponds to the solution of the following integer program:
    { \begin{subequations}\label{eq: baseline}
        \begin{align}
          \min \;\; &  \sum_{(s,c)\in \pairs} r_{s,c}\cdot x_{s,c}  \notag\\
          s.t. \;\;
          &  q_{c}^{\ell(s)}\cdot \Bigg(1- \sum_{\substack{c'\in\schools:\\c'\succeq_s c}} x_{s,c'}\Bigg) \leq  \sum_{\substack{s'\in \students^{\ell(s)}:\\ s'\succ_c s}} x_{s', c}\;, \qquad\forall (s,c) \in \students\times \schools, \label{eq: constraint stability}\\
          & \mathbf{x}\in \cP, 
        \end{align}
    \end{subequations}}
    where 
    \[
        \cP = \Bigg\{
                \mathbf{x}\in \lrl{0,1}^{\pairs} : 
                \sum_{\substack{c\in\schools\cup\{\emptyset\}: \\(s,c)\in \pairs }}x_{s,c} =  1,  \; \forall \ s\in  \students, \quad
                \sum_{\substack{s\in \students^\ell: \\ (s,c)\in \pairs}} x_{s,c} \leq q_{c}^\ell, \; \forall \ c\in  \schools, \; \ell \in \levels
        \Bigg\}    
    \]
    is the set of feasible assignments, i.e., $\mathbf{x}$ encodes the matching of students to schools and $\mathbf{x} \in \cP$ ensures that each student is assigned to at most one school and that each school does not exceed its capacity in each level.
    The objective is to minimize the preference of assignment of each student, and the set of constraints~\eqref{eq: constraint stability} guarantees that student $s$ has no initial justified-envy  in school $c$.

    {Formulation~\eqref{eq: baseline} does not account for contingent priorities. To incorporate them, we extend this formulation 
    {as follows. With a slight abuse of notation, let $f(s,c)=\lrl{s'\in f(s)\;:\; (s',c) \in \pairs}$, i.e., $f(s,c)$ captures the family members of $f(s)$ who apply to school $c$ (i.e., prefer it over being unassigned). Then, we define the }
    set of variables $z_{s,c} \in \lrl{0,1}$ for all $(s,c)\in \pairs$ with $|f(s,c)|\geq 2$ and $c\in\schools$  (i.e., student $s$ has siblings also participating in the admissions process and who prefer $c$ over being unassigned), where $z_{s, c}$ is equal to 1 if student $s$ is an effective priority provider in school $c$ and one of their siblings is also assigned to $c$, and zero otherwise. Given a set of decision variables $\mathbf{x}$ (i.e., a matching), the set of variables $\mathbf{z}$ that are consistent with the model discussed in Section~\ref{sec: model} can be characterized as follows: 
    {\begin{subequations}\label{eq:constraints_priority_z}
        \hspace{-1cm}\begin{align}
            \cR(\mathbf{x}) = \Bigg\{
            &\mathbf{z}  \in\lrl{0,1}^\pairs:\; 
            z_{s,c} \leq x_{s,c}, \hspace{5.0cm} \forall\; s\in \students: |f(s,{c})|\geq 2, \; c \in \schools \label{eq: provider condition (i)} \\
             & {z_{s,c} \leq \sum_{s'\in f(s)\setminus \lrl{s}} x_{s',c}, \hspace{5.5cm} \forall\; s \in \students: |f(s,{c})|\geq 2,\; c \in \schools\;} \label{eq: provider condition (ii)}  \\             
            &
            {\sum_{\substack{s'\in\students^{\ell(s)}\\s' \succ_c s} } \sum_{\substack{c'\in\schools\cup \lrl{\emptyset}\\c\succeq_{s'} c'} }} x_{s',c'} \leq \lrp{q_c^{\ell(s)}-1} + \lra{\students^{\ell(s)}}\cdot (1-z_{s,c}), \;\forall\; s \in \students: |f(s,{c})|\geq 2, \; c \in \schools  \label{eq: provider condition (iii)}    \\
            &\sum_{s\in f} z_{s,c}\leq 1, \hspace{6.5cm}\forall\; f\in \families: \lra{f(s,{c})} \geq 2, \; c \in \schools   \label{eq: provider condition (iv)}
             \Bigg\}.      
        \end{align}
    \end{subequations} 
    }   

    Following Definition~\ref{def: provider of contingent priority} and the definition of $\mathbf{z}^\mu$ given in Section~\ref{sec: refining contingent priorities}, $s$ provides priority to their siblings in school $c$ if (i) they get matched to it (captured by the set of constraints~\eqref{eq: provider condition (i)}), {(ii) at least one of their siblings potentially benefits from it (captured by the set of constraints~\eqref{eq: provider condition (ii)}), and (iii)} there are at most $q_c^{\ell(s)}-1$ students who initially have higher priority in $c$ and weakly prefer $c$ over their matching, captured by the set of constraints~\eqref{eq: provider condition (iii)}. Finally, the set of constraints~\eqref{eq: provider condition (iv)} ensures that at most one member of each family provides contingent priority to their siblings in each school.

    To capture the action of receiving contingent priority and benefiting from it, we consider the set of variables $y_{s, s', c} \in \lrl{0,1}$ for all $s\in \students$ with $|f(s,{c})|\geq 2$, {and} $s'\in f(s)\setminus\{s\}$, 
    where $y_{s, s', c}$ is equal to 1 if student $s'$ gets matched to $c$ having $s$ as their effective priority provider, and zero otherwise. Thus, given a set of decision variables $\mathbf{x} \in \lrl{0,1}^\pairs$ and $\mathbf{z} \in \lrl{0,1}^{\pairs}$, the set of variables $\mathbf{y}$ can be fully characterized as follows:
    {\begin{subequations}\label{eq:constraints_priority_y}
        \begin{align}
            \cQ(\mathbf{x}, \mathbf{z}) = \Bigg\{ \mathbf{y} \in \{0,1\}^{\students\times \pairs} \;:\; 
              & y_{s,s' ,c} \leq x_{s',c}, \hspace{1.2cm} \forall\; s \in \students: |f(s,{c})|\geq 2, \; s' \in f(s),\; c \in \schools\; \label{eq: definition of y 4} \\         
              & y_{s,s',c} \leq  z_{s,c}, \hspace{1.2cm} \forall\; s \in \students: |f(s,{c})|\geq 2, \; s' \in f(s),\; c \in \schools\;  \label{eq: definition of y 1} \\           
              & y_{s',s,c} \leq 1 - z_{s,c}, \hspace{0.6cm} \forall\; s \in \students: |f(s,{c})|\geq 2, \; s' \in f(s),\; c \in \schools\;  \label{eq: definition of y 2} 
              \Bigg\}    .      
        \end{align}
    \end{subequations} 
    } 
    The sets of constraints~(\ref{eq: definition of y 4}) and~(\ref{eq: definition of y 1}) guarantee that student $s'$ can receive contingent priority from $s$ in school $c$ only if the former gets assigned and the latter is their family's provider of priority in that school, respectively. Finally, the set of constraints~(\ref{eq: definition of y 2}) ensures that students do not simultaneously provide and receive contingent priority.   
    }

    \begin{remark}\label{remark: single initial group}
        To ease exposition, in the remainder of the paper, we assume that $\lra{\groups} = 1$, i.e., there is initially a single group and, thus, the initial order in each school $c$, $\succ_c$, is solely determined by the initial random tie-breakers. In Appendix~\ref{app: extensions}, we discuss how to incorporate other groups.
    \end{remark}

    \begin{remark}
        We omit constraints that explicitly require $z_{s,c} = 1$ for the student $s$ with the most favorable lottery among the family members in $f(s)$ satisfying Definition~\ref{def: provider of contingent priority} since such constraints are unnecessary in both the absolute and partial cases, as there always exists a feasible solution leading to the same allocation that inherently satisfies this condition (see Lemma~\ref{lemma: ensure highest lottery provider} in Appendix~\ref{app: formulations technical lemmas}). More precisely, given a feasible solution, we can construct another feasible solution with the same matching, i.e., variables $\mathbf{x}$, and modified variables $\mathbf{z}$ (and $\mathbf{y}$) to respect the condition.\label{remark:effective_priority_provider}
    \end{remark}

\subsection{Absolute Priority}\label{subsec: formulation absolute}
    {As discussed in Definition~\ref{def: absolute contingent priority}, a student $s$ who either provides or receives absolute contingent priority can displace any other student who does not hold such priority, regardless of their random tie-breakers. However, when two students belong to the same priority group, they are ranked according to their tie-breakers. Consequently, we can formulate the problem of finding a rank-optimal stable matching with absolute contingent priority as follows:}
    {
    \begin{subequations}\label{eq: absolute formulation}
        \begin{align}
             \min & \quad \sum_{(s,c)\in \pairs} r_{s,c}\cdot x_{s,c} \notag \\
            s.t.  
            & \quad q_{c}^{\ell(s)} \cdot\Bigg(1- \sum_{\substack{c'\in\schools:\\c'\succeq_s c}} x_{s,c'}\Bigg) \leq \sum_{\substack{a\in \students^{\ell(s)}:\\ a\succ_c s}} x_{a, c}
            + \sum_{\substack{f\in\mathcal{F}:\\ |f|\geq 2}}\sum_{\substack{\lrl{a,a'}\subseteq f: \\ a\in  \students^{\ell(s)}, \\ a\prec_c s } } y_{a',a,c}+\sum_{\substack{a\in  \students^{\ell(s):} \\ a\prec_c s}}z_{a,c} \; ,\hspace{1em} \forall (s,c) \in \pairs, \label{eq: constraint stability abs indep} \\
            & \quad x_{s',c} + \Bigg(1-\sum_{\substack{c'\in\schools:\\c'\succeq_s c}} x_{s,c'}\Bigg) \leq 2 - x_{a,c} + \indicator{a\succ_c s}\cdot\Bigg(z_{a,c} + \sum_{a'\in f(a)\setminus\{a\} } y_{a',a,c}\Bigg), \notag\\ &\hspace{18.5em} \forall c\in \schools, f\in\cF, \{s,s'\}\subseteq f, a\in \students^{\ell(s)}\setminus f, \label{eq: tie-breaking contingent abs indep}\\
            &\quad \mathbf{x}\in \cP, \; \mathbf{z} \in \cR(\mathbf{x}),\; \mathbf{y} \in \cQ(\mathbf{x}, \mathbf{z}). 
        \end{align}
    \end{subequations}}

    The first set of constraints~(\ref{eq: constraint stability abs indep}) extends~(\ref{eq: constraint stability}) to incorporate absolute contingent priorities. Specifically, suppose that student $s$ is not matched with school $c$ or better. This set of constraints implies that there are at least $q_{c}^{\ell(s)}$ students matched to school $c$ at level $\ell(s)$ who are either (i) initially more preferred than student $s$ (first term on the right-hand side), (ii) initially less preferred than $s$ but receive contingent priority from one of their siblings (second term in the right-hand side), or (iii) initially less preferred than $s$ but provide contingent priority to their siblings (third term in the right-hand side). 
    {The latter two terms capture that, according to Definition~\ref{def: absolute contingent priority}, receivers and providers of priority keep their initial group, while students with no contingent priority moved to the new (lowest) priority group $\lra{\groups}+1$ and, thus, are less preferred under the contingent matching $\mathbf{x}$. }
    The second set of constraints~(\ref{eq: tie-breaking contingent abs indep}) ensures that ties among prioritized students are broken according to their initial random tie-breakers (or, equivalently, using the initial priority order due to Remark~\ref{remark: single initial group}). 
    Namely, if student $s$ has a sibling $s'$ matched to $c$ (potentially in a different level) and $s$ is not matched to $c$ or better, then no other student $a\in \students^{\ell(s)} \setminus f(s)$ can get matched to $c$ unless their random-tie breaker is better than that of $s$ and either they receive or effectively provide priority. Moreover, this set of constraints ensures that prioritized students are never displaced by students without priority. To see this, note that if $a$ has no siblings, the sum on the right-hand side is zero and, consequently, if $s'$ is assigned to $c$ and $s$ is not assigned to $c$ or better, then $x_{a,c}$ is forced to be zero.
    In Theorem~\ref{thm:correctness_absolute},  we formally show that the feasible region~\eqref{eq: absolute formulation} correctly captures the notion of contingent absolute priorities. We defer the proof of this result to Appendix~\ref{app:proof_correcteness_absolute}.
    \begin{theorem}\label{thm:correctness_absolute}
        Feasible region~\eqref{eq: absolute formulation} is a valid formulation of the set of contingent stable matchings under absolute priorities.
    \end{theorem}

\subsubsection{Hard vs. Soft Absolute Contingent Priorities.} 

We now explore how to incorporate hard and soft absolute priorities in Formulation~\eqref{eq: absolute formulation}. As discussed in Section~\ref{subsec: hard vs soft}, these priorities are captured by the way we assign values to $z^\mu(s,c)\in\lrl{0,1}$ for each pair $(s,c)\in\pairs$ under a given matching $\mu$. Recall that, regardless of whether priorities are hard or soft, we assign $z^\mu(s,c)=0$ for every student $s$ such that $f(s)=\{s\}$. Hence, the differences between hard and soft arise when we consider students with siblings.

\paragraph{Hard Priorities.}
In this case, $z^\mu(s,c)=1$ whenever $s$ is an effective priority provider in school $c$, i.e., $s$ has the most favorable random tie-breaker among their family members satisfying the conditions in Definition~\ref{def: provider of contingent priority} at school $c$. 
{In Theorem~\ref{thm:correctness_absolute}, we show that the set of constraints~\eqref{eq: absolute formulation} ensures that $z_{s,c} = 1$ for all such pairs satisfying $\lra{f(s)\cap \mu(c)} \geq 2$ (if a feasible solution exists). This result relies on  Lemma~\ref{lemma:existence_of_provider} (see Appendix~\ref{app:formulations}), which guarantees that whenever a family of size at least two has one of their members matched to a school and another to a less preferred one, then it \emph{must} have an effective priority provider in that school.  Key to ensure this result and enforce absolute priorities is the set of constraints~\eqref{eq: tie-breaking contingent abs indep}. To see this, suppose there is a family $f=\lrl{s,s'}$ where $x_{s',c}=1$ and $x_{s,c'} = 0$ for all $c'\succeq_s c$, so that the left-hand side is equal to two. Then,~\eqref{eq: tie-breaking contingent abs indep} can be re-written as 
\[
x_{a,c} \leq \indicator{a\succ_c s}\cdot \lrp{z_{a,c} + \sum_{a' \in f(a)\setminus \lrl{a}} y_{a',a,c}}, \quad \forall a\in \students^{\ell(s)}\setminus f.
\]
Note that if $a$ is such that $a\prec_c s$, then the right-hand side is zero, forcing $x_{a,c} = 0$ and granting that $s$ has no contingent justified envy towards $a$. Conversely, we could have $x_{a,c} = 1$ if $a$ either provides or receives priority, ensuring that the family $f(a)$ has an effective provider of priority and guaranteeing that $s$ has no contingent justified envy (as $a\succ_c s$).
}

\paragraph{Soft Priorities.}
In this case, the clearinghouse can ignore some effective priority providers and set $z^\mu(s,c)=0$ depending on its goals. To capture this, we relax the set of constraints~\eqref{eq: tie-breaking contingent abs indep} as follows
\begin{equation}\label{eq: tie-breaking contingent abs indep soft}
z_{s',c} + \Bigg(1-\sum_{\substack{c'\in\schools:\\c'\succeq_s c}} x_{s,c'}\Bigg) \leq 2 - x_{a,c} + \indicator{a\succ_c s}\cdot\Bigg(z_{a,c} + \sum_{a'\in f(a)\setminus\{a\} } y_{a',a,c}\Bigg). \tag{7d}   
\end{equation}
{Note that the only change comes from replacing $x_{s',c}$ with $z_{s',c}$ on the left-hand side, which means that having a sibling $s'$ matched to a school $c$ is not enough for student $s$ to displace another student $a\notin f(s)$, i.e., the discussion for hard absolute priorities above does not apply here. Instead, the displacement will occur if and only if $z_{s',c}=1$. In Proposition~\ref{prop:soft_absolute}, we show that the set of constraints~\eqref{eq: tie-breaking contingent abs indep soft} allow us to capture soft absolute priorities and, furthermore, an entire spectrum of solutions ranging from the initially stable matchings to those with hard absolute priorities. We defer the proof to the Appendix~\ref{app:hard_vs_soft_absolute}.}
\begin{proposition}\label{prop:soft_absolute}
{For every point $(\mathbf{x},\mathbf{z},\mathbf{y})$ feasible for Formulation~\eqref{eq: absolute formulation}, we can construct $(\mathbf{x}',\mathbf{z}',\mathbf{y}')$ feasible for Formulation~\eqref{eq: absolute formulation} with~\eqref{eq: tie-breaking contingent abs indep} replaced by \eqref{eq: tie-breaking contingent abs indep soft}. Moreover, this modified formulation is always feasible as it contains the set of initially stable matchings.} 
\end{proposition}

\subsection{Partial Priority}\label{subsec: formulation partial}
     The key difference between absolute and partial is that, in the former case, a prioritized student can move any other student without siblings matched to the school. In the latter, in contrast, prioritized students can only take over those who have no siblings assigned to the school if their contingent tie-breaker (the most favorable between theirs and that of their effective priority provider) is better. 
     
     
    To capture this, we modify the set of constraints in~\eqref{eq: absolute formulation} in three important ways. 
    {First, we remove the terms involving $z_{a,c}$ from both~\eqref{eq: constraint stability abs indep} and~\eqref{eq: tie-breaking contingent abs indep} since effective priority providers keep their initial group under partial priorities. }
    Second, we modify the set of constraints~\eqref{eq: constraint stability abs indep} adding the condition $a' \succ_c s$ in the second summation of the right-hand side to ensure that the student $a'$ providing the siblings' priority has a better tie-breaker than the student $s$ who gets displaced by their prioritized sibling. 
    {Finally, we modify the set of constraints~(\ref{eq: tie-breaking contingent abs indep}) by multiplying $x_{a,c}$ and $y_{a',a,c}$ by the indicators $\indicator{a \prec_c\; s\wedge s'}$ and $\indicator{a\wedge a' \succ_c \; s\wedge s'}$, respectively, where $s\wedge s'$ denotes the best initial priority order among the students $s,s'$.\footnote{Concretely, $s \wedge s' \succ_c a$ if and only if $s\succ_c a$ or $s' \succ_c a$, and $s \wedge s' \succ_c a \wedge a'$ if and only if $s \wedge s' \succ_c a $ and $s \wedge s' \succ_c a' $.} To see the role played by these indicators, let $f=\lrl{s,s'}$ be a family with $x_{s',c = 1}$ and $x_{s,c'} = 0$ for all $c'\succeq_s c$, and consider an additional student $a \in \students^{\ell(s)}\setminus f$. Then, 
    \begin{itemize} 
        \item If $s\succ_c a$, then the constraint ensures that $a$ can only be matched to $c$ (i.e., $x_{a,c} = 1$) if they receive contingent priority from a sibling $a'$ (i.e., $y_{a', a, c} = 1$) with higher initial priority than both $s$ and $s'$ (i.e., $\indicator{a'\wedge a \succ_c s' \wedge s} = \indicator{a' \succ_c s'\wedge s} = 1$). Otherwise, the constraint forces $x_{a,c} = 0$.
        \item If $s'\succ_c a\succ_c s$, then the constraint ensures that $a$ can only be matched to $c$ if they receive contingent priority from a sibling $a'$ with higher initial priority than both $s$ and $s'$. Otherwise, the terms $y_{a',a,c}$ are equal to zero, which forces $x_{a,c} = 0$.
    \end{itemize}
    }

    As a result, we formulate the problem of finding a stable assignment with partial priorities as:
    {
    \begin{subequations}\label{eq:partial_formulation}
        \begin{align}
          \min & \quad \sum_{(s,c)\in \cV} r_{s,c}\cdot x_{s,c} \notag \\
          s.t.  
          & \quad q_{c}^{\ell(s)} \cdot\Bigg(1- \sum_{\substack{c'\in\schools:\\c'\succeq_s c}} x_{s,c'}\Bigg) \leq 
          \sum_{\substack{a\in \students^{\ell(s)}:\\ a\succ_c s}} x_{a, c}
          + \sum_{\substack{f\in\mathcal{F}:\\ |f|\geq 2}} \sum_{\substack{\lrl{a,a'}\subseteq f: \\ a\in  \students^{\ell(s)} \\ a\prec_c \; s\; \prec_c \; a' }}y_{a',a,c},\hspace{4em} \forall (s,c) \in \pairs, \label{eq: constraint stability par indep} \\
          & x_{s',c} + \Bigg(1- \sum_{\substack{c'\in\schools:\\c'\succeq_s c}} x_{s,c'}\Bigg) \leq 2 - x_{a,c}\cdot \indicator{a \prec_c s \wedge s'} +\sum_{a'\in f(a)\setminus\{a\} } y_{a',a,c}\cdot \indicator{a'\wedge a\; \succ_c s'\wedge s}, \notag \\ &\hspace{18em}\forall c\in \schools, f\in\cF, \{s,s'\}\subseteq f, a\in \students^{\ell(s)}\setminus f,  \label{eq: tie breaking contingent par indep}\\
        &\quad \mathbf{x}\in \cP,\; \mathbf{z} \in \cR(\mathbf{x}),\; \mathbf{y} \in \cQ(\mathbf{x}, \mathbf{z}). 
        \end{align} 
    \end{subequations}}

    In Theorem~\ref{thm:correctness_partial},  we show that the feasible region~\eqref{eq:partial_formulation}  correctly captures the notion of contingent partial priorities. We defer the proof of this result to Appendix~\ref{app:proof_correcteness_partial}.
    \begin{theorem}\label{thm:correctness_partial}
    The feasible region~\eqref{eq:partial_formulation} is a valid formulation of the set of contingent stable matchings under partial priorities.
    \end{theorem}


\subsubsection{Hard vs. Soft Partial Contingent Priorities.} 
    On the one hand, Formulation~\eqref{eq:partial_formulation} guarantees that the optimal solution adheres to hard partial priorities. We omit the formal proof here, as it parallels the reasoning used in the absolute case. On the other hand, to account for soft partial priorities, we replace constraint~\eqref{eq: tie breaking contingent par indep} with the following set of constraints:
    \begin{equation}\label{eq: tie breaking contingent par indep soft}
    z_{s',c} + \Bigg(1- \sum_{\substack{c'\in\schools:\\c'\succeq_s c}} x_{s,c'}\Bigg) \leq 2 - x_{a,c}\cdot \indicator{a \prec_c s \wedge s'} +\sum_{a'\in f(a)\setminus\{a\} } y_{a',a,c}\cdot \indicator{a'\wedge a\; \succ_c s'\wedge s}. \tag{8d}
    \end{equation}
    As in the absolute case, note that the only difference between \eqref{eq: tie breaking contingent par indep} and \eqref{eq: tie breaking contingent par indep soft} is the use of $z_{s',c}$ instead of $x_{s',c}$. In Proposition~\ref{prop:soft_partial}, we show that this change guarantees that the problem is always feasible and that we can recover a range of stable matchings with partial priorities varying $\mathbf{z}$.
    \begin{proposition}\label{prop:soft_partial}
        {For every point $(\mathbf{x},\mathbf{z},\mathbf{y})$ feasible for Formulation~\eqref{eq:partial_formulation}, we can construct $(\mathbf{x}',\mathbf{z}',\mathbf{y}')$ feasible for Formulation~\eqref{eq:partial_formulation} with \eqref{eq: tie breaking contingent par indep} replaced \eqref{eq: tie breaking contingent par indep soft}. Moreover, this modified formulation is always feasible as it contains the set of initially stable matchings.}
    \end{proposition}

\section{Application to School Choice in Chile}\label{sec: application}

To illustrate the benefits of our framework, we use data from the Chilean school choice system. This system was introduced in 2016 and currently serves more than half a million students across all regions and levels (i.e., from Pre-K to 12th grade).

{The school choice system in Chile works as follows.} 
Families report a strict preference list for each of their children participating in the system, and each school sorts students according to five groups:
(i) Students whose siblings are already enrolled for the next year (i.e., static siblings' priority, as described in Appendix~\ref{app: extensions});
(ii) Students with older siblings also participating in the admissions process and assigned to that school;
(iii) Students with parents working at the school;
(iv) Former students returning to the school;
(v) Students who do not satisfy any of the former priority groups.
These priorities are processed in strict order, with students with siblings enrolled given the highest priority, followed by students with older siblings assigned to the school, and so on down the list. 
To break ties within each priority group, the system uses a multiple tie-breaking rule at the family level, i.e., each family gets a different random tie-breaker for each school that is slightly perturbed to break ties among siblings while keeping the order across families.

\subsection{Benchmarks}
    We compare our framework with the algorithm currently used to solve the Chilean school choice problem.
    After collecting families' preferences and sorting students in each school, the clearinghouse runs an algorithm that processes levels in decreasing order, i.e., from the highest (12th grade) down to the lowest (Pre-K). For each level $\ell$, the algorithm:
    \begin{enumerate}
        \item Updates schools' priorities to account for the {students who may benefit from having an older sibling previously processed and assigned to the school.}
        \item Runs the student-proposing Deferred Acceptance algorithm considering the updated preferences and priorities among students and schools that belong to level $\ell$.
    \end{enumerate}
    We refer to this algorithm as \emph{Descending}. Note that this algorithm limits siblings' priorities to be \emph{one-directional} in the sense that only older students can provide priority to their younger siblings. As a result, this approach does not lead to a stable matching~\citep{Correa_2022}.
    Furthermore, this suggests another natural benchmark which is the
    \emph{Ascending} algorithm, i.e., processing grades in ascending order starting from the lowest level. 
    Finally, we also compare our approaches with (i) the student-optimal stable matching (SOSM) obtained by the Deferred Acceptance algorithm (i.e., assuming no one can benefit from having contingent priorities) and (ii) the \emph{family-oriented stable matching} (FOSM), which corresponds to the standard stable matching that maximizes the number of family members assigned to the same school; we include the integer linear programming formulation to obtain FOSM in Appendix~\ref{app:baselineFormulation}. 

\subsection{Data and Simulation Setting}
    \subsubsection{Data.}
    We use data from the admissions process in 2018 and we consider all students who applied to the system in the southernmost region of the country (Magallanes).\footnote{Specifically, we use actual families, students' reported preferences, and schools' capacities. All data is publicly available and can be downloaded from this \href{https://datosabiertos.mineduc.cl/sistema-de-admision-escolar-sae/}{website}.} We focus on this region for three reasons: (i) it is the region where all policy changes are first evaluated, (ii) it is isolated from the rest of the country so that every student who applies to local schools does not include schools in other regions, and (iii) the composition of students and schools is representative of the rest of the country, while the size of the instance allows us to speed up computations.

\begin{figure}[ht]
      \begin{minipage}[t]{0.49\textwidth}
        \centering
        \captionof{table}{Instance for Evaluation}\label{tab: instance for school choice}
        \vspace{0.6cm}
        \begin{tabular}{lcc}
            \toprule
             & Magallanes & Overall \\
            \midrule
            Students & 5113 & 274990 \\
            Siblings & 1300 & 44810 \\
            Schools & 61 & 6421 \\
            Applications &  15426 & 874565 \\
            \bottomrule
        \end{tabular}    
      \end{minipage}
      \begin{minipage}[t]{0.49\textwidth}
        \centering
        \captionof{figure}{Students per level (Magallanes)}\label{fig: distribution of students per level}
        \includegraphics[scale=0.5]{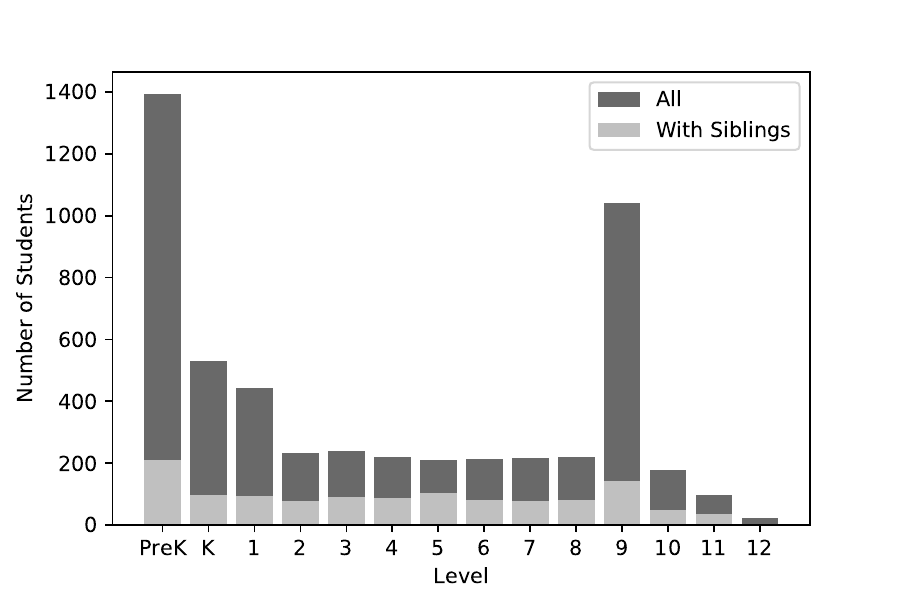}
      \end{minipage}%
    \end{figure}

   In Table~\ref{tab: instance for school choice}, we report summary statistics about the instance, and we compare it with the values nationwide for the same year.\footnote{In our simulations, we consider a total of 5257 students. The difference is due to students who are not from the Magallanes region but only apply to schools in that region.} In addition, in Figure~\ref{fig: distribution of students per level}, we plot the distribution of students across levels, highlighting in each case the number of students with siblings. Note that most of the students who participate in the system apply to one of the five entry levels (Pre-K, K, 1st, 7th, and 9th grade), but the distribution of siblings is relatively uniform across levels.

\subsubsection{Setup.} 
    We perform 100 simulations for each tie-breaking rules, i.e., single and multiple tie-breakers at the individual level (STB and MTB, respectively) and the family level (STB-F and MTB-F, respectively). Specifically, in each simulation,  
    we first draw the random tie-breakers and then solve each benchmark using the resulting priorities and students' preferences.
    In addition, {for the solving of the mathematical programming formulations, we use Gurobi 11.0.\footnote{\url{www.gurobi.com}} with} a MipGap tolerance of 0.1\% with no time limit, and {we use} a penalty parameter for unassigned students $r_{s, \emptyset} = \lra{\succ_s} + 1$ in the formulations of absolute priorities  \eqref{eq: absolute formulation}, partial priorities~\eqref{eq:partial_formulation} and FOSM~\eqref{eq:baseline2}.\footnote{The results are similar if we consider $r_{s, \emptyset} = \lra{\schools} + 1$, i.e., assuming large penalties for having unassigned students. As discussed in~\cite{bobbio22}, considering large penalties reduces the number of unassigned students, while considering small penalties improves the assignment of more students.} Furthermore, we simulate both hard and soft contingent priorities for each type and tie-breaking rule. Finally, to simplify the analysis and exposition of the numerical results, we assume that all students initially belong to the same priority group (i.e., every student initially has no priority). In Appendix~\ref{app: extensions}, we discuss how to incorporate other priority groups.

\subsection{Results}
\begin{figure}
        \caption{Preference of Assignment by Group}\label{fig: distribution preference of assignment}
        \vspace{-0.5cm}
        \begin{subfigure}{.5\textwidth}
            \centering
            \caption{Siblings}\label{fig: distribution of students per level sib}
            \includegraphics[width=\textwidth]{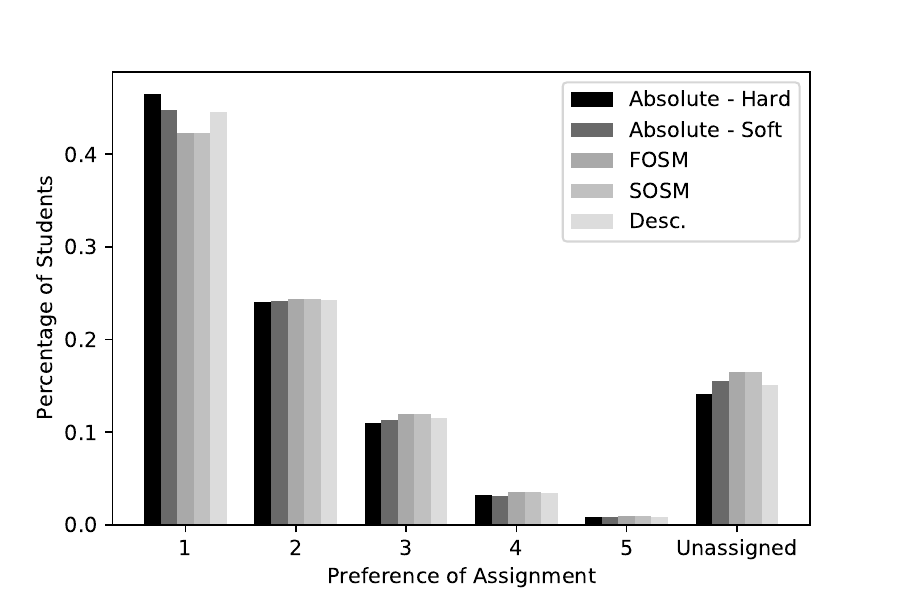}
        \end{subfigure}%
        \begin{subfigure}{.5\textwidth}
            \centering
          \caption{No siblings}\label{fig: distribution of students per level no sib}
          \includegraphics[width=\textwidth]{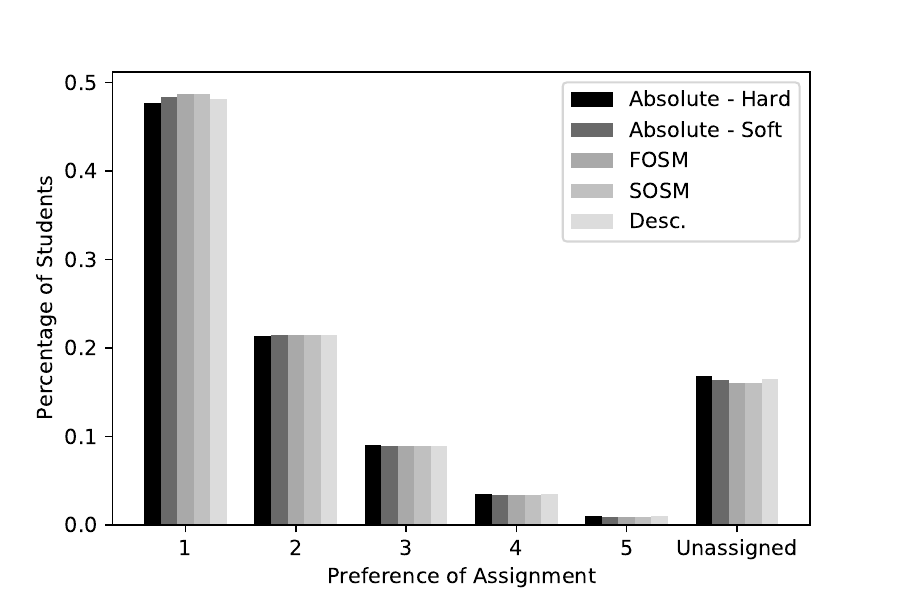}
        \end{subfigure}
    \end{figure}

    In Figure~\ref{fig: distribution preference of assignment}, we report the distribution of the preference of assignment for students with (Figure~\ref{fig: distribution of students per level sib}) and without (Figure~\ref{fig: distribution of students per level no sib}) siblings participating in the admissions process. To facilitate the comparison, we report the results considering multiple tie-breakers at the family level, and we only plot the results for (i) {Absolute-Hard}, (ii) {Absolute-Soft}, (iii) FOSM, (iv) SOSM, and (v) Descending. We focus on MTB-F and Descending because these are the features currently used to solve the Chilean school choice problem. Moreover, we add SOSM to isolate the effects of the siblings' priority, and we include FOSM as an alternative approach. We skip the results for {Partial} because they are equivalent to those obtained by {SOSM} for MTB-F, as shown in Proposition~\ref{thm: existence partial}. Finally, we omit students assigned to their 6th or lower preference because they represent less than 0.5\% of students across all the simulations performed. We report the detailed results for other methods and tie-breaking rules in Appendix~\ref{app: additional results}.
    First, we find that the problem of finding a stable assignment with hard absolute contingent priorities is feasible for roughly a third of the simulations considered (see Table~\ref{tab: effect on siblings}), while it is always feasible in the soft case (as shown in Proposition~\ref{prop:soft_absolute}).
    Second, we observe that the results for Absolute-Soft and Descending are relatively similar for both students with and without siblings, suggesting that these two approaches lead to similar distributions of assignment.
    Third, among the solved instances, we observe that Absolute-Hard leads to a higher proportion of students assigned to their top preference for students with siblings, while it is slightly smaller for students without siblings. Moreover, the number of unassigned students with siblings is significantly lower in Absolute-Hard, while it is slightly higher for students without siblings. The latter two results suggest that Absolute-Hard, whenever feasible, is effective at prioritizing students with siblings while it has no large effect on students without siblings. 
    Finally, we observe that FOSM leads to similar results than SOSM. One potential explanation is that the core of initial stable matchings  tends to be small for large markets~\citep{ashlagi_2017} and, thus, there are not many feasible solutions that prioritize the joint assignment of families.

    We now discuss the results in Table~\ref{tab: effect on siblings}.
    The first column shows the number of instances successfully solved (out of 100).\footnote{This means that Gurobi did not declare the problem infeasible and output a feasible solution with a 0.1\% MipGap.} The following two sets of columns (under \emph{Top Pref.} and \emph{Unassigned}) present the average and standard error for the number of students assigned to their top preference and those left unassigned, respectively. The \emph{Together} columns report the average and standard error for the number of students matched to the same school than at least one other sibling. In the \emph{Separated} columns, we then examine cases where siblings who applied to at least one school in common ended up separated. The \emph{None} columns detail the average and standard error for students whose siblings all fail to secure a match.\footnote{In other words, no member in a family is matched to a school.} The \emph{One} columns provide the same statistics for students where only one sibling is matched, while there is at least one school on both siblings' lists that is more preferred than the one matched. Finally, the \emph{Both} columns show these statistics for students where both siblings are matched to different schools, despite having a third school they both prefer.\footnote{{Note that, in the last three cases, we may double count in cases of families with more than two applicants. Nevertheless, only 69 out of 571 families with multiple applicants involve three or more students.}} As before, we focus on MTB-F, absolute priorities, and a subset of the benchmarks; the full results with all the methods and tie-breaking rules are reported in Appendix~\ref{app: additional results}.

    \begin{table}
        \caption{Comparison of Benchmarks}\label{tab: effect on siblings}
        \centerline{\scalebox{0.85}{\begin{tabular}{lcccccccccccccc}
            \toprule
            & \multicolumn{7}{c}{} & \multicolumn{6}{c}{Separated} \\
            \cmidrule(lr){9-14}
            &  & \multicolumn{2}{c}{Top Pref.} & \multicolumn{2}{c}{Unassigned} & \multicolumn{2}{c}{Together} & \multicolumn{2}{c}{None} & \multicolumn{2}{c}{One} & \multicolumn{2}{c}{Both} \\
            \cmidrule(lr){3-4}\cmidrule(lr){5-6}\cmidrule(lr){7-8}\cmidrule(lr){9-10}\cmidrule(lr){11-12}\cmidrule(lr){13-14}
            & Solved & Mean & SE & Mean & SE & Mean & SE & Mean & SE & Mean & SE & Mean & SE \\
            \midrule
                 Absolute - Hard & 37 & 2492.46 & 3.11 & 848.97 & 1.72 & 604.19 & 2.01 & 85.41 & 1.19 & 76.65 & 1.52 & 106.27 & 1.29\\
                 Absolute - Soft & 100 & 2504.51 & 1.79 & 847.02 & 1.16 & 513.8 & 1.51 & 87.29 & 0.72 & 114.55 & 1.24 & 157.98 & 1.19\\
                 FOSM & 100 & 2486.04 & 1.82 & 849.17 & 1.21 & 427.29 & 1.56 & 88.89 & 0.76 & 146.27 & 1.36 & 219.65 & 1.23\\
                 SOSM & 100 & 2486.38 & 1.79 & 849.17 & 1.21 & 427.35 & 1.56 & 88.89 & 0.76 & 146.27 & 1.36 & 219.57 & 1.23\\
                 Descending & 100 & 2489.80 & 1.69 & 847.74 & 1.20 & 528.90 & 1.44 & 85.87 & 0.71 & 106.23 & 1.10 & 155.57 & 0.99\\
            \bottomrule
        \end{tabular}}}

        \vspace{0.2cm}
        {\footnotesize \emph{Note:} Results obtained considering a multiple tie-breaking rule at the family level (MTB-F).}
    \end{table}

    First, we observe that both Absolute-Hard and Absolute-Soft increase the average overall number of students matched to their top preference, and Absolute-Soft also leads to the smallest number of unassigned students.
    Second, we observe that Absolute-Hard leads to the highest number of siblings matched together and, consequently, to the lowest average number of students who got separated. 
    Third, the largest difference between Absolute-Hard and Descending (the second most beneficial for students with siblings) is for siblings where at least one of them ends up unassigned (columns One and Both). Intuitively, the absolute contingent priority allows students with low lottery numbers---who would most likely end up being unassigned---to increase substantially their chances of admission, thus decreasing the overall number of families with members being unassigned.
    Fourth, we observe that Absolute-Soft improves upon considering no priorities (i.e., SOSM and FOSM) but leads to fewer students matched together compared to Descending, but the differences are relatively small.
    Finally, we observe that FOSM improves the number of applicants matched with their siblings compared to SOSM, but only marginally. FOSM, a tailored method for families, actually fails to match more families together than Absolute-Hard and Absolute-Soft.
    
    Overall, these results suggest that absolute priorities effectively prioritize students with siblings and increase their joint assignments, while the standard notion of stability (i.e., initial stability) has a limited impact on keeping families together.

    \subsubsection{Sensitivity to ``Softness''.}
        One limitation of soft priorities is that they must balance prioritizing students with siblings against potentially harming the objective function. For instance, by receiving contingent priority, a student may displace others and have a negative overall effect on the distribution of assignment preferences, leading to a higher objective function value. Hence, considering soft priorities alone may skip too many effective priority providers amid a small benefit in the distribution of assignment preferences.

        To explore this potential issue and the sensitivity of the solution to the ``softness'' of priorities, one approach is to add a minimum requirement in the number of effective providers to consider by adding the following set of constraints:
        \begin{equation}\label{eq: minimum effective providers considered}
            \begin{aligned}
            \sum_{c\in \schools} \sum_{s\in \students} z_{s,c}&\geq \zeta,     
            \end{aligned}
        \end{equation}
        where $\zeta$ is the minimum number of effective providers we force the solution to incorporate. Then, by varying $\zeta$, we can evaluate the trade-off between benefiting students with siblings and the overall distribution of assignment preference.

        In Table~\ref{tab: sensitivity to softness}, we report simulation results similar to those reported before,\footnote{Specifically, we perform 100 simulations for each value of $\zeta$ considering the same simulation setup discussed above, and then we compute the average across these simulations for each outcome of interest.} but now adding the set of constraints~\eqref{eq: minimum effective providers considered} in the formulation of absolute-soft priorities (i.e., Formulation~\eqref{eq: absolute formulation} with~\eqref{eq: tie-breaking contingent abs indep} replaced by~\eqref{eq: tie-breaking contingent abs indep soft}) and varying the value of $\zeta$. For comparison, we also add the results of Absolute-Hard and Absolute-Soft (i.e., considering $\zeta = 0$).
        First, we observe that the problem is feasible for all instances considered when $\zeta \leq 280$, while the problem is always infeasible with $\zeta > 300$. 
        Second, we observe no impact on the objective of the problem when increasing the number of siblings matched together (through increasing $\zeta$).
        On the one hand, increasing $\zeta$ and forcing the solution to incorporate more effective priority providers has no significant effect on the distribution of assignment preferences, as the number of students matched to their top preference or unassigned remains almost the same. On the other hand, the number of siblings who benefit from contingent priorities significantly increases as we increase $\zeta$ (see the Together column). For instance, when comparing the results with $\zeta = 0$ (Absolute-Soft) and $\zeta = 300$, we observe that the number of students matched with their siblings increases roughly by 20\% (from 515.57 to 616.03). 
        Third, we observe that the primary source of improvement in the joint assignment of siblings comes from reducing the number of cases where both siblings are unassigned (see the Both column), which drops by more than 30\% (from 154.66 when $\zeta = 0$ to 108.35 when $\zeta = 300$). 
        Furthermore, we observe that the number of siblings who benefit from a sufficiently high $\zeta$ that keeps the feasibility of the problem (e.g., $\zeta=280$) is similar to those who benefit from hard priorities.
        {Finally, comparing the results in Table~\ref{tab: sensitivity to softness} for $\zeta = 280$ with those for Descending (in Table~\ref{tab: effect on siblings}), we observe that the former dominates the latter in every dimension, as the number of students assigned to their top preference is higher, those unassigned is lower, and absolute-soft also leads to more siblings assigned together. This result is not surprising, since Absolute-Soft contains the feasible solutions to Descending.}
        
        \begin{table}
            \caption{Sensitivity to Softness}\label{tab: sensitivity to softness}
            \centerline{\scalebox{0.85}{\begin{tabular}{ccccccccccccccc}
                \toprule
                & \multicolumn{7}{c}{} & \multicolumn{6}{c}{Separated} \\
                \cmidrule(lr){9-14}
                &  & \multicolumn{2}{c}{Top Pref.} & \multicolumn{2}{c}{Unassigned} & \multicolumn{2}{c}{Together} & \multicolumn{2}{c}{None} & \multicolumn{2}{c}{One} & \multicolumn{2}{c}{Both} \\
                \cmidrule(lr){3-4}\cmidrule(lr){5-6}\cmidrule(lr){7-8}\cmidrule(lr){9-10}\cmidrule(lr){11-12}\cmidrule(lr){13-14}
                $\zeta$ & Solved & Mean & SE & Mean & SE & Mean & SE & Mean & SE & Mean & SE & Mean & SE \\
                \midrule
                 150 & 100 & 2503.5 & 1.82 & 846.11 & 1.3 & 515.57 & 1.47 & 86.53 & 0.9 & 114.7 & 1.27 & 154.66 & 1.1\\
                 175 & 100 & 2503.0 & 1.82 & 846.07 & 1.32 & 517.08 & 1.45 & 86.46 & 0.9 & 113.73 & 1.22 & 153.95 & 1.1\\
                 200 & 100 & 2503.4 & 1.85 & 845.93 & 1.33 & 518.68 & 1.51 & 86.45 & 0.92 & 113.12 & 1.23 & 153.33 & 1.17\\
                 225 & 100 & 2502.81 & 1.87 & 845.4 & 1.33 & 521.34 & 1.34 & 86.24 & 0.93 & 111.22 & 1.21 & 152.39 & 1.1\\
                 250 & 100 & 2503.56 & 1.83 & 845.63 & 1.31 & 532.46 & 1.02 & 86.47 & 0.9 & 107.08 & 1.17 & 145.53 & 1.01\\
                 275 & 100 & 2502.17 & 1.85 & 845.38 & 1.32 & 567.05 & 0.31 & 85.03 & 0.86 & 92.71 & 1.09 & 127.64 & 1.07\\
                 280 & 100 & 2501.41 & 1.87 & 845.65 & 1.32 & 576.53 & 0.3 & 84.86 & 0.86 & 89.0 & 1.06 & 122.86 & 1.03\\
                 285 & 97 & 2500.14 & 1.91 & 845.31 & 1.36 & 586.35 & 0.27 & 84.26 & 0.85 & 86.06 & 1.08 & 117.91 & 1.03\\
                 290 & 89 & 2497.62 & 2.02 & 845.2 & 1.45 & 596.34 & 0.28 & 83.4 & 0.88 & 82.81 & 1.03 & 112.72 & 1.08\\
                 295 & 60 & 2495.8 & 2.3 & 844.68 & 1.78 & 606.3 & 0.3 & 81.63 & 1.01 & 80.08 & 1.11 & 109.53 & 1.31\\
                 300 & 31 & 2494.94 & 3.39 & 843.29 & 2.73 & 616.03 & 0.37 & 79.94 & 1.29 & 77.23 & 1.56 & 108.35 & 1.55\\
                \midrule
                 Hard & 37 & 2492.46 & 3.11 & 848.97 & 1.72 & 604.19 & 2.01 & 85.41 & 1.19 & 76.65 & 1.52 & 106.27 & 1.29\\
                 Soft & 100 & 2504.51 & 1.79 & 847.02 & 1.16 & 513.8 & 1.51 & 87.29 & 0.72 & 114.55 & 1.24 & 157.98 & 1.19\\            
                \bottomrule                
            \end{tabular}}}

            \vspace{0.2cm}
            {\footnotesize \emph{Note:} Results obtained considering absolute contingent priorities and a multiple tie-breaking rule at the family level (MTB-F).}
        \end{table}   

        These results suggest that combining a soft implementation of contingent priorities with constraints that enforce the solution to incorporate a minimum number of effective priority providers achieves a good balance between feasibility, benefiting students with siblings, and not harming the distribution of preference of assignment.

        \subsubsection{Sensitivity to Tie-Breaking Rule.}
           In this section, we analyze the sensitivity of our results to the different tie-breaking rules previously discussed, i.e., single and multiple tie breaking at the individual and family levels.
            To accomplish this, we follow a similar approach to that discussed above, i.e., we perform 100 simulations for each method and tie-breaking rule, and then compute the average across simulations for each outcome of interest. To simplify exposition, the methods considered are absolute hard, soft and hybrid (i.e., the approach described above with $\zeta = 280$), SOSM, and Descending.

            \begin{figure}
                \caption{Sensitivity to Tie-Breaking Rule}\label{fig: tie-breaking rule}
                \vspace{-0.5cm}
                \begin{subfigure}{.5\textwidth}
                    \centering
                    \caption{Top Preference}\label{fig: top preference}
                    \includegraphics[width=\textwidth]{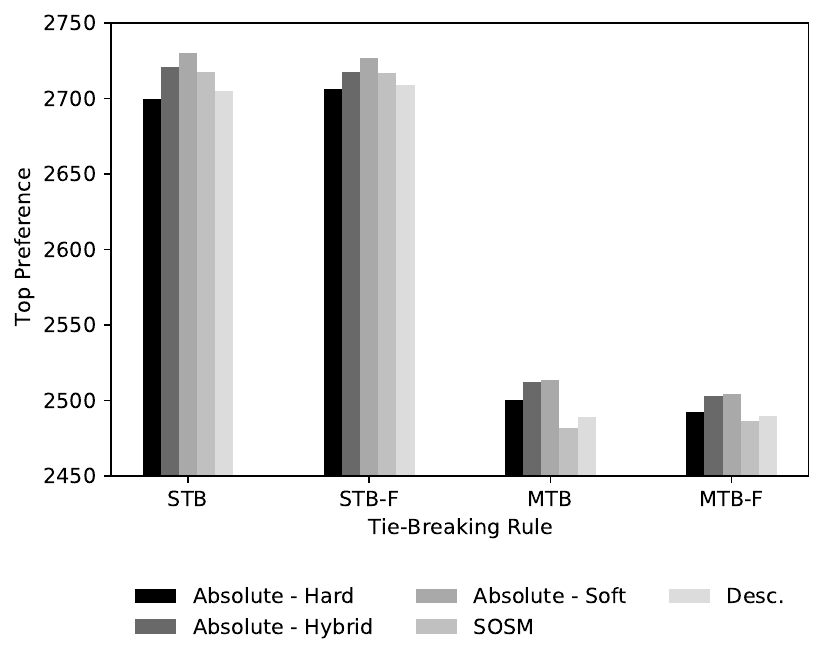}
                \end{subfigure}%
                \begin{subfigure}{.5\textwidth}
                    \centering
                  \caption{Together}\label{fig: together}
                  \includegraphics[width=\textwidth]{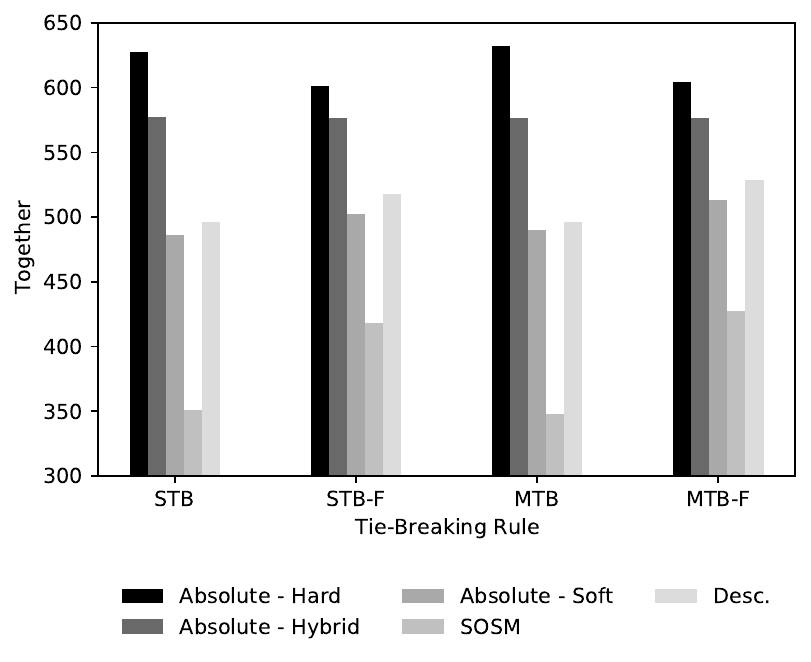}
                \end{subfigure}
            \end{figure}   

            In Figure~\ref{fig: tie-breaking rule}, we report (i) the number of students assigned to their top preference (\emph{Top Preference}; Figure~\ref{fig: top preference}), and (ii) the number of siblings assigned together (\emph{Together}; Figure~\ref{fig: together}) obtained by each approach (with detailed results in Appendix~\ref{app: additional results}).             
            Consistent with the results for MTB-F, we find that the hybrid approach solves all the instances for each tie-breaking rule and consistently outperforms Descending in both dimensions, leading to more students assigned to their top choice and more siblings assigned together. Moreover, we observe that the number of students assigned to their top preference is quite sensitive to the tie-breaking rule, with more students assigned to it under single tie-breaking for all methods considered. In contrast, the number of students assigned together is relatively consistent for each method across the different tie-breaking rules. {These insights are crucial for policymakers aiming to select an appropriate approach (combination of a tie-breaking rule and a type of contingent priority) depending on their goals.}


\section{Conclusions}\label{sec: conclusions}

Motivated by the context of school choice with sibling priorities, we study the problem of finding a stable matching under contingent priorities, i.e., students get prioritized if they have siblings participating in the process and who are currently assigned. We introduce a model of a matching market where siblings may apply together to potentially different levels, and we define a series of guidelines for implementing contingent priorities. Based on these, we propose two different approaches to account for contingent priorities: (i) absolute, whereby a prioritized student can displace any other student without priority, and (ii) {partial}, whereby a prioritized student can only displace others who have a lower tie-breaker than the provider of the priority. In each case, we characterize properties of the corresponding mechanism and provide mathematical programming formulations to find a stable matching under such contingent priorities (if they exist). Furthermore, we introduce a soft version of these contingent priorities, whereby the clearinghouse can skip some effective priority providers to ensure the feasibility of the problem.
Finally, we use data from the Chilean school choice system to illustrate the benefits of adopting our framework.

Even though it lacks some desirable properties, such as guaranteed existence and strategy-proofness, our results show that considering {absolute} contingent priorities can significantly improve the outcomes for students with siblings (e.g., preference of assignment and probability of getting assigned together with their siblings), while it has no sizable negative effect on students without siblings. Moreover, we find that {absolute} significantly outperforms other benchmarks specially designed to target students with siblings, such as the algorithm currently used in Chile and the stable matching that maximizes siblings assigned together. Finally, most of the drawbacks of our proposed framework can be alleviated using a hybrid approach that combines the soft version of absolute priorities with a minimum number of effective providers, as it ensures the existence of stable matchings and benefits a large number of students.
Therefore, clearinghouses focused on the joint assignment of siblings may largely benefit from implementing this hybrid approach.

Our work also illustrates the importance of carefully studying different approaches to achieve a specific outcome (e.g., increasing the number of siblings assigned together), as seemingly irrelevant choices may play a substantial role. For instance, our results show that varying the extent to which prioritized students can displace non-prioritized ones leads to entirely different outcomes. Similarly, the choice of tie-breaking rule and whether contingent priorities are hard or soft can have important effects on some properties of the mechanism, such as the existence of a stable matching. 
Finally, although we focus on school choice as a motivating example, there are many other settings where participants may care about the assignment of others and where clearinghouses may benefit from their joint assignment, including daycare and refugee resettlement. We believe the guidelines and insights derived from our work may help design policies to achieve those outcomes.

\bibliographystyle{informs2014}
\bibliography{bib}

\ACKNOWLEDGMENT{This work was funded by FRQ-IVADO Research Chair in Data Science for Combinatorial Game Theory, and the NSERC grant 2019-04557 and 2024-04051.}
\renewcommand{\theHchapter}{A\arabic{chapter}}
\begin{APPENDICES}
\section{Appendix to Section~\ref{sec: model}}\label{app: model}

\subsection{Contingent Stability and Other Definitions}\label{appendix:stability_comparisons}

In this section, we outline how our definitions of contingent stability compare to other definitions of stability with complementarities. Some of these definitions pertain to a single-level model, where applicants with complementarities apply, with the main application of interest being the market for matching residents to hospitals~\citep{klaus2005stable,mcdermid2010keeping,biro2011stable,marx2011stable,kojima13}.\footnote{For the work by~\citet{mcdermid2010keeping}, we consider the definition of stable matching used for \emph{consistent couples}. {Note that~\citet{csaji2023couples} use the same definition of blocking pair, so the same considerations as in~\citet{mcdermid2010keeping} apply.}} Some other definitions of stable matching with complementarities are given when there are siblings applying at different levels of the school system~\citep{Correa_2022,Dur_2022,sun2023daycare}. 

Next, we provide four illustrative instances of the matching problem with complementarities to display the key differences between our definitions of contingent stability and those given in the literature. The first three instances are drawn from the survey by~\cite{biro2013matching}. For all the instances, we assume that (i) there is only one school $c$ to which all the students apply to, and (ii) all the students are in group $|\groups|=1$. 

\begin{itemize}
    \item Instance 1: There is a family $f=\{ f_1, f_2\} $ and a single student $s$. All the students apply to the same level. School $c$ has two seats and the following preference list: $f_1 \succ s \succ f_2$.
    
    \item Instance 2:  There are two families, $f=\{ f_1, f_2\} $ and $h=\{ h_1, h_2\} $. All the students apply to the same level. School $c$ has two seats and the following preference list: $f_1 \succ h_1 \succ h_2 \succ f_2$. 
    
    \item Instance 3: There are two families, $f=\{ f_1, f_2\} $ and $h=\{ h_1, h_2\} $. All the students apply to the same level. School $c$ has two seats and the following preference list: $f_1 \succ h_1  \succ f_2 \succ h_2$.
    
    \item Instance 4: There is a family $f=\{ f_1, f_2\} $ and a single student $s$, where students $s_1$ and $f_1$ apply to level $l_1$, and student $f_2$ applies to level $l_2$.  School $c$ has one seat at level $l_1$ and at level $l_2$, and the following preference list: $s \succ f_1 \succ f_2  $. 
    
\end{itemize}


\begin{table}[ht!]
    \centering \footnotesize
\begin{tabular}{lcl}
            \toprule
           Instance  & Matching & \multicolumn{1}{c}{Stability} \\
            \midrule
             & \{$s$\}  & \citet{biro2011stable,kojima13} \quad\quad\quad\quad\quad\quad\quad \\
             \cmidrule(lr){2-3}
            1 & \{$f_1, f_2$\}  & Absolute, Partial \\
            \cmidrule(lr){2-3}
             & \{$s,f_1$\}  &  \citet{klaus2005stable,mcdermid2010keeping,marx2011stable,Correa_2022}\\
             \midrule
             & \{$f_1, f_2$\}  &  \citet{mcdermid2010keeping,marx2011stable}, \\
             & & \cite{kojima13}, Absolute, Partial \\
             \cmidrule(lr){2-3}
            2 & \{$h_1, h_2$\}  & \citet{marx2011stable,kojima13,biro2011stable}, Absolute \\
            \cmidrule(lr){2-3}
             & \{$f_1, h_1$\}  & \citet{klaus2005stable,mcdermid2010keeping,Correa_2022}, Absolute  
             \\
             \midrule
             & \{$f_1, f_2$\}  &  \citet{marx2011stable,kojima13,biro2011stable}, Absolute, Partial \\
             \cmidrule(lr){2-3}
            3 & \{$h_1, h_2$\}  & \citet{kojima13}, Absolute  \\
            \cmidrule(lr){2-3}
             & \{$f_1, h_1$\}  & \citet{klaus2005stable,mcdermid2010keeping,Correa_2022}, Absolute \\
             \midrule
            & \{$s$\}  & \cite{Dur_2022} \\
            \cmidrule(lr){2-3}
           4  & \{$f_1, f_2$\}  & \cite{Correa_2022},  Absolute \\
           \cmidrule(lr){2-3}
            & \{$s,f_2$\}  & \cite{sun2023daycare}, Partial  \\
            \bottomrule
\end{tabular}\caption{Definitions under which each feasible matching is considered stable}\label{Table:stability_definitions_comparisons}
\end{table}

In Table~\ref{Table:stability_definitions_comparisons}, we present the matchings that are considered stable according to the different definitions given in the literature and ours. In particular, in the third column, Absolute and Partial stand for contingent stable matching under absolute and partial priorities, respectively. Note also that in instances 1, 2 and 3, the school has only one level, while in Instance 4 the school has two levels. For (single-level) instances 1 to 3, the definitions by~\cite{mcdermid2010keeping} and~\cite{sun2023daycare} coincide, so we use only the first reference as a label in the table. 

The first observation is that our definitions of stability, along with the ones employed by~\cite{Correa_2022} and~\cite{sun2023daycare}, are the only ones that identify at least a stable matching for all the instances. Second, the definition given by~\cite{Dur_2022} can only be used for siblings that are not applying to the same level, while the remaining definition, with exception of~\cite{Correa_2022,sun2023daycare} and ours, are not given for matching markets with multiple levels.  
Third, our definitions align with the majority of the existent definitions. In instance 1, this is because our definition allows family $f$ to be matched together, thus encoding as stable a matching in which priorities play a critical role. Instances 2 and 3 display cases in which two families, $f$ and $h$, are competing for the same positions: In both instances 2 and 3, our definitions of contingent stability consider the matching \{$f_1, f_2$\} stable; moreover, Absolute identifies as stable all the feasible matchings deemed stable by some other definition from the literature. Therefore, our definition of contingent stable matching under absolute priorities dramatically expands the set of matchings deemed stable in comparison to the other definitions from the literature. Finally, in instance 4, our definition of contingent stable matching under absolute priorities identifies as stable the matching  \{$f_1, f_2$\}, in line with the definition by~\cite{Correa_2022}. \cite{sun2023daycare} treat schools with multiple levels as independent which prevents them to identify as stable the matching \{$f_1, f_2$\}.

\section{Appendix to Section~\ref{sec:existence}: Existence}\label{app:existence}
        \begin{proof}{Proof of Proposition~\ref{thm: inexistence absolute envy}.}
            It is enough to show the result for a single tie-breaking rule at the family level since all the other tie-breakers can be obtained through small perturbations.
            Consider an instance with four schools, $c_1, c_2, c_3,$ and $ c_4$, and two levels $\ell_1$ and $\ell_2$. School $c_1$ has only one seat at level $\ell_2$; schools $c_2$ and $c_4$ have one seat in each level; school $c_3$ has only one seat at level $\ell_1$.   There are four families of students, $f_a = \{ a_1, a_2\} $, $f_x = \{ x_1\} $, $f_d = \{ d_1, d_2\} $, $f_y = \{ y_2 \} $. Students $a_1, x_1, d_1$ apply at level $\ell_1$, and students $a_2, d_2, y_2,$ apply at level $\ell_2$. The preferences of the families (and of each student) are the following, $f_a: c_3 \succ c_4 $; $f_x: c_2 $; $f_d: c_1 \succ c_2 \succ c_3 $; $f_y: c_4 \succ c_1 $. Every school has the same tie-breaker which leads to the following initial order
            $y_2\succ x_1\succ d_1\succ d_2\succ a_1\succ a_2$. Finally, $|\groups| =1$.
            
            Note there is only one initial stable matching: \[\mu=\{(a_1,c_4),(a_2, \emptyset),(x_1,c_2),(d_1,c_3),(d_2,c_1),(y_2,c_4)\}.\] 
            Notice that the only matchings that may be contingent stable are those that would match $a_1, a_2$ in school $c_4$ or $d_1, d_2$ in school $c_2$. We study each case separately.

            \begin{enumerate}
                \item Suppose that there is a matching $\mu'$ such that $\mu'(c_4) = \{a_1,a_2\}$. In this case, $a_2$ cannot be an effective priority provider because condition (iii) in Definition~\ref{def: provider of contingent priority} is not satisfied since $y_2 \succ_{c_4} a_2$ and $y_2$ is either unassigned or matched to $c_1$ in $\mu'$, both of which are less preferred than $c_4$. Therefore, only $a_1$ could be an effective priority provider (no one applying to $\ell_1$ ranks $c_4$) and $a_2$ is the receiver. If $y_2$ is unmatched, that means $c_1$ is full at level $\ell_2$, which happens only if $d_2$ is matched there, but in this case $y_2$ would have contingent justified-envy towards $d_2$ (unless $d_2$ is an effective priority provider which cannot happen because $d_2$ does not satisfy condition (iii)). Therefore, $y_2$ is matched to $c_1$. Here we have two cases: (i)  $\mu'(c_2)=\{x_1,d_2\}$ and $\mu'(c_3) = \{d_1\}$; (ii) $\mu'(c_2)=\{d_1,d_2\}$, $\mu'(c_3) = \emptyset$, $x_1$ is unmatched. In (i), note that $d_2$ is an effective priority provider, so $d_1$ would have contingent justified envy towards $x_1$. In (ii), note that $c_3$ is empty which implies $\mu'$ is wasteful as $a_1$ prefers $c_3$ more than $c_4$.

                \item Assume that there is a matching $\mu'$ such that $\mu'(c_2) = \{d_1,d_2\}$. In this case, $d_1$ cannot be an effective priority provider because $x_1$ is more preferred than $d_1$ and it is unassigned, so $d_1$ does not satisfy condition (iv) in Definition~\ref{def: provider of contingent priority}. On the other hand $d_2$ is an effective priority provider since no other student applying to level $\ell_2$ ranks $c_2$, consequently, $d_1$ is receiving priority. Since $x_1$ is unassigned then, from the previous point $a_1$ and $a_2$ cannot be together, so $a_1$ is matched to $c_3$. Consequently, $\mu'(c_4) =\{y_2\}$ since $a_2$ cannot compete with $y_2$. Therefore, $c_1$ is empty which would be wasteful as $d_2$ prefers $c_1$ over $c_2$. \hfill\Halmos
            \end{enumerate}

        
        
        \end{proof}
    
        \begin{proof}{Proof of Proposition~\ref{thm: existence partial}.}
         First, we show that a contingent stable matching may not always exist under individual initial tie-breakers. Then, we show that the contingent stability in the partial case coincides with initial stability when we consider family level tie-breakers and, thus, existence is guaranteed.

            \textbf{Individual Level Tie-breakers.} There are four schools, $c_1, c_2, c_3,$ and $c_4$, and two levels $\ell_1$ and $\ell_2$.  At level $\ell_1$, schools $c_1$ and $c_3$ have one seat, and all the other schools have two seats. At level $\ell_2$, $c_1$ has one seat, and all the other schools have zero seats. There are five families of students, $f_a = \{ a_1, a_1'\} $, $f_x = \{ x_1\} $, $f_e = \{ e_1\} $, $f_d = \{ d_1, d_1'\} $, $f_h = \{ h_1, h_2 \} $. All the students, except for $h_2$, apply to level $\ell_1$. The preferences of the students (which are the same for both levels) are the following, $f_a:  c_3 \succ c_4 $; $f_x: c_2 $; $f_e: c_2 $; $f_d: c_1 \succ c_2 \succ c_3 $; $f_h: c_4 \succ c_1 $. The random tie-breakers are  the same for all schools and lead to the following initial order
            $h_2\succ d_1\succ x_1\succ e_1\succ d_1'\succ a_1\succ h_1\succ a_1'$. Finally, assume $|\groups|=1$.
    
            In this instance, there is only one initial stable matching: \[\mu=\{(a_1,c_4),(a_2,\emptyset),(x_1,c_2),(e_1,c_2),(d_1,c_1),(d_2,c_3),(h_1,c_4),(h_2, c_1)\}.\] 
        
            Notice that the only matchings that may be contingent stable are those that would match $a_1, a_1'$ in school $c_4$, $d_1, d_1'$ in school $c_2$, or $h_1, h_2$ in school $c_1$. We now study each case separately.
            \begin{enumerate}
                \item Let $\mu'$ be a matching such that $\mu'(c_4) = \{a_1,a_1'\}$. It is easy to see that $a_1$ is the effective priority provider and $a_1'$ is the priority receiver. Due to partial priorities $h_1$ is either unmatched or matched to $c_1$. If $h_1$ is unmatched is because $c_1$ is full at level $\ell_1$ which means $d_1$ and $d_1'$ are matched there. However, to satisfy contingent stability $h_2$ must also be matched to $c_1$ at level $\ell_2$ (otherwise it would be wasteful). Clearly, $h_2$ is an effective priority provider there, so $h_1$ receives $h_2$'s tie-breaker displacing $d_1'$. Consequently, $d_1'$ is matched to $c_3$ (because $d_1'$ cannot compete with $x_1$ and $e_1$ in $c_2$). This leaves one seat free in $c_3$ which is wasteful because $a_1$ or $a_1'$ prefer $c_3$ more than $c_4$.
                \item Let $\mu'$ be a matching such that $\mu'(c_2)=\{d_1,d_1'\}$. Note that the only option for this is that $h_1$ and $h_2$ are matched to $c_1$ (if $h_1$ is matched to $c_4$ or unmatched is the same reasoning), however, that leaves an empty seat in $c_1$ which is wasteful as $d_1$ prefers $c_1$ over $c_2$.
                \item Let $\mu'$ be a matching such that $\mu'(c_1)\supseteq\{h_1,h_2\}$. The only interesting case here is that $d_1$ is matched to $c_1$ as well. This case can be analyzed as the first point since we already know that $a_1$ and $a_1'$ cannot be together.
            \end{enumerate}

            \textbf{Family Level Tie-breakers.} This result easily follows by noting  from Definition~\ref{def: partial contingent priority} that: (i) contingent groups coincide with the initial groups; (ii) $\lotteries^\mu =\lotteries$ for any $\mu$ when tie-breakers are at the family level. Therefore, the set of contingent stable matchings under partial priorities coincide with the set of initial stable matchings.\hfill\Halmos 

        \end{proof}
    
\section{Appendix to Section~\ref{sec:existence}: Complexity}\label{app:complexity}

\subsection{Proof of Theorem~\ref{theorem:absolute_complexity}}
In this section, we use a reduction from the complete stable matching problem with ties and incomplete preference lists (COM-SMTI). This problem consists of two sides (men and women), where each preference list is of size at most 3. Men have strict preferences lists and women have either a strict preference list of size at most 3 or a tie of size 2 as a preference list. The decision problem of whether there exists a complete stable matching in this setting is known to be NP-complete~\citep{mcdermid2010keeping}.
\begin{proposition}\label{prop:comsti_absolute}
COM-SMTI can be reduced in polynomial time to Absolute.
\end{proposition}
For this reduction, our main gadget is the example that we use for non-existence in Proposition~\ref{thm: inexistence absolute envy}. 
\begin{remark}\label{remark:contingent_example}
There are two main observations from the example in Proposition~\ref{thm: inexistence absolute envy}:
\begin{enumerate}
\item Suppose there is an extra school $\hat{c}$ (which offers only level $\ell_2$ and has one seat) such that  $\hat{c}\succ_{d_2}c_1$. Then, in this new instance, the following is a contingent stable matching under absolute priorities: \(\mu(\hat{c}) = \{d_2\}, \mu(c_1)=\{y_2\}, \mu(c_2) = \{x_1\}, \mu(c_3) = d_1, \mu(c_4) = \{a_1,a_2\}\). In words{, relative to the initially stable matching given in the proof of Proposition~\ref{thm: inexistence absolute envy}}, $a_2$ displaced $y_2$, who was moved to $c_1$ thanks to the space that $d_2$ left (no cycling occurs as in the example).
\item Suppose we add an extra seat to school $c_4$ in level $\ell_2$. In this setting, $a_2$ can be matched to $c_4$ which gives a contingent stable matching under absolute priorities.
\end{enumerate}
\end{remark}
\begin{proof}{Proof of Proposition~\ref{prop:comsti_absolute}.}
Let $I$ be an instance of COM-SMTI composed by $U$ the set of men, $W = W^s\cup W^t$ the set of women where $W^s$ is the set of women with a strict preference list and $W^t$ the set of women with a tie of length 2 as a preference list. {Without loss of generality, we assume that $|U|=|W|$.} We denote by $P(u)$ the preference list for $u\in U$, similarly $P(w)$ for $w\in W$. For a women with a tie as a preference list we denote by $u$ the man in the first position and $u'$ the man in the second position (this become relevant in the preference lists in Figure~\ref{fig:preference_list}).

\textbf{Instance Construction.} We construct an instance $I'$ from a given instance $I$ as follows:
\begin{enumerate}
\item For every man $u\in U$, we create 4 dummy schools $c^1_u,c^2_u, c^3_u, c^4_u$, 4 dummy families $f^a_u = \{a_u^1,a^2_u\},f^d_u = \{d_u^1,d^2_u\}, f_u^x = \{x^1_u\}, f^y_u = \{y^2_u\}$ where the superscripts of the students indicate the level in which the students are applying to. The preference lists and capacity of the schools are  in Figure~\ref{fig:preference_list}. In particular, the preference list of the schools $c^j_u$ with $j\in[4]$ are all the same.
Note that this construction follows the counter-example in Proposition~\ref{thm: inexistence absolute envy}. 
\item For every women $w\in W^s$ we create a school $c_w$ that only offers level 2 with capacity 1.
\item For every women $w\in W^t$, we create two dummy schools $h^1_w,h^2_w$ that only offer one seat in level 2, 4 dummy schools $c^1_w,c^2_w, c^3_w, c^4_w$, 5 dummy families $f^a_w = \{a_w^1,a^2_w\},f^d_u = \{d_w^1,d^2_w\}, f_w^x = \{x^1_u\}, f^{y}_w = \{y^2_w\}, f^{\hat{y}}_w = \{\hat{y}^2_w\}$ where superscript indicates the level the students are applying to. We present the preference lists and capacity of the schools in Figure~\ref{fig:preference_list}. 
\end{enumerate}
Clearly the construction above can be done in polynomial time. In Figure~\ref{fig:preference_list}, $P(u)$ corresponds to the original preference list of $u\in U$ with women $w\in W^s$ replaced by $c_w$ and women $w\in W^t$ replaced by either $h^1_w$ or $h^2_w$ depending on the position of $u$ in the tie (consider an arbitrary order that applies to all the ties). On the other hand, for each $w\in W$, $P(w)$ corresponds to the original preference list of $w$ with men replaced by $d_u^2$. \\

\noindent{\bf Remark:} The main idea of the construction is that the instance of COM-SMTI is embedded in level 2 of schools $c_w$, $h^1_w$, $h^2_w$ (women) and students $d^2_u$ (men).

\begin{figure}[htpb]
{\small\begin{align*}
&\{a^1_u,a^2_u\}:\; c^3_u \succ c^4_u & {\text{for all} \; u\in U}\\
&x^1_u:\; c^2_u & {\text{for all} \; u\in U}\\
&\{d^1_u,d^2_u\}:\; P(u)\succ c^1_u\succ c^2_u\succ c^3_u & {\text{for all} \; u\in U}\\
&\{y^2_u\}:\; c^4_u\succ c^1_u& {\text{for all} \; u\in U}\\
&c^1_u (0,1), c^2_u (1,1), c^3_u (1,0),c^4_u (1,1):\; y^2_u\succ x^1_u \succ  d^1_u \succ d^2_u \succ a^1_u \succ a^2_u & {\text{for all} \; u\in U}\\
& \\
&c_w (0,1):\; P(w) & \text{for all} \; w\in W^s\\
&\{a^1_w,a^2_w\}:\; c^3_w \succ c^4_w & \text{for all} \; w\in W^t\\
&x^1_w:\; c^2_w &\text{for all} \; w\in W^t\\
&\{d^1_w,d^2_w\}:\; c^1_w\succ c^2_w\succ c^3_w &\text{for all} \; w\in W^t\\
&\{y^2_u\}:\; h^1_w\succ  c^4_w\succ c^1_w &\text{for all} \; w\in W^t\\
&\{\hat{y}^2_u\}:\; h^2_w\succ c^4_w\succ c^1_w &\text{for all} \; w\in W^t\\
&h^1_w (0,1):\; d^2_u\succ y^2_w &\text{for all} \; w\in W^t\\
&h^2_w (0,1):\; d^2_{u'}\succ \hat{y}^2_w  &\text{for all} \; w\in W^t\\
&c^1_w (0,1), c^2_w (1,1), c^3_w (1,0),c^4_w (1,2):\; \hat{y}^2_w\succ y^2_w \succ x^1_w \succ  d^1_w \succ d^2_w \succ a^1_w \succ a^2_w &\text{for all} \; w\in W^t\\
\end{align*}}
\caption{Capacities are next to each school in parenthesis where each component indicates the level's capacity.}
\label{fig:preference_list}
\end{figure}


\begin{claim}
If $M$ is a {complete} stable matching in $I$, then there exists a contingent stable matching in $I'$.
\end{claim}
\begin{proof}{Proof.}
Given a {complete} matching $M$ in $I$, we construct a matching $\mu$ in $I'$ as follows: For every man $u\in U$ and a women $w\in W$ such that $M(w) = u$, then we match $\mu(c_w) = d^2_u$ if $w\in W^s$ or $\mu(h^1_w) = d^2_u$ if $w\in W^t$ and $u$ is in the first position of tie or, otherwise $\mu(h^2_w) = d^2_u$. The remaining matches are as follows:
\begin{align*}
\mu(c^1_u) &= \{y^2_u\}, \quad \mu(c^2_u)= \{x^1_u\}, \quad \mu(c^3_u) = \{d^1_u\}, \quad\mu(c^4_u) = \{a^1_u,a^2_u\}, 
\end{align*}
If $d^2_u$ is matched to $h^1_w$ then we add the following pairs
\begin{align*}
\mu(h^2_w) &= \hat{y}^2_w \quad \mu(c^1_w) =  d^2_w \quad \mu(c^2_w) = \{x^1_w\} \quad \mu(c^3_w) = \{d^1_w\} \quad \mu(c^4_w) = \{a^1_w,a^2_w,y^2_w\}
\end{align*}
Otherwise, if $d^2_u$ is matched to $h^2_w$ then we add the following pairs
\begin{align*}
\mu(h^1_w) &= y^2_w \quad \mu(c^1_w) =  d^2_w \quad \mu(c^2_w) = \{x^1_w\} \quad \mu(c^3_w) = \{d^1_w\} \quad \mu(c^4_w) = \{a^1_w,a^2_w,\hat{y}^2_w\}
\end{align*}
{Note that since matching $M$ is complete, no two distinct $d^2_u$, $d^2_{u'}$ can be tentatively matched to the same $c_w$ where $w\in W^s$; moreover, no students in $I'$ are unmatched.} Clearly this matching is contingent stable under absolute priorities thanks to the observations in Remark~\ref{remark:contingent_example}. \hfill\Halmos
\end{proof}
\begin{claim}
Given a contingent stable matching under absolute priorities $\mu$ in $I'$, we can construct a {complete} stable matching $M$ in $I$.
\end{claim}
\begin{proof}{Proof.}
First, we show that in $\mu$ a student $d^2_u$ must be either matched to $c_w$, $h^1_w$ or $h^2_w$ in $P(u)$. Suppose otherwise, then it is matched to $c^1_u$ but this will contradict the contingent stability of $\mu$ in that gadget (in other words, this would lead to the counterexample in Proposition~\ref{thm: inexistence absolute envy}).

Now we show that no school created from a woman $w\in W$  will be matched to more than one student of type $d^2_u$ in $\mu$. First note that for $c_w$ this is trivial as it has capacity 1. Now we need to show that at most one school, between $h^1_w$ and $h^2_w$, is matched to a student of the type $d^2_u$. Suppose the opposite in which $h^1_w$ is matched to $d^2_u$ and $h^2_w$ is matched to $d^2_{u'}$. Then $\hat{y}^2_w$ and $y^2_w$ would be match to $c^4_w$. However, $a^2_w$ can displace $y^2_w$ by claiming contingent priority thanks to $a^1_w$. Then $y^2_w$ would displace $d^2_w$ (e.g. currently matched to $c^1_w$) which then will be matched to $c^2_w$. This will create the same ``cycle'' that we found in Proposition~\ref{thm: inexistence absolute envy}.

In conclusion, every $d^2_u$ is matched to $c_w$, $h^1_w$ or $h^2_w$. Moreover, schools of type $h^1_w$ and $h^2_w$ receive at most one student of the form $d^2_u$. Matching $M$  is constructed as 
\[
M = \{(u,w)\in U\times W: \ (d_u^2,c_w)\in \mu \text{ or }(d_u^2,h^1_w)\in \mu \text{ or } (d_u^2,h^2_w)\in \mu \}, 
\]
whose stability result from the contingent stability of $\mu$ in those schools which is essentially one level and no siblings apply to those schools (siblings of type $d^1_u$ only applies to level 1 and those schools do not have that level), therefore, the standard stability. {Finally,  since every $d_u$ corresponds to a man in $I$, and every $d_u$ is matched, then $M$ is complete.}\hfill\Halmos
\end{proof}
Since Absolute is clearly in NP, we conclude the proof of Proposition~\ref{prop:comsti_absolute} and Theorem~\ref{theorem:absolute_complexity} from both claims. \hfill \Halmos
\end{proof}

\subsection{Proof of Theorem~\ref{theorem:partial_complexity}}

\begin{proposition}\label{prop:comsti_partial}
COM-SMTI can be reduced in polynomial time to Partial.
\end{proposition}
\begin{proof}{Proof.}
    The proof follows a similar reasoning as the one for absolute priorities, while using as a pivot model the one from Proposition~\ref{thm: existence partial}. The main differences are the following:

    \begin{enumerate}
        \item If $u\in U$, then we create dummy schools and students as in Proposition~\ref{thm: existence partial}, where the preference list of family $f(h)$ is $P(u)\succ c^4_u\succ c^1_u$. 
        \item If $w\in W^t$ is such that it has preference list $(u,u')$, then we create dummy schools and students as in Proposition~\ref{thm: existence partial} with the following modifications: (1) we also add two students $y^2_w, \Bar{y}_w^2$ and two schools $z^2_w, \Bar{z}_w^2$ (each with one spot at level 2); the preference of $y^2_w$ is $z^2_w\succ c^1_w$ (similar for $\Bar{y}^2_w$); the preference of $z^2_w $ is $d_u \succ y^2_w$ (similar for $\Bar{z}^2_w$). (2) The preference list of the schools  $c^1_w, c^2_w, c^3_w, c^4_w$ is $ y^2_w \succ \Bar{y}^2_w \succ h^2_w $ $ \succ d^1_w \succ x^1_w  \succ e^1_w $ $\succ d^{1'}_w   $ $ \succ a^1_w \succ h^1_w \succ a^{1'}_w  $.
    \end{enumerate}
    \hfill \Halmos
\end{proof}

   \section{Appendix to Section~\ref{sec:incentives}: Incentives}\label{app:incentives}
    
        \begin{proof}{Proof of Proposition~\ref{prop: incentives absolute}.}
            First, note that it is enough to show the result for a single tie-breaking rule at the family level since we can construct examples for all the other tie-breaking rules by simply adding a small perturbation to this counter-example.
            
            Consider an instance of the problem with two single students $s_1, s_2$ and two families $f = \lrl{f_1, f_2}$, $f' = \lrl{f_1', f_2'}$, with students $s_1, f_1, f_1'$ applying to level $\ell_1$ and $s_2, f_2, f_2'$ applying to level $\ell_2$. In addition, suppose there are four schools $\schools = \lrl{c_1, c_2, c_3, c_4}$ each offering one seat in each level except for $c_2$ in $\ell_1$ and $c_3$ in $\ell_2$, for which $q_{c_2}^{\ell_1} = q_{c_3}^{\ell_2} = 0$. Suppose that students' preferences are:
            \begin{align*}
                f_1&: c_1 \succ c_3, \hspace{1.8cm} f_2: c_1 \succ c_2 \\
                f_1'&: c_3 \succ c_4, \hspace{1.8cm} f_2': c_3 \succ c_4  \\
                s_1&: c_1 \succ c_2, \hspace{1.8cm} s_2: c_4 \succ c_1.
            \end{align*}
            Finally, the clearinghouse uses a single tie-breaker at the family level (i.e., $p_{s,c} = p_{f(s)}$ for all $s\in \students$ and $c\in \schools$) whose realized values are:
            \[
                p_{s_2} < p_{s_1} < p_{f} < p_{f'} \quad\Leftrightarrow\quad s_2\succ s_1\succ f \succ f'
            \]
            Note that, if every family reports their preferences truthfully, then there is a unique \emph{rank-optimal} contingent stable matching under absolute priorities: 
            \begin{align}\label{eq: assignment proof incentives truthful}
                \mu &= \lrl{(s_1, \emptyset), (f_1, c_1), (f_1', c_3), (s_2, c_4), (f_2,c_1), (f'_2,\emptyset)}.
            \end{align}
            In this case, four students get assigned to their top choice and two of them get unassigned. Note that $f_2'$ may unilaterally improve their assignment by adding more schools to their reported list. To see this, suppose that $f_2'$ reports the following preference order: 
            \[
            f_2': c_4 \succ c_3 \succ c_1.
            \]
            Based on these new preferences, the matching $\mu$ is no longer rank-optimal as it is dominated by the matching  
            \begin{equation}\label{eq: assignment proof incentives misreport}
                \mu' = \lrl{(s_1, c_1), (f_1, c_3), (f_1', c_4), (s_2, c_1), (f_2, c_2), (f_2', c_4)},
            \end{equation}
            since four students ($\lrl{f_1, f_1', s_2, f_2}$) get assigned to their second choice, two ($\lrl{s_1, f_2'}$) get their top choice, and no student results unassigned. Hence, the assignment $\mu'$ leads to a strict improvement over the sum of the students' rankings and, thus, $f_2'$ can improve their assignment by misreporting their preferences.
        
        \end{proof}
        
        \begin{proof}{Proof of Proposition~\ref{prop: incentives partial}.}
            We first show that the mechanism is not strategy-proof under individual tie-breakers. As before, it is enough to show that this is the case under single tie-breakers, as the result for multiple tie-breakers can be obtained by adding a small perturbation.
        
            Consider the same market as described in the proof of Proposition~\ref{prop: incentives absolute} but with a small variation in the tie-breakers. Specifically, suppose they are given by:
            \[
            p_{f_2} < p_{s_1} < p_{f_1} < p_{f_1'} < p_{s_2} < p_{f_2'} \quad \Leftrightarrow \quad f_2\succ s_1\succ f_1\succ f_1'\succ s_2\succ f_2'.
            \]
            Then, the assignment in~\eqref{eq: assignment proof incentives truthful} is also the only {rank-optimal} contingent stable matching under {partial} priorities. Moreover, as before, $f_2'$ can improve its assignment by misreporting their preferences by including all the schools in the following order:
            \[
            f_2': c_4 \succ c_1 \succ c_3.
            \]
            In this case, the assignment $\mu'$ in~\eqref{eq: assignment proof incentives misreport} that $f_2'$ strictly prefers is also feasible and leads to an overall better assignment if the goal is to find a rank-optimal contingent stable matching under partial priorities.
            
            Finally, the fact that the mechanism returning a rank-optimal contingent stable is strategy-proof under family level tie-breakers is a corollary of Proposition~\ref{thm: existence partial} and, specifically, of the equivalence between {partial} contingent priorities and initial stability. 
        \end{proof}
        
        \begin{proof}{Proof of Proposition~\ref{prop: SP-L}.}
            We prove the result for the mechanism to find a contingent stable matching with {absolute} priorities with a single tie-breaking rule. The proof for the other cases follows similarly. ~\citet{azevedo18} show that a sufficient condition for a mechanism to be strategy-proofness in the large is to be (i) \emph{semi-anonymous} and (ii) \emph{envy-freeness but for ties} (EF-TB). Hence, it is enough to show that our mechanism satisfies these two properties.
        
            \textbf{Semi-anonimity.} As defined in~\cite{azevedo18}, a mechanism is semi-anonymous if there is a partition $\Theta$ of the set of students and, within each element of the partition \(\theta\in \Theta\), there is a finite set of types \(T_\theta\) that specifies the set of possible actions for a student with that type. Specifically, if student $s$ belongs to $\theta$ and $t\in T_\theta$ is their type, then the set of possible actions that $s$ can take is defined as $A_{\theta,t} \subseteq A_\theta$. 
            In our school choice setting, we can (i) set $\Theta = \lrl{1, \ldots, \lra{\groups}+1}^\schools$, i.e., the partition given by the vector of contingent groups that students belong to in each school, (ii) define the types as students' preferences \(\succ_s\), and (iii) set the actions to be the list of preferences that students can submit. Then, two students \(s\) and \(s'\) such that $g(s,c) = g(s',c)$ for all $c \in \schools$ (i.e., belong to the same group in each school and, thus, to the same partition element $\theta \in \Theta$) and $\succ_s\; =\; \succ_{s'}$ (belong to the same type as they share the same preferences) differ only because of their tie-breakers. Note that \(\Theta\) has finite cardinality (since the number of schools and groups are finite) and that there is a finite set of preference lists $\succ_s$ that a student $s$ can potentially report since the number of schools is finite. Therefore, we know that the number of groups, the number of types, and the set of possible actions for each group and type are finite, so the mechanism is semi-anonymous.
        
            \textbf{EF-TB.} Given a market with \(n\) students, a direct mechanism is a function \(\Phi^n:\; T^n\rightarrow \Delta(C \cup \lrl{\emptyset})^n \)
            that receives a vector of types \(T\) (the application list of each student) and returns a (potentially randomized) feasible allocation. In addition, let \(u_t(\tilde{c})\) be the utility that a student with type \(t\in T_\theta, \theta\in \Theta\) gets from the lottery over assignments \(\tilde{c}\in \Delta(C\cup \lrl{\emptyset})\) (note that, by assumption, two students belonging to the same type have exactly the same preferences and, thus, get the same utility in each school \(c\in C\cup \lrl{\emptyset}\)).
            Then, a semi-anonymous mechanism is envy-free but for tie-breaking if for each \(n\)
            there exists a function \(x^n: (T\times [0,1])^n \rightarrow \Delta(C\cup\lrl{\emptyset})^n\) such that
            \[
            \Phi^n(t) = \int_{l\in [0,1]^n} x^n(t,l) dl
            \]
            and, for all \(i,j,n,t\) and \(l\) with \(l_i \geq l_j\), and if \(t_i\) and \(t_j\) belong to the same type, then
            \[
            u_{t_i}\lrc{x_i^n(t,l)} \geq u_{t_i}\lrc{x_j^n(t,l)}.
            \]
            In plain words, to show that our mechanism is EF-TB, we need to show that whenever two students that belong to the same type differ in their tie-breakers, then the assignment of the student with the higher lottery cannot be worse than that of the other student. This follows directly from the definition of the contingent priority order $\succ^\mu$, since for each group, we know that the clearinghouse breaks ties within each group using the tie-breaking rule. As a result, if two students $s$ and $s'$ belong to the same group, we know that the resulting matching \(\mu\) satisfies \(\mu(s) \succ_s \mu(s')\) if \(s\succ_c s'\) for all \(c\in C\). Then, for any function $x$, it is direct that \(u_{t_s}\lrc{x_{s}^{n} (t,l)} \geq u_{t_s}\lrc{x_{s'}^{n} (t,l)}\). Hence, we conclude that our mechanism is EF-TB, and therefore it is strategy-proof in the large.
        \end{proof} 
        \begin{remark}
            Note that our notions of absolute/partial and hard/soft only differ on how they update students' contingent groups and lotteries. Thus, once these are defined, the contingent priority order remains fixed and known. For this reason, the proof of strategy-proofness in the large is analogous for any combination of these elements.
        \end{remark}
\section{Appendix to Section~\ref{sec: formulations}}\label{app:formulations}

\subsection{Technical Lemmas}\label{app: formulations technical lemmas}

\begin{lemma}\label{lemma: ensure highest lottery provider} 
    Given feasible $\mathbf{x}^\star, \mathbf{z}^\star, \mathbf{y}^\star$ for Formulation~\ref{eq: absolute formulation} (resp. Formulation~\ref{eq:partial_formulation}), there exists another feasible solution $\mathbf{x}^\blacktriangledown, \mathbf{z}^\blacktriangledown, \mathbf{y}^\blacktriangledown$ for Formulation~\ref{eq: absolute formulation} (resp. Formulation~\ref{eq:partial_formulation}) where the effective provider of priority of each family is the student satisfying Definition~\ref{def: provider of contingent priority} with the most favorable initial tie-breaker at the school. Moreover, $\mathbf{x}^\star=\mathbf{x}^\blacktriangledown$, i.e., the students' ranking has the same value.    
\end{lemma}
\begin{proof}{Proof.}
     Let $\mathbf{x}^\star, \mathbf{z}^\star, \mathbf{y}^\star$ be a solution (for either formulation), and suppose that $\exists f\in \families$ with $\lrl{a,a'} \subseteq f$ where both $a,a'$ satisfy Definition~\ref{def: provider of contingent priority}, $z^\star_{a,c} = 1$,  $z^\star_{a',c} = 0$, and $a'\succ_c a$. Consider the case when $x^\star_{a',c} = 1$ (the converse case is direct). Then, we can always set $y^\star_{a, a'', c} = 1$ for all $a'' \in f(a)\setminus \lrl{a}$ such that $x^\star_{a'',c} = 1$, as it would not change the objective and would still satisfy all constraints. Then, let $\mathbf{x}^\blacktriangledown, \mathbf{z}^\blacktriangledown, \mathbf{y}^\blacktriangledown$ be such that $\mathbf{x}^\blacktriangledown = \mathbf{x}^\star$, 
        $z_{a,c}^\blacktriangledown = 0$, $z_{a',c}^\blacktriangledown = 1$, $z_{s,c'}^\blacktriangledown = z_{s,c'}^\star$ for all $(s,c') \notin\lrl{(a,c), (a',c)}$,  
        $y_{a,a',c}^\blacktriangledown = 0$, $y_{a', a'',c}^\blacktriangledown = 1$ for all $a'' \in f(a') \setminus \lrl{a'}$ with $x^\blacktriangledown_{a'',c} = 1$, and $y_{s, s',c'}^\blacktriangledown = y_{s, s',c'}^\star$ for all $s,s'\notin f, \; c'\in \schools$. 
        Since $\mathbf{x}^\star = \mathbf{x}^\blacktriangledown$, the objective function value remains unchanged. 
        Also, it is easy to see that $\mathbf{x}^\blacktriangledown \in \mathcal{P}$, $\mathbf{z}^\blacktriangledown \in \mathcal{R}(\mathbf{x}^\blacktriangledown)$, and $\mathbf{y}^\blacktriangledown \in \mathcal{R}(\mathbf{x}^\blacktriangledown, \mathbf{z}^\blacktriangledown)$. Moreover, note that 
        \[
        z^\star_{s,c'} + \sum_{s'\in f(s)\setminus \lrl{s}} y^\star_{s',s,c'} = z^\blacktriangledown_{s,c'} + \sum_{s'\in f(s)\setminus \lrl{s}} y^\blacktriangledown_{s',s,c'}, \quad \forall s\in \students, \;s'\in f(s), \; c'\in \schools.
        \]
        
        In the absolute case, this implies that both sets of constraints~\eqref{eq: constraint stability abs indep} and~\eqref{eq: tie-breaking contingent abs indep} are directly satisfied by $\mathbf{x}^\blacktriangledown, \mathbf{z}^\blacktriangledown, \mathbf{y}^\blacktriangledown$. 
        
        In the partial case, note that the right-hand side of~\eqref{eq: constraint stability par indep} is always at least as high under $\mathbf{x}^\blacktriangledown, \mathbf{z}^\blacktriangledown, \mathbf{y}^\blacktriangledown$ compared to $\mathbf{x}^\star, \mathbf{z}^\star, \mathbf{y}^\star$ for pairs $(s, c)$ with  $\ell(s)=\ell(a)$ (since $y_{a',a,c}^\star = 0$ and $y_{a',a,c}^\blacktriangledown = 1$), while they are equal if we consider the constraint for $(s,c)$ with $\ell(s)=\ell(a')$ since $\indicator{a' \prec_c s \prec_c a} = 0$ as $a'\succ_c a$.  
        Hence, the right-hand side of~\eqref{eq: constraint stability par indep} is always at least as high under $\mathbf{x}^\blacktriangledown, \mathbf{z}^\blacktriangledown, \mathbf{y}^\blacktriangledown$ compared to $\mathbf{x}^\star, \mathbf{z}^\star, \mathbf{y}^\star$. 
        Finally, note that the right-hand side of the set of constraints~\eqref{eq: tie breaking contingent par indep} is also always at least as high under $\mathbf{x}^\blacktriangledown, \mathbf{z}^\blacktriangledown, \mathbf{y}^\blacktriangledown$ compared to $\mathbf{x}^\star, \mathbf{z}^\star, \mathbf{y}^\star$. To see this, note first that $a \succ_c s\wedge s'$ implies $\indicator{a \prec_c s\wedge s'} = \indicator{a' \prec_c s\wedge s'} = 0$, so these constraints are always satisfied. On the other hand, if $a \prec_c s \wedge s'$, we have two cases:
        \begin{itemize}
            \item If $s$ is such that $\ell(s) = \ell(a)$, then 
            \(
            y^\star_{a',a,c}\cdot \indicator{a\wedge a' \succ_c s \wedge s'} = 0 \leq y^\blacktriangledown_{a',a,c}\indicator{a\wedge a' \succ_c s \wedge s'},
            \)
            where the equality holds since $y^\star_{a',a,c} = 0$ and $a \prec_c s\wedge s'$, while the inequality follows since $y^\blacktriangledown_{a',a,c} = 1$ and potentially $a' \succ_c s\wedge s'$ (which would imply strict inequality).
            \item If $s$ is such that $\ell(s) = \ell(a')$, then~\eqref{eq: tie breaking contingent par indep} follows because  $a'\succ_c a$ implies that $\indicator{a\wedge a' \succ_c s \wedge s'} = \indicator{a' \succ_c s \wedge s'} $ 
            so, if $a'\succ_c s \wedge s'$, then $x_{a',c}$ is multiplied by zero, and if $a'\prec_c s \wedge s'$ would imply that there is no term in the rightmost sum, leading to $x^\star_{a,c}=0$ which is a contradiction to the contingent stability of $\mathbf{x}^\star$.
        \end{itemize}
        Thus, the right hand side of the set of constraints~\eqref{eq: tie breaking contingent par indep} is always at least as high under $\mathbf{x}^\blacktriangledown, \mathbf{z}^\blacktriangledown, \mathbf{y}^\blacktriangledown$ compared to $\mathbf{x}^\star, \mathbf{z}^\star, \mathbf{y}^\star$, so these constraints are always satisfied for the former. 
\end{proof}

\begin{lemma}\label{lemma:existence_of_provider}
{Consider a contingent stable matching $\mu$ (either under absolute or partial priorities) and a school $c$. Then, every family of size at least 2 with some member $s$ matched to $c$ and another member $s'\in f(s)\setminus\{s\}$ either matched to $c$ or a less preferred school must have at least one member satisfying the conditions of priority provider given in Definition~\ref{def: provider of contingent priority} for school $c$.}
\end{lemma}
\begin{proof}{Proof.}
Fix a contingent stable matching $\mu$. Suppose by contradiction that there exists a school $c$ and a family $f$ with $|f|\geq 2$ such that {$\exists s'\in f(s)$ with $\mu(s') \preceq_{s'} c$ and} every sibling $s\in f$ with $\mu(s)=c$ does not satisfy the conditions of priority provider given in Definition~\ref{def: provider of contingent priority} for school $c$. In particular, each $s\in f$ with $\mu(s)=c$ does not satisfy condition (iii) in Definition~\ref{def: provider of contingent priority}, which means that $\lra{\lrl{\hat{s} \in \students^{\ell(s)}\;:\; \hat{s} \succ_c s, \; c \succeq_{\hat s} \mu(\hat{s}) }} \geq q_c^{\ell(s)}$. From this, we note that if $\lra{\lrl{\hat{s} \in \students^{\ell(s)}\;:\; \hat{s} \succ_c s, \;c \succ_{\hat{s}} \mu(\hat{s}) }} = 0$, then $\lra{\lrl{\hat{s} \in \students^{\ell(s)}\;:\; \hat{s} \succ_c s, \; c = \mu(\hat{s}) }} \geq q_c^{\ell(s)}$, which leads to a contradiction due to the capacity of school $c$ in level $\ell(s)$ and the fact that $\mu(s)=c$. Therefore, we must have $\lra{\lrl{\hat{s} \in \students^{\ell(s)}\;:\; \hat{s} \succ_c s, \; c \succ_{\hat{s}} \mu(\hat{s}) }} >0$, in other words, there exists a student $\hat{s}$ such that $\hat{s} \succ_c s, \; c \succ_{\hat{s}}\mu(\hat{s})$. Since there is no effective priority provider in school from family $f(s)$ then $\hat{s}\succ_c^\mu s$ (either for absolute or partial priorities) which, along with $c \succ_{\hat{s}}\mu(\hat{s})$, implies that $\hat{s}$ has contingent justified-envy towards $s$. This contradicts the contingent stability of $\mu$.\hfill\Halmos
\end{proof}

\subsection{Absolute Priorities}\label{app:proof_correcteness_absolute}
In this section, we focus on the proof of Theorem~\ref{thm:correctness_absolute}. Let us denote by $\mathcal{A}$ the set of points $(\mathbf{x},\mathbf{z},\mathbf{y})$ that satisfy~\eqref{eq: absolute formulation} and by $\mathcal{M}$ the set of contingent stable matchings under absolute priorities. We separate the proof in two parts: In the first part, we show that, for any feasible solution $(\mathbf{x},\mathbf{y},\mathbf{z})\in\mathcal{A}$, we can construct a matching $\mu\in\mathcal{M}$ (Appendix~\ref{app:proof_correcteness_absolute_partI}). In the second part, we show the opposite direction, i.e., for any matching $\mu\in\mathcal{M}$, there exists a feasible point $(\mathbf{x},\mathbf{y},\mathbf{z})\in\mathcal{A}$ (Appendix~\ref{app:proof_correcteness_absolute_partII}). Note that we can safely assume that both sets, $\mathcal{A}$ and $\mathcal{M}$, are not empty (if both are empty the theorem follows and if only one of them is empty, then one of the proof parts below would lead to a contradiction). 


\subsubsection{Proof of Theorem~\ref{thm:correctness_absolute}, Part I.}\label{app:proof_correcteness_absolute_partI}
Consider a feasible solution $(\mathbf{x},\mathbf{y},\mathbf{z})\in\mathcal{A}$. Recall that variables $\mathbf{y}$ and $\mathbf{z}$ are defined only for families of size at least 2 {in schools where at least two of their members apply}. Construct a matching $\mu$ as follows: For every feasible pair $(s,c)\in\pairs$, set $\mu(s)=c$ if and only if $x_{s,c}=1$. Since $\mathbf{x}\in\mathcal{P}$, then $\mu$ is a matching. For every student $s\in\students$ and school $c\in\schools$ we set values $z^\mu(s,c)\in\{0,1\}$ according to the rules given after Definition~\ref{def: provider of contingent priority} for matching $\mu$. 
Now, we show that $\mu$ is contingent stable under absolute priorities.

First, let us prove that no student has contingent justified-envy. To find a contradiction, suppose that there exists a student $s$ that has contingent justified-envy towards student $a$  matched to school $c$ (in the same level), i.e., $\mu(a)=c$ (i.e., $x_{a,c}=1$), $c\succ_s\mu(s)$ (i.e., $x_{s,c'}=0$ for $c'\succeq_s c$) and $s\succ^\mu_c a$. Let us assume that $a\notin f(s)$ as the other case is straightforward. Constraint~\eqref{eq: constraint stability abs indep} for pair $(s,c)$ corresponds to
\begin{equation}\label{eq:7a_proof_absolute}
q_{c}^{\ell(s)} \leq \sum_{\substack{a\in \students^{\ell(s)}:\\ a\succ_c s}} x_{a, c} +  \sum_{\substack{f\in\mathcal{F}:\\ |f|\geq 2}}\sum_{\substack{\lrl{a,a'}\subseteq f: \\ a\in  \students^{\ell(s)}, \\ a\prec_c s } } y_{a',a,c}+\sum_{\substack{a\in  \students^{\ell(s):} \\ a\prec_c s}}z_{a,c}.
\end{equation}
Since $s\succ^\mu_c a$, then either (a) $g^\mu(s,c)<g^\mu(a,c)$ or (b) $g^\mu(s,c)=g^\mu(a,c)$ and  $s\succ_c a$ (i.e., $p_{s,c}<p_{a,c}$).\footnote{Recall that for absolute priorities $\lotteries^\mu=\lotteries$.} We study each case separately:
\begin{enumerate}
    \item [(a)] Suppose that $g^\mu(s,c)<g^\mu(a,c)$. Then, we know that $s$ has a sibling $s'$ that is an effective priority provider (i.e., $z^\mu(s',c)=1$) and $a$ either has no siblings {assigned to $c$} (i.e., $\lra{f(a)\cap \mu(c)} \leq 1$) or has no effective priority provider ($z^\mu(a',c)=0$ for all $a'\in f(a)\setminus \lrl{a}$). 
    This implies that $x_{s,c'}=0$ for $c'\succeq_s c$, $x_{s',c} = 1$, $x_{a,c}=1$, {$z_{a,c} = 0$ and $y_{a',a,c} = 0$ for all $a' \in f(a)\setminus \lrl{a}$}, so constraint~\eqref{eq: tie-breaking contingent abs indep} leads to $2\leq 1$, contradicting the feasibility of $(\mathbf{x},\mathbf{y},\mathbf{z})$.
    \item [(b)] Assume $g^\mu(s,c)=g^\mu(a,c)$ and $s\succ_c a$. Then, we have two sub-cases. First, consider the case when both students have no effective priority provider in their families, i.e., $g^\mu(s,c)=g^\mu(a,c) = |\groups|+1$. Since $s\succ_c a$, then $x_{a,c} =1$ is not accounted for in any sum in~\eqref{eq:7a_proof_absolute}, which is a contradiction because it would violate the capacity constraint of school $c$ at level $\ell(s)$. Second, consider the case when both have an effective priority provider (i.e., $g^\mu(s,c)=g^\mu(a,c)=|\groups|$),\footnote{Recall Remark~\ref{remark: single initial group}.} say $s'$ and $a'$. Since $x_{s',c}=x_{a,c}=1$ and $x_{s,c'}=0$ for $c'\succeq_s c$ then constraint~\eqref{eq: tie-breaking contingent abs indep} would result in $2\leq 1$ because $s\succ_s a$.
\end{enumerate}
Now, we show that $\mu$ is non-wasteful. Suppose the contrary, namely there exists a student $s$ and a school $c$ such that $c\succ_s\mu(s)$ and $|\mu^{\ell(s)}(c)|<q^{\ell(s)}_c$. From $c\succ_s\mu(s)$, we can conclude that constraint~\eqref{eq: constraint stability abs indep} for pair $(s,c)$ results in constraint~\eqref{eq:7a_proof_absolute} which contradicts $|\mu^{\ell(s)}(c)|<q^{\ell(s)}_c$. This finishes the first part of the proof of Theorem~\ref{thm:correctness_absolute}.\hfill\Halmos

\subsubsection{Proof of Theorem~\ref{thm:correctness_absolute}, Part II.}\label{app:proof_correcteness_absolute_partII}
Consider a contingent stable matching under absolute priorities $\mu\in\mathcal{M}$. We now construct a point $(\mathbf{x},\mathbf{y},\mathbf{z})$ and show that it is in $\mathcal{A}$. For $\mathbf{x}$, we set values $x_{s,c}=1$ if and only if $\mu(s) = c$ for every $(s,c)\in\pairs$. For every effective priority provider $s$ in school $c$ {with at least one of their siblings assigned to $c$} (i.e., $z^\mu(s,c)=1$ and $\lra{f(s)\cap \mu(c)} \geq 2$), we set $z_{s,c}=1$. Finally, for every sibling $s'\in f(s)\setminus\{s\}$ such that $\mu(s')=c$, we set $y_{s,s',c}=1$. Clearly $\mathbf{x}\in\mathcal{P}$ because $\mu$ is a matching, $\mathbf{z}\in\mathcal{R}(\mathbf{x})$ because $z_{s,c}=1$ only for effective priority providers (satisfying the conditions in Definition~\ref{def: provider of contingent priority}) {with at least one other sibling assigned to $c$}, and one can easily check that $\mathbf{y}\in\mathcal{Q}(\mathbf{x},\mathbf{z})$. We now prove that $(\mathbf{x},\mathbf{y},\mathbf{z})$ satisfies constraints~\eqref{eq: constraint stability abs indep} and~\eqref{eq: tie-breaking contingent abs indep}.

First, suppose that there exists a pair $(s,c)\in\pairs$ such that constraint~\eqref{eq: constraint stability abs indep} is not satisfied, i.e.,
\begin{equation}\label{eq:7a_proof_absolute_part2}
q_{c}^{\ell(s)}\cdot\Bigg(1- \sum_{\substack{c'\in\schools:\\c'\succeq_s c}} x_{s,c'}\Bigg) > \sum_{\substack{a\in \students^{\ell(s)}:\\ a\succ_c s}} x_{a, c} +  \sum_{\substack{f\in\mathcal{F}:\\ |f|\geq 2}}\sum_{\substack{\lrl{a,a'}\subseteq f: \\ a\in  \students^{\ell(s)}, \\ a\prec_c s } } y_{a',a,c}+\sum_{\substack{a\in  \students^{\ell(s):} \\ a\prec_c s}}z_{a,c}.
\end{equation}
Since variables are non-negative, \eqref{eq:7a_proof_absolute_part2} implies that $x_{s,c'}=0$ for $c'\succeq_sc$ (i.e., $c\succ_s\mu(s)$) and the right hand side sums strictly less than $q_{c}^{\ell(s)}$. There are two cases: First, if $|\mu^{\ell(s)}(c)|<q_{c}^{\ell(s)}$, then matching $\mu$ would be wasteful (because $c\succ_s\mu(s)$) which contradicts its contingent stability. Second, if $|\mu^{\ell(s)}(c)|=q_{c}^{\ell(s)}$, \eqref{eq:7a_proof_absolute_part2} implies that the right-hand side does not count all the students in school $c$. In other words, there is a student $a$ such that $x_{a,c}=1$ and $a\prec_c s$ who is not receiving priority and is not an effective priority provider {with another sibling assigned to $c$}. That means $g^\mu(s,c) = g^\mu(a,c)$ and $p^\mu_{a,c} = p_{a,c} > p_{s,c} = p^{\mu}(s,c)$, in other words $s$ has justified envy towards $a$ which is a contradiction to the contingent stability of $\mu$.

We now focus on showing that constraints~\eqref{eq: tie-breaking contingent abs indep} are satisfied. Suppose there exist a school $c\in\C$, family $|f|\geq 2$, siblings $\{s,s'\}\subseteq f$ and student $a\in\students^{\ell(s)}\setminus f$ such that constraint~\eqref{eq: tie-breaking contingent abs indep} is not satisfied, i.e.,
\begin{equation}\label{eq:7b_proof_absolute_part2}
x_{s',c} + \Bigg(1-\sum_{\substack{c'\in\schools:\\c'\succeq_s c}} x_{s,c'}\Bigg) > 2 - x_{a,c} + \indicator{a\succ_c s}\cdot\Bigg(z_{a,c} + \sum_{a'\in f(a)\setminus\{a\} } y_{a',a,c}\Bigg).
\end{equation}
We have several cases: 
\begin{enumerate}
    \item[(a)] If $x_{s',c}\in\{0,1\}$ and $x_{s,c'}=1$ for some $c'\succeq_sc$, then~\eqref{eq:7b_proof_absolute_part2} would result in  
    \[
    x_{s',c} > 2-x_{a,c}  + \indicator{a\succ_c s}\cdot\Bigg(z_{a,c} + \sum_{a'\in f(a)\setminus\{a\} } y_{a',a,c}\Bigg),
    \]
    which is not {feasible} under any variable assignment.
    \item[(b)] If $x_{s',c}=0$ and $x_{s,c'}\in \lrl{0,1}$, we achieve the same conclusion as in (a).
    \item[(c)] If $x_{s',c}=1$ and $x_{s,c'}=0$ for all $c'\succeq_s c$,  Lemma~\ref{lemma:existence_of_provider} and $x_{s',c}=1$ imply that there is an effective priority provider in family $f(s)=f(s')$ matched to $c$.
    We obtain the following
    \[
    x_{a,c} > \indicator{a\succ_c s}\cdot\Bigg(z_{a,c} + \sum_{a'\in f(a)\setminus\{a\} } y_{a',a,c}\Bigg).
    \]
    Since $x_{a,c}\in\{0,1\}$, 
    the right-hand side must be equal to  0 {to ensure that the constraint is feasible}, 
    {so we must have that either $a \prec_c s$ or the term in parenthesis is equal to zero}: 
    {\begin{itemize}
        \item If $a \prec_c s$, then we know that $g^\mu(s,c) \leq g^\mu(a,c)$ (since $s$ has an effective priority provider) and $p(s,c) < p(a,c)$ (since $a\prec_c s$), so we know that $s\succ_c^\mu a$. Hence, $s$ has justified-envy towards $a$, which contradicts the contingent stability of $\mu$.
        \item If $z_{a,c} + \sum_{a'\in f(a)\setminus\{a\} } y_{a',a,c} = 0$, then we know that $g^\mu(s,c) < g^\mu(a,c)$, which in turn implies that $s \succ_c^\mu a$, resulting in the same conclusion as before.
    \end{itemize}    }
    
\end{enumerate}

This finishes the second, and final, part of the proof of Theorem~\ref{thm:correctness_absolute}.\hfill\Halmos


\subsection{Partial Priorities}\label{app:proof_correcteness_partial}
As we did for absolute priorities in Appendix~\ref{app:proof_correcteness_absolute}, we divide the analysis in two parts. Let us denote by $\mathcal{A}$ the set of points $(\mathbf{x},\mathbf{z},\mathbf{y})$ that satisfy~\eqref{eq:partial_formulation} and by $\mathcal{M}$ the set of contingent stable matchings under partial priorities.

\subsubsection{Proof of Theorem~\ref{thm:correctness_partial}, Part I.}\label{app:proof_correctness_partial_partI}
The proof is along the same lines than the proof for absolute priorities, but for completeness we include every argument. Consider a feasible solution $(\mathbf{x},\mathbf{y},\mathbf{z})\in\mathcal{A}$. Construct a matching $\mu$ as follows: For every feasible pair $(s,c)\in\pairs$, set $\mu(s)=c$ if and only if $x_{s,c}=1$. Since $\mathbf{x}\in\mathcal{P}$, then $\mu$ is a matching. For every student $s\in\students$ and school $c\in\schools$ we set values $z^\mu(s,c)\in\{0,1\}$ according to the rules given after Definition~\ref{def: provider of contingent priority} for matching $\mu$. 
Now we show that $\mu$ is contingent stable under partial priorities.

First, let us prove that no student has contingent justified-envy. To find a contradiction, suppose that there exists a student $s$ that has contingent justified-envy towards student $a$  matched to school $c$ (in the same level) which means that: $\mu(a)=c$ (i.e., $x_{a,c}=1$), $c\succ_s\mu(s)$ (i.e., $x_{s,c'}=0$ for $c'\succeq_s c$) and $s\succ^\mu_c a$. Constraint~\eqref{eq: constraint stability par indep} for pair $(s,c)$ corresponds to
\begin{equation}\label{eq:8a_proof_priority}
q_{c}^{\ell(s)} \leq \sum_{\substack{a\in \students^{\ell(s)}:\\ a\succ_c s}} x_{a, c} +  \sum_{\substack{f\in\mathcal{F}:\\ |f|\geq 2}} \sum_{\substack{\lrl{a,a'}\subseteq f: \\ a\in  \students^{\ell(s)} \\ a\prec_c \; s\; \prec_c \; a' }}y_{a',a,c} .
\end{equation}
Since $s\succ^\mu_c a$, then 
$g^\mu(s,c)=g^\mu(a,c)$ and   $p^\mu_{s,c}<p^\mu_{a,c}$.\footnote{Recall that under partial priorities a student receiving contingent priority obtains the most favorable random tie-breaker between the provider and the receiver.} 
First, note that $s$ cannot be an effective priority provider in $c$ because $x_{s,c} = 0$. 
Therefore, we have the following cases: 
\begin{itemize}
    \item Student $a$ is an effective priority provider. Then, $p^\mu_{s,c} <p^\mu_{a,c} = p_{a,c}-\epsilon$ which means either: $s\succ_c a$ or $s$ has an effective priority provider in $c$, say $s'$, such that $s\wedge s'\succ_c a$. First, if $s\succ_c a$ then we would have a capacity contradiction in school $c$ because the term $x_{a,c}=1$ is not accounted for in~\eqref{eq:8a_proof_priority}. If $s'\succ_c a\succ_c s$, then let us analyze constraint~\eqref{eq: tie breaking contingent par indep}. Since $x_{s,c'} =0$ for $c'\succeq_s c$, $x_{s',c}=1$, $x_{a,c}=1$ and $s'\succ_c a\succ_c s$ we obtain \(
    2 \leq 2 - x_{a,c} + \sum_{a'\in f(a)\setminus \{ a\} } y_{a',a,c}\cdot\indicator{a'\wedge a\succ_c s'\wedge s},
    \)
    which is not satisfied since $x_{a,c}=1$ and $a$ is not a priority receiver so the right-hand side is 1. 
\end{itemize}
For the next cases, we assume that $a$ is not an effective priority provider:
\begin{itemize}
    \item Both $s$ and $a$ are not receiving priority in $c$. In this case, $p^\mu_{s,c}<p^\mu_{a,c}$ is equivalent to $p_{s,c}<p_{a,c}$, i.e., $s\succ_c a$ according to initial orders. This means that $x_{a,c}=1$ is not accounted in any sum term in~\eqref{eq:8a_proof_priority} as $a$ is not receiving or providing priority, which contradicts the capacity of school $c$ in level $\ell(s)$.
    \item Student $s$ is receiving priority (assume from $s'$) and $a$ is not receiving. In this case, $p^\mu_{s,c}<p^\mu_{a,c}=p_{a,c}$ which leads to $s\wedge s'\succ_c a$. We can conclude similarly to the first point.

    \item Both $s$ and $a$ are receiving priority, say from $s'$ and $a'$ respectively. Then, $p^\mu_{s,c}<p^\mu_{a,c}$ means that $s\wedge s'\succ_c a\wedge a'$. Note that this implies $s\wedge s'\succ_c a$ which means that constraint~\eqref{eq: tie breaking contingent par indep} results in  \(
    2 \leq 2 - x_{a,c} + 0,
    \)
    which is not {feasible} because $x_{a,c}=1$.
    
    \item Student $s$ is not receiving priority and $a$ is receiving from $a'$. Then, $p^\mu_{s,c}<p^\mu_{a,c}$ means that $s\succ_c a\wedge a'$ which leads to the same contradiction as above.
\end{itemize}

From the above, we conclude that no students in $\mu$ have justified-envy.

Now, we show that $\mu$ is non-wasteful. Suppose the contrary, namely there exists a student $s$ and a school $c$ such that $c\succ_s\mu(s)$ and $|\mu^{\ell(s)}(c)|<q^{\ell(s)}_c$. From $c\succ_s\mu(s)$, we can conclude that constraint~\eqref{eq: constraint stability par indep} for pair $(s,c)$ results in constraint~\eqref{eq:8a_proof_priority} which contradicts $|\mu^{\ell(s)}(c)|<q^{\ell(s)}_c$. This finishes the first part of the proof of Theorem~\ref{thm:correctness_partial}.\hfill\Halmos

\subsubsection{Proof of Theorem~\ref{thm:correctness_partial}, Part II.}\label{app:proof_correctness_partial_partII}
We proceed in the same way as we did for absolute priorities. Consider a contingent stable matching under partial priorities $\mu\in\mathcal{M}$. We now construct a point $(\mathbf{x},\mathbf{y},\mathbf{z})$ and show that it is in $\mathcal{A}$. For $\mathbf{x}$, we set values $x_{s,c}=1$ if and only if $\mu(s) = c$ for every $(s,c)\in\pairs$. For every effective priority provider $s$ in school $c$ (i.e., $z^\mu(s,c)=1$) {with at least one sibling assigned to $c$ (i.e., $|f(s)\cap \mu(c)|\geq2$)}, we set $z_{s,c}=1$. Finally, for every sibling $s'\in f(s)\setminus\{s\}$ such that $\mu(s')=c$, we set $y_{s,s',c}=1$. Clearly $\mathbf{x}\in\mathcal{P}$ because $\mu$ is a matching, $\mathbf{z}\in\mathcal{R}(\mathbf{x})$ because $z_{s,c}=1$ only for effective priority providers (they satisfy the conditions in Definition~\ref{def: provider of contingent priority}), and one can easily check that $\mathbf{y}\in\mathcal{Q}(\mathbf{x},\mathbf{z})$. We now prove that $(\mathbf{x},\mathbf{y},\mathbf{z})$ satisfies constraints~\eqref{eq: constraint stability par indep} and~\eqref{eq: tie breaking contingent par indep}.

First, suppose that there exists a pair $(s,c)\in\pairs$ such that constraint~\eqref{eq: constraint stability par indep} is not satisfied, i.e.,
\begin{equation}\label{eq:8a_proof_partial_part2}
q_{c}^{\ell(s)}\cdot\Bigg(1- \sum_{\substack{c'\in\schools:\\c'\succeq_s c}} x_{s,c'}\Bigg) > \sum_{\substack{a\in \students^{\ell(s)}:\\ a\succ_c s}} x_{a, c} +  \sum_{\substack{f\in\mathcal{F}:\\ |f|\geq 2}} \sum_{\substack{\lrl{a,a'}\subseteq f: \\ a\in  \students^{\ell(s)} \\ a\prec_c \; s\; \prec_c \; a' }}y_{a',a,c} .
\end{equation}
Since variables are non-negative, \eqref{eq:8a_proof_partial_part2} implies that $x_{s,c'}=0$ for $c'\succeq_sc$ (i.e., $c\succ_s\mu(s)$) and the right hand side sums strictly less than $q_{c}^{\ell(s)}$. There are two cases: First, if $|\mu^{\ell(s)}(c)|<q_{c}^{\ell(s)}$, then matching $\mu$ would be wasteful (because $c\succ_s\mu(s)$) which contradicts its contingent stability. Second, if $|\mu^{\ell(s)}(c)|=q_{c}^{\ell(s)}$ then \eqref{eq:8a_proof_partial_part2} implies that the right-hand side does not count all the students in school $c$. In other words, there is a student $a$ such that $x_{a,c}=1$ and $s\succ_c a$ (i.e., $p_{a,c}>p_{s,c}$) who is not receiving priority. 
This means that $g^\mu(s,c) = g^\mu(a,c)$ and $p^\mu_{a,c} = p_{a,c} > p_{s,c} = p^{\mu}_{s,c}$. Therefore, we conclude that $s$ has justified envy towards $a$ (recall that $c\succ_s\mu(s)$) which is a contradiction to the contingent stability of $\mu$.

We now focus on showing that constraints~\eqref{eq: tie breaking contingent par indep} are satisfied. Suppose there exist a school $c\in\C$, family $|f|\geq 2$, siblings $\{s,s'\}\subseteq f$ and student $a\in\students^{\ell(s)}\setminus f$ such that constraint~\eqref{eq: tie breaking contingent par indep} is not satisfied, i.e.,
\begin{equation}\label{eq:8b_proof_partial_part2}
x_{s',c} + \Bigg(1- \sum_{\substack{c'\in\schools:\\c'\succeq_s c}} x_{s,c'}\Bigg) > 2 - x_{a,c}\cdot \indicator{a \prec_c s \wedge s'} +\sum_{a'\in f(a)\setminus\{a\} } y_{a',a,c}\cdot \indicator{a'\wedge a\; \succ_c s'\wedge s}.
\end{equation}
We have several cases: 
\begin{enumerate}
    \item[(a)] If $x_{s',c}\in\{0,1\}$ and $x_{s,c'}=1$ for some $c'\succeq_sc$, then~\eqref{eq:8a_proof_partial_part2} would result in  
    \[
x_{s',c}  > 2 - x_{a,c}\cdot \indicator{a \prec_c s \wedge s'} +\sum_{a'\in f(a)\setminus\{a\} } y_{a',a,c}\cdot \indicator{a'\wedge a\; \succ_c s'\wedge s},
    \]
    \item[(b)] If $x_{s',c}=0$ and $x_{s,c'}\in \lrl{0,1}$, we obtain the same conclusion as in (a).
    \item[(c)] If $x_{s',c}=1$ and $x_{s,c'}=0$ for all $c'\succeq_s c$, we know from Lemma~\ref{lemma:existence_of_provider} that $s$ has at least one effective priority provider. Suppose that $s'$ is $s$'s effective priority provider. Therefore, \eqref{eq:8b_proof_partial_part2} results in
    \[
x_{a,c}\cdot \indicator{a \prec_c s \wedge s'} > \sum_{a'\in f(a)\setminus\{a\} } y_{a',a,c}\cdot \indicator{a'\wedge a\; \succ_c s'\wedge s}.
    \]
    Since $x_{a,c}\in\{0,1\}$, this constraint is valid if and only if $x_{a,c} = 1$, $a \prec_c s'\wedge s$ and the right-hand side is 0, i.e., $a$ is not a priority receiver or $\indicator{a'\wedge a\; \succ_c s'\wedge s}=0$. Either one implies that 
    $p^\mu_{s,c}<p^\mu_{a,c}=p_{a,c}$, i.e., $s\succ^\mu_c a$, which leads to a contradiction because $s$ would have contingent justified-envy towards $a$ which is not possible due to the contingent stability of $\mu$.
\end{enumerate}
This finishes the second, and final, part of the proof of Theorem~\ref{thm:correctness_partial}.\hfill\Halmos

\subsubsection{Hard vs. Soft Absolute Priorities.}\label{app:hard_vs_soft_absolute}

\begin{proof}{Proof of Proposition~\ref{prop:soft_absolute}.}
Consider Formulation~\eqref{eq: absolute formulation} with \eqref{eq: tie-breaking contingent abs indep} replaced by \eqref{eq: tie-breaking contingent abs indep soft}. If we set $\mathbf{z} = \mathbf{0}$, then $\mathbf{y}=\mathbf{0}$ and constraints~\eqref{eq: tie-breaking contingent abs indep soft} are always satisfied, no matter $\mathbf{x}$. Then, constraints~\eqref{eq: constraint stability abs indep} reduce to constraints~\eqref{eq: constraint stability} which captures initial stability.
On the other hand, to capture contingent stable matchings with hard absolute priorities we set values $\mathbf{z}$ as we did in Appendix~\ref{app:proof_correcteness_absolute_partII}.\hfill\Halmos 
\end{proof}
\section{Appendix to Section~\ref{sec: application}: Additional Results}\label{app: additional results}

    \begin{table}[htbp!]
        \caption{Results: Single Tie-breaking at Individual Level (STB)}\label{tab: results stb}
        \centerline{\scalebox{0.85}{\begin{tabular}{lcccccccccccccc}
            \toprule
            & \multicolumn{7}{c}{} & \multicolumn{6}{c}{Separated} \\
            \cmidrule(lr){9-14}
            &  & \multicolumn{2}{c}{Top Pref.} & \multicolumn{2}{c}{Unassigned} & \multicolumn{2}{c}{Together} & \multicolumn{2}{c}{None} & \multicolumn{2}{c}{One} & \multicolumn{2}{c}{Both} \\
            \cmidrule(lr){3-4}\cmidrule(lr){5-6}\cmidrule(lr){7-8}\cmidrule(lr){9-10}\cmidrule(lr){11-12}\cmidrule(lr){13-14}
            & Solved & Mean & SE & Mean & SE & Mean & SE & Mean & SE & Mean & SE & Mean & SE \\
            \midrule
         Absolute - Hard & 27 & 2699.93 & 2.61 & 881.44 & 2.36 & 627.85 & 3.03 & 77.41 & 1.41 & 67.15 & 1.45 & 87.63 & 1.71\\
         Absolute - Soft & 100 & 2730.25 & 1.31 & 880.69 & 1.18 & 486.06 & 1.59 & 76.0 & 0.74 & 126.08 & 1.24 & 172.21 & 1.45\\
         Partial - Hard & 90 & 2715.51 & 1.4 & 886.17 & 1.28 & 488.29 & 1.37 & 76.76 & 0.76 & 126.24 & 1.18 & 173.91 & 1.36\\
         Partial - Soft & 100 & 2725.9 & 1.25 & 885.01 & 1.19 & 417.52 & 1.38 & 76.4 & 0.69 & 154.35 & 1.35 & 218.42 & 1.42\\
         FOSM & 100 & 2717.4 & 1.26 & 885.73 & 1.16 & 351.08 & 1.42 & 76.67 & 0.74 & 180.33 & 1.35 & 263.08 & 1.51\\
         SOSM & 100 & 2717.4 & 1.26 & 885.73 & 1.16 & 351.08 & 1.42 & 76.67 & 0.74 & 180.33 & 1.35 & 263.08 & 1.51\\
         Descending & 100 & 2704.64 & 1.37 & 882.86 & 1.2 & 496.2 & 1.46 & 75.11 & 0.76 & 120.92 & 1.15 & 171.98 & 1.37\\
         Ascending & 100 & 2707.57 & 1.38 & 885.27 & 1.17 & 483.05 & 1.6 & 77.99 & 0.76 & 131.25 & 1.16 & 175.99 & 1.3\\
            \bottomrule
        \end{tabular}}}
    \end{table}
    
    \begin{table}[htbp]
        \caption{Results: Single Tie-breaking at Family Level (STB-F)}\label{tab: results stbf}
        \centerline{\scalebox{0.85}{\begin{tabular}{lcccccccccccccc}
            \toprule
            & \multicolumn{7}{c}{} & \multicolumn{6}{c}{Separated} \\
            \cmidrule(lr){9-14}
            &  & \multicolumn{2}{c}{Top Pref.} & \multicolumn{2}{c}{Unassigned} & \multicolumn{2}{c}{Together} & \multicolumn{2}{c}{None} & \multicolumn{2}{c}{One} & \multicolumn{2}{c}{Both} \\
            \cmidrule(lr){3-4}\cmidrule(lr){5-6}\cmidrule(lr){7-8}\cmidrule(lr){9-10}\cmidrule(lr){11-12}\cmidrule(lr){13-14}
            & Solved & Mean & SE & Mean & SE & Mean & SE & Mean & SE & Mean & SE & Mean & SE \\
            \midrule
         Absolute - Hard & 23 & 2706.35 & 3.72 & 876.96 & 2.55 & 600.96 & 2.37 & 92.48 & 1.62 & 76.17 & 1.14 & 94.17 & 1.95\\
         Absolute - Soft & 100 & 2726.5 & 1.47 & 879.59 & 1.2 & 502.77 & 1.28 & 95.84 & 0.88 & 113.06 & 1.01 & 150.8 & 1.25\\
         Partial - Hard & 100 & 2717.12 & 1.52 & 883.21 & 1.15 & 417.76 & 1.28 & 98.9 & 0.91 & 138.93 & 1.01 & 214.37 & 1.25\\
         Partial - Soft & 100 & 2717.12 & 1.52 & 883.21 & 1.15 & 417.76 & 1.28 & 98.9 & 0.91 & 138.93 & 1.01 & 214.37 & 1.25\\
         FOSM & 100 & 2717.12 & 1.52 & 883.21 & 1.15 & 417.76 & 1.28 & 98.9 & 0.91 & 138.93 & 1.01 & 214.37 & 1.25\\
         SOSM & 100 & 2717.12 & 1.52 & 883.21 & 1.15 & 417.76 & 1.28 & 98.9 & 0.91 & 138.93 & 1.01 & 214.37 & 1.25\\
         Descending & 100 & 2708.61 & 1.52 & 880.34 & 1.18 & 517.55 & 1.13 & 94.66 & 0.82 & 103.38 & 0.99 & 149.03 & 1.06\\
         Ascending & 100 & 2710.48 & 1.47 & 883.1 & 1.24 & 504.84 & 1.29 & 98.62 & 0.88 & 109.52 & 0.85 & 154.19 & 1.27\\
            \bottomrule
        \end{tabular}}}
    \end{table}
    
    \begin{table}[htpb]
        \caption{Results: Multiple Tie-breaking at Individual Level (MTB)}\label{tab: results mtb}
        \centerline{\scalebox{0.85}{\begin{tabular}{lcccccccccccccc}
            \toprule
            & \multicolumn{7}{c}{} & \multicolumn{6}{c}{Separated} \\
            \cmidrule(lr){9-14}
            &  & \multicolumn{2}{c}{Top Pref.} & \multicolumn{2}{c}{Unassigned} & \multicolumn{2}{c}{Together} & \multicolumn{2}{c}{None} & \multicolumn{2}{c}{One} & \multicolumn{2}{c}{Both} \\
            \cmidrule(lr){3-4}\cmidrule(lr){5-6}\cmidrule(lr){7-8}\cmidrule(lr){9-10}\cmidrule(lr){11-12}\cmidrule(lr){13-14}
            & Solved & Mean & SE & Mean & SE & Mean & SE & Mean & SE & Mean & SE & Mean & SE \\
            \midrule
         Absolute - Hard & 29 & 2500.14 & 3.38 & 852.24 & 1.93 & 632.14 & 2.11 & 75.28 & 1.4 & 71.17 & 1.66 & 96.48 & 2.04\\
         Absolute - Soft & 100 & 2513.93 & 1.6 & 849.33 & 1.13 & 490.33 & 1.6 & 75.0 & 0.78 & 128.16 & 1.19 & 180.4 & 1.33\\
         Partial - Hard & 95 & 2487.02 & 1.68 & 851.94 & 1.2 & 487.27 & 1.7 & 76.22 & 0.74 & 126.23 & 1.19 & 191.19 & 1.59\\
         Partial - Soft & 100 & 2496.01 & 1.6 & 849.77 & 1.13 & 415.21 & 1.47 & 75.84 & 0.72 & 154.64 & 1.22 & 236.44 & 1.59\\
         FOSM & 100 & 2481.16 & 1.59 & 849.87 & 1.12 & 348.06 & 1.37 & 76.15 & 0.72 & 173.88 & 1.35 & 286.28 & 1.48\\
         SOSM & 100 & 2481.44 & 1.59 & 849.87 & 1.12 & 348.18 & 1.37 & 76.15 & 0.72 & 173.88 & 1.35 & 286.16 & 1.48\\
         Descending & 100 & 2488.85 & 1.59 & 850.37 & 1.16 & 495.83 & 1.57 & 74.55 & 0.7 & 119.25 & 1.21 & 187.0 & 1.26\\
         Ascending & 100 & 2485.71 & 1.7 & 851.8 & 1.15 & 476.38 & 1.55 & 77.86 & 0.75 & 128.03 & 1.13 & 198.25 & 1.45\\
            \bottomrule
        \end{tabular}}}
    \end{table}
    
    \begin{table}[htpb]
        \caption{Results: Multiple Tie-breaking at Family Level (MTB-F)}\label{tab: results mtbf}
        \centerline{\scalebox{0.85}{\begin{tabular}{lcccccccccccccc}
            \toprule
            & \multicolumn{7}{c}{} & \multicolumn{6}{c}{Separated} \\
            \cmidrule(lr){9-14}
            &  & \multicolumn{2}{c}{Top Pref.} & \multicolumn{2}{c}{Unassigned} & \multicolumn{2}{c}{Together} & \multicolumn{2}{c}{None} & \multicolumn{2}{c}{One} & \multicolumn{2}{c}{Both} \\
            \cmidrule(lr){3-4}\cmidrule(lr){5-6}\cmidrule(lr){7-8}\cmidrule(lr){9-10}\cmidrule(lr){11-12}\cmidrule(lr){13-14}
            & Solved & Mean & SE & Mean & SE & Mean & SE & Mean & SE & Mean & SE & Mean & SE \\
            \midrule
         Absolute - Hard & 37 & 2492.46 & 3.11 & 848.97 & 1.72 & 604.19 & 2.01 & 85.41 & 1.19 & 76.65 & 1.52 & 106.27 & 1.29\\
         Absolute - Soft & 100 & 2504.51 & 1.79 & 847.02 & 1.16 & 513.8 & 1.51 & 87.29 & 0.72 & 114.55 & 1.24 & 157.98 & 1.19\\
         Partial - Hard & 100 & 2486.38 & 1.79 & 849.17 & 1.21 & 427.35 & 1.56 & 88.89 & 0.76 & 146.27 & 1.36 & 219.58 & 1.23\\
         Partial - Soft & 100 & 2486.38 & 1.79 & 849.17 & 1.21 & 427.35 & 1.56 & 88.89 & 0.76 & 146.27 & 1.36 & 219.58 & 1.23\\
         FOSM & 100 & 2486.04 & 1.82 & 849.17 & 1.21 & 427.29 & 1.56 & 88.89 & 0.76 & 146.27 & 1.36 & 219.65 & 1.23\\
         SOSM & 100 & 2486.38 & 1.79 & 849.17 & 1.21 & 427.35 & 1.56 & 88.89 & 0.76 & 146.27 & 1.36 & 219.57 & 1.23\\
         Descending & 100 & 2489.8 & 1.69 & 847.74 & 1.2 & 528.9 & 1.44 & 85.87 & 0.71 & 106.23 & 1.1 & 155.57 & 0.99\\
         Ascending & 100 & 2489.7 & 1.83 & 850.21 & 1.19 & 508.22 & 1.43 & 88.79 & 0.79 & 116.32 & 1.22 & 164.59 & 1.09\\
            \bottomrule
        \end{tabular}}}
    \end{table}

\newpage
\section{Additional Examples}\label{app: additional examples}
\begin{example}\label{ex: no unique priority order}
    Consider an instance with a two families $f = \lrl{f_1, f_2}$, $f' = \lrl{f_1', f_2'}$, two levels $\levels = \lrl{\ell_1,\ell_2}$ with $\students^{\ell_1} = \lrl{f_1, f_1'}$ and $\students^{\ell_2} = \lrl{f_2, f_2'}$ and a single school $c$ offering one seat in each level. Moreover, suppose that every student prefers $c$ over being unassigned, every student belongs to the same initial group (i.e., \(g(s,c) = 1\) for all $s\in \students$), and that the initial vector of random tie-breakers is such that $p_{f_1,c} < p_{f_2',c} < p_{f_1',c} < p_{f_2,c} $. Then, consider the matchings   
    \[
    \mu = \lrl{
            (f_1, c), (f_2, c), 
            (f_1', \emptyset), (f_2', \emptyset)
        }
    \]
    and 
    \[
        \mu' = \lrl{
            (f_1, \emptyset), (f_2, \emptyset), 
            (f_1', c), (f_2', c)
        }.
    \]
    Neither $\mu$ nor $\mu'$ are initially stable (i.e., using the {initial} priority order $\succ_c$ induced by the random tie-breakers and omitting contingent priorities). However, if the school considers {contingent} priorities as an additional group that has higher priority than students with no siblings, then $g^\mu(f_2,c) < g^\mu(f_2',c)$ and, thus, $f_2 \succ_c^\mu f_2'$. Similarly, $g^{\mu'}(f_1',c) < g^{\mu'}(f_1,c)$ and, thus, $f_1' \succ_c^{\mu'} f_1$. As a result, both $\mu$ and $\mu'$ would be contingent stable given the contingent priority orders $\succ_c^\mu$ and $\succ_c^{\mu'}$, respectively. 
    \hfill \(\square\)
\end{example}

    \begin{example}\label{ex: provider of priority}
        Suppose there are two schools $\schools  = \lrl{c_1, c_2}$, each offering one seat in each level $\ell\in \levels = \lrl{\ell_1,\ell_2}$. In addition, suppose there are two families $f = \lrl{f_1, f_2}$, $f' = \lrl{f_1', f_2'}$ and two single child $s_1$, $s_2$, where subscripts denote the level to which the student belongs. Finally, suppose that preferences are such that $f_1, f_2, s_2: c_1 \succ c_2$, $f_1', f_2', s_1: c_2 \succ c_1$, and initial priorities are such that 
        \begin{align*}
            c_1: \; & \; s_2 \succ_{c_1} f_2 \succ_{c_1} f_2' \succ_{c_1} f_1 \succ_{c_1} f_1' \succ_{c_1} s_1 \\
            c_2: \; & \; f_2 \succ_{c_2} f_2' \succ_{c_2} s_2 \succ_{c_2} s_1 \succ_{c_2} f_1' \succ_{c_2} f_1
        \end{align*}
        With no contingent priorities, the unique {initially} stable matching is 
        $$\mu = \lrl{(s_2, c_1), (f_2, c_2), (f_1, c_1), (s_1, c_2), (f_1', \emptyset),(f_2', \emptyset) }.$$
        In contrast, with contingent priorities, the alternative matching 
        $$\mu' = \lrl{(f_1, c_1), (f_2, c_1), (f_1', c_2), (f_2', c_2), (s_1, \emptyset),(s_2, \emptyset) }$$
        is stable as long as $g^{\mu'}(f_2, c_1) < g^{\mu'}(s_2, c_1)$ and $g^{\mu'}(f_1', c_2) < g^{\mu'}(s_1, c_1)$. This condition can hold because $f_1$ may grant contingent priority to $f_2$ in $c_1$, allowing $f_2'$ to be assigned to $c_2$ and, consequently, provide contingent priority to $f_1'$ at that school. However, it is important to note that neither $(f_1', c_2)$ nor $(f_2', c_2)$ belong to any {stable} matching under the initial priority orders $\lrl{\succ_c}_{c\in\schools}$.
    \end{example}

            \begin{example}\label{ex: non-uniqueness}
            Consider an instance with two schools $\schools = \lrl{c_1, c_2}$, three levels $\levels = \lrl{\ell_1,\ell_2,\ell_3}$, two single students $\lrl{s, s'}$ applying to level $\ell_3$, two families $f=\lrl{f_1, f_2}$ and $f' = \lrl{f_1', f_2', f_3'}$, with students $f_1, f_1'$ applying to level $\ell_1$, $f_2, f_2'$ applying to $\ell_2$, and $f_3'$ applying to $\ell_3$. In addition, suppose that preferences are $f_1, f_2, f_1', f_2', f_3': c_1 \succ \emptyset$ and $s, s': c_1 \succ c_2$.            
            Finally, suppose that school $c_1$ offers one seat in each level, school $c_2$ offers two seats in level $\ell_3$ (and zero in the other levels), and that the clearinghouse uses a single tie-breaking rule at the individual level (i.e., $p_{s,c} = p_s$ for all $c\in \schools$) with realized values:
            \(
            p_{s} < p_{s'} < p_{f_1} < p_{f_2'} < p_{f_2} < p_{f_1'} < p_{f_3'}. 
            \)
            In this case, there are two contingent stable matching under absolute priorities:
            \begin{align*}
                &\mu = \lrl{
                    (f_1, c_1), (f_2, c_1), (s,c_1), (s',c_2) 
                    (f_1', \emptyset), (f_2', \emptyset), , (f_3', \emptyset)
                },\\
                &\mu' = \lrl{
                        (f_1, \emptyset), (f_2, \emptyset), (s,c_2), (s',c_2) 
                        (f_1', c_1), (f_2', c_1), (f_3', c_1)
                }.                
            \end{align*}
            These two matchings are weakly optimal for students, i.e., {there is at least one student who strictly prefers $\mu$ over any other contingent stable matching under absolute priorities; similarly for $\mu'$}. 
            Moreover, the set of students matched in each case (and even its cardinality) differs.\hfill \(\square\)
        \end{example}
\section{Extensions}\label{app: extensions}

    \subsection{Additional Groups}\label{app: additional groups}
        We discuss how the model presented in Section~\ref{sec: model}
         can be easily extended to incorporate \emph{secured-enrollment} and \emph{static} siblings priority.
        
        \subsubsection{Secured Enrollment.}
        Let $\enrollees \subseteq \students$ be the subset of students who were enrolled in some school in $\schools$ during the most recent academic year, who can continue in the same school, and who are not seeking to switch schools in the current one. Note that a student $s \in \students \setminus \enrollees$ may also have been enrolled in some school during the most recent academic year; however, we assume that they are seeking to switch schools and would only prefer to be assigned to their current school if they cannot be assigned to a more preferred one.
        
        To capture this, let $\bar{\mu}(s) \in \schools \cup \lrl{\emptyset}$ be the school where student $s \in \students$ was enrolled during the most recent academic year, with $\bar{\mu}(s) = \emptyset$ meaning that student $s$ was not enrolled in any school during that time (e.g., $s$ is applying to an entry-level or was enrolled in a private school that does not belong to $\schools$). Similarly, let $\bar{\mu}(c) \subseteq \students$ be the set of students who were enrolled in school $c \in \schools \cup \lrl{\emptyset}$ during the most recent academic year, and $\bar{\mu}^\ell(c) = \bar{\mu}(c) \cap \students^\ell$ for each $\ell \in \levels$.
        Then, for each student $s \in \enrollees$, we assume that $\bar{\mu}(s) \in \schools$ and that they prefer $\bar{\mu}(s)$ over any other school in $\schools \cup \lrl{\emptyset} \setminus \lrl{\bar{\mu}(s)}$. In contrast, for each student $s \in \students \setminus \enrollees$, we assume that they strictly prefer at least one school in $\schools \cup \lrl{\emptyset} \setminus \lrl{\bar{\mu}(s)}$ over $\bar{\mu}(s)$. We formalize this in Assumption~\ref{assumption: preferences}.\footnote{If $\succ_s$ is such that $\emptyset \succ_s \bar{\mu}(s)$, then student $s$ prefers to be unassigned over continue attending school $\bar{\mu}(s)$.}
        
        \begin{assumption}\label{assumption: preferences}
            For each student $s\in \enrollees$, $\bar{\mu}(s) \succ_s c$ for all $c\in \schools \cup \lrl{\emptyset} \setminus \lrl{\bar{\mu}(s)}$. In contrast, for each student $s\in \students \setminus \enrollees$, there exists at least one school  $\schools \cup \lrl{\emptyset} \setminus \lrl{\bar{\mu}(s)}$ such that $c \succ_s \bar{\mu}(s)$.
        \end{assumption}
        
        Another desirable property in any school choice system is that students who are currently enrolled have their seats secured and, consequently, cannot be displaced by other applicants unless assigned to a more preferred school. We refer to this property as \emph{secured enrollment}.
        
        \begin{definition}[Secured enrollment]
            A matching $\mu$ satisfies \emph{secured enrollment} if $\mu(s) \succeq_s \bar{\mu}(s)$ for all $s\in \students$.
        \end{definition}
        
        In the remainder of this section, we assume that the clearinghouse aims to find a stable matching that satisfies \emph{secured enrollment}. As a result, we assume that (i) each school $c$ offers enough seats at each level to accommodate all students who have secured enrollment and that (ii) each student $s$ with $\bar{\mu}(s) = c$ has higher priority than any student $s'$ with $\bar{\mu}(s') \neq c$, i.e., $g(s,c) = 1$ if and only if $\bar{\mu}(s) = c$. We formalize this in Assumption~\ref{assumption: secured enrollment}.
        
        \begin{assumption}\label{assumption: secured enrollment}
            For each school $c\in \schools$ and level $\ell\in \levels$, $q_c^\ell \geq \lra{\bar{\mu}^\ell(c)} $. Moreover, $g(s,c) = 1$ if and only if $\bar{\mu}(s) = c$ and, thus, $s \succ_c s'$ for all $(s,s') \in \bar{\mu}^\ell(c) \times \lrp{\students^\ell \setminus \bar{\mu}(c)}$.
        \end{assumption}
        
        \subsubsection{Static Priorities.}
        As discussed in Section~\ref{sec: introduction}, some school districts consider \emph{static} sibling priorities, whereby students get prioritized in schools where they have a sibling who was enrolled during the most recent academic year and is not seeking to switch schools.\footnote{When clear from the context, we will remove the word ``sibling'' from \emph{static} (\emph{contingent}, resp.) sibling priorities and simply refer to it as \emph{static} (\emph{contingent}, resp.) priorities. } 
        
        \begin{definition}[Static Siblings Priority]\label{def: static priority}
            A student $s\in \students$ has \emph{static} siblings priority in school $c\in \schools$ if $c \succ_s \bar{\mu}(s)$ and $s$ has a sibling $s'\in f(s) \cap \enrollees$ such that $\bar{\mu}(s') = c$.
        \end{definition}
        
        If there are no contingent priorities, incorporating secured enrollment and static priorities is straightforward, as the clearinghouse can preprocess the instance ensuring that (i) $g(s,c) = 1$ if and only if $\bar{\mu}(s) = c$, (ii) $g(s,c) = 2$ if and only if $\exists s'\in f(s) \setminus \lrl{s}$ such that $\bar{\mu}(s') = c$, and (iii) $g(s,c) = 3$ otherwise. Combined with a vector of random tie-breakers $\lotteries$, these priority groups define a (unique) priority order $\succ_c$ for each school $c$, such that all students with secured enrollment have the highest priority, followed by students with static siblings priority, and finally, students with no priority. Then, the clearinghouse can proceed with the standard deferred acceptance algorithm~\cite{gale1962college} to find a stable matching, as suggested below by Corollary~\ref{corollary: static priorities}.
        Note that Assumptions~\ref{assumption: preferences} and~\ref{assumption: secured enrollment} ensure that all students with secured enrollment in $c$ and who can continue in $c$ and are not seeking to switch schools will get assigned to $c$ for sure.
        
        \begin{corollary}[\cite{gale1962college}]\label{corollary: static priorities}
            If school priorities are given by the combination of \emph{static} priorities and random tie-breakers (with no contingent priorities), then a stable matching exists.
        \end{corollary}
        
        The key distinction between \emph{static} and \emph{contingent} priorities is that the former depends on $\bar{\mu}$---which is invariant and known before the assignment---while the latter depends on the assignment $\mu$. 
        Given that students assigned to some school may decide not to enroll and, thus, the priority given may not be effective, it is natural to assume that students with static priorities prevail over students with contingent priorities. Indeed, this is the case in the Chilean school choice system, where siblings with static priority have the highest priority, and then students with contingent priority are considered only if there are vacancies left. We formalize this in Assumption~\ref{ass: static priority over contingent priority}.

        \begin{assumption}\label{ass: static priority over contingent priority}
            Students with static priority have a higher priority than students with contingent priority.
        \end{assumption}
        
        The formulations provided in Section~\ref{sec: formulations} can be easily extended to account for static priorities under Assumption~\ref{ass: static priority over contingent priority}. In particular, the formulation for absolute priorities in~\eqref{eq: absolute formulation} can be updated to account for more priority groups as follows:

        {
            \begin{subequations}\label{eq: absolute formulation with static}
                \begin{align}
                     \min & \quad \sum_{(s,c)\in \pairs} r_{s,c}\cdot x_{s,c} \notag \\
                    s.t.  
                     & \quad q_{c}^{\ell(s)} \cdot\Bigg(1- \sum_{\substack{c'\in\schools:\\c'\succeq_s c}} x_{s,c'}\Bigg) \leq \sum_{\substack{a\in \students^{\ell(s)}:\\ a\succ_c s}} x_{a, c}, \hspace{6em} \forall (s,c) \in \pairs, \; g(s,c) < \lra{\groups}
                    \label{eq: constraint stability abs indep with static 2} \\
                    & \quad q_{c}^{\ell(s)} \cdot\Bigg(1- \sum_{\substack{c'\in\schools:\\c'\succeq_s c}} x_{s,c'}\Bigg) \leq \sum_{\substack{a\in \students^{\ell(s)}:\\ a\succ_c s}} x_{a, c}
                    + \sum_{\substack{f\in\mathcal{F}:\\ |f|\geq 2}}\sum_{\substack{\lrl{a,a'}\subseteq f: \\ a\in  \students^{\ell(s)}, \\ a\prec_c s } } y_{a',a,c}+\sum_{\substack{a\in  \students^{\ell(s):} \\ a\prec_c s}}z_{a,c} \notag \\
                    & \hspace{22em} \forall (s,c) \in \pairs, \; g(s,c) = \lra{\groups} \label{eq: constraint stability abs indep with static 1} \\                  
                    & \quad x_{s',c} + \Bigg(1-\sum_{\substack{c'\in\schools:\\c'\succeq_s c}} x_{s,c'}\Bigg) \leq 2 - x_{a,c} + \indicator{a\succ_c s}\cdot\Bigg(z_{a,c} + \sum_{a'\in f(a)\setminus\{a\} } y_{a',a,c}\Bigg), \notag\\ &\hspace{8.5em} \forall c\in \schools, f\in\cF, \{s,s'\}\subseteq f, a\in \students^{\ell(s)}\setminus f,\; g(s,c) = g(a,c) = \lra{\groups} \label{eq: tie-breaking contingent abs indep with static}\\                      
                    &\quad \mathbf{x}\in \cP, \; \mathbf{z} \in \cR(\mathbf{x}),\; \mathbf{y} \in \cQ(\mathbf{x}, \mathbf{z}). 
                \end{align}
            \end{subequations}
            }

        Relative to~\eqref{eq: absolute formulation}, there two changes worth noticing. On the one hand, since students with contingent priority can only displace students with no priority, we update the constraints~\eqref{eq: constraint stability abs indep with static 1} and~\eqref{eq: tie-breaking contingent abs indep with static} to only consider students who initially have no priority, i.e., $g(s,c) = \lra{\groups}$. On the other hand, we know that students who initially belong to a higher priority group (i.e.., $g(s,c) < \lra{\groups}$), can only be displaced by students who have higher priority according to the initial order $\succ_c$, i.e., by students in a higher priority group ($g(s',c) < g(s,c)$) or students in the same group with a more favorable tie-breaker ($g(s',c) = g(s,c)$ and $p_{s',c} < p_{s,c}$). Thus, we add the set of constraints in~\eqref{eq: constraint stability abs indep with static 2} to ensure that students with higher priority are not displaced by students with contingent priority. Note that this set of constraints is the analogous of the initial stability constraints in~\eqref{eq: constraint stability}.

        \begin{remark}
            We can easily incorporate secured enrollment or any other group (e.g., students in the walk-zone, with parents working at the school, etc.) by redefining the initial groups. Furthermore, we can update the partial formulation and the soft counterparts analogously.
        \end{remark}
        
\section{Further Discussion}\label{app:further_discussion}

\subsection{Extra discussion on how to process levels and others}\label{sec:extra_discussion}

    As proposed in~\cite{Correa_2022}, one option to handle contingent priorities is to define an order in which levels are processed and sequentially solve the assignment of each level using the {student-proposing} variant of the Deferred Acceptance (DA) algorithm. More specifically, the algorithm in~\cite{Correa_2022} starts processing the highest level (i.e., 12th grade). Then, before moving to the next level, the sibling priorities are updated, considering the assignment of the levels already processed. After processing the final level (i.e., Pre-K), this procedure finishes.
    Notice that this heuristic obtains a stable assignment if the preferences of families satisfy \emph{higher-first}, i.e., each family prioritizes the assignment of their oldest member (see Proposition 2 in~\cite{Correa_2022}). However, this is not the case if some families' preferences do not satisfy this condition. In addition, as Example~\ref{ex: order of grades matters} illustrates, the order in which levels are processed matters.
    \begin{example}\label{ex: order of grades matters}
        Consider an instance with two levels \(\ell_1 < \ell_2\), {two schools $c_1$ and $c_2$ with one seat in each level}, one family \(f = \lrl{f_1,f_2}\), and two additional students, $a_1$ and $b_2$. Students \(f_1\) and \(a_1\) apply to level \(\ell_1\),
        and \(f_2\) and \(b_2\) apply to level \(\ell_2\). Finally, we assume that the clearinghouse uses a single tie-breaking rule at the individual level (i.e., $p_{s,c} = p_s$ for all $s$ and $c$) with 
        \(p_{a_1} < p_{f_1} < p_{b_2} < p_{f_2}\),
        and students' preferences are:
        \[
        \begin{aligned}
            c_2 \succ_{f_1} c_1 \quad & \quad c_1 \succ_{f_2} c_2 \\
            c_2  \succ_{a_1} c_1 \quad & \quad c_1  \succ_{b_2} c_2
        \end{aligned}        
        \]

        We observe that, if levels are processed in decreasing order (as in Chile), we obtain the matching \(\mu = \lrl{(f_1, c_2), (a_1, c_1), (f_2, c_2), (b_2, c_1)} \). In contrast, if we process levels in increasing order, we obtain the matching \(\mu' = \lrl{(f_1, c_1), (a_1, c_2), (f_2, c_1), (b_2, c_2)} \). 
        \hfill $\square$
    \end{example}

\subsection{Family-Oriented Formulation}\label{app:baselineFormulation}
    A natural benchmark for comparing our approaches is to find the stable matching that maximizes
    the number of families whose members are assigned to the same school. The following mathematical programming formulation aims to model this baseline:
    \begin{subequations}\label{eq:baseline2}
            \begin{align}
              \max_{\mathbf{t} \in \{0,1\}^{\cF \times \schools},\mathbf{x}\in \cP} \quad & \quad \sum_{f \in \cF} \sum_{c \in \schools} \left(\sum_{s \in f} x_{s,c} - |f| \cdot t_{f,c}\right)  \\
              st. \quad 
              & \quad \text{Constraint} \ \eqref{eq: constraint stability} \nonumber\\
              & \frac{\sum_{s \in f} x_{s,c}}{|f|}\leq t_{f,c} \leq \sum_{s \in f} x_{s,c} \; ,\; \forall f \in \cF, \forall c \in \schools. \label{eq:family_in_c}
            \end{align}
        \end{subequations}
    This formulation is similar to Formulation~\eqref{eq: baseline}. However, in Formulation~\eqref{eq:baseline2}, we have a new binary variable $t_{f,c}$ which is 1 if and only if family $f$ has at least one sibling in school $c$, and zero otherwise; this is enforced through constraint~\eqref{eq:family_in_c}. In addition, in the objective, we maximize the number of family members in the same school.

\end{APPENDICES}


\end{document}